\documentclass[a4paper,preprint,prd,showpacs,preprintnumbers,amsmath,amssymb,floatfix]{revtex4}
\usepackage{amsmath}
\usepackage{graphicx} 
\usepackage{dcolumn} 
\usepackage{bm} 
\usepackage{ulem} 
\usepackage[usenames]{color}
\usepackage{epstopdf}
\usepackage{float}
\usepackage{subcaption} 
\usepackage{multirow}
\usepackage{array}
\usepackage{setspace}
\usepackage{caption}
\usepackage{hyperref}
\usepackage{rotating}
\usepackage{xcolor} 
\usepackage{ulem}

\renewcommand\arraystretch{0.64}

\graphicspath{{fig/}} 

\newcommand{\nc}{\newcommand} 
\nc{\vc}[1]{\mbox{\boldmath $#1$}} 
\nc{\del}{\partial} 
\nc{\bra}{\langle} 
\nc{\ket}{\rangle} 
\nc{\bras}[1]{\langle #1|} 
\nc{\kets}[1]{|#1\rangle} 
\nc{\mapleft}[1]{\smash{\mathop{\,\hbox to 1.5cm{\rightarrowfill}\,}\limits_{#1}}}
\nc{\nn}{\\\nonumber}
\nc{\vs}{\vspace{-0.275cm}}
\nc{\fra}{\frac{1}{2}}
\nc{\mb}{\mathbf}

\begin{document}
	
	\preprint{}
	\title{
	The equation of state for neutron stars with speed of sound constraints via Bayesian inference
	} 
	
	\author{Xieyuan Dong}
	\affiliation{School of Physics, Nankai University, Tianjin 300071,  China}
	\author{Jinniu Hu}
	\email{hujinniu@nankai.edu.cn}
	\affiliation{School of Physics, Nankai University, Tianjin 300071,  China}
	\affiliation{Shenzhen Research Institute of Nankai University, Shenzhen 518083, China}
	\author{Ying Zhang}
	\email{yzhangjcnp@tju.edu.cn}
	\affiliation{Department of Physics, Faculty of Science, Tianjin University,Tianjin 300072, China}
	\author{Hong Shen}
	\affiliation{School of Physics, Nankai University, Tianjin 300071,  China}
	\date{\today} 
	\begin{abstract}
		
	The parametrized equation of state (EOS) of neutron stars is investigated by Bayesian inference method with various constraints from both nuclear physics and modern astronomical observations, where the energy per particle is expanded as a polynomial function of the density. The expansion coefficients correspond to the properties of symmetric nuclear matter and the density dependence of the symmetry energy. The empirical values of the symmetry energy at subsaturation density and the density of crust-core phase transition are considered to limit the low-density behavior of EOS, i.e. $L_{\mathrm{sym}} $, while the speed of sound of neutron star matter and mass-radius observations of millisecond pulsars PSR J0030+0451 and PSR J0740+6620 are adopted to eliminate the high-order expansion coefficients, such as $Q_{\mathrm{sat}}$ and $Q_{\mathrm{sym}} $. Finally, our analysis reveals that the skewness coefficient $Q_{\mathrm{sat}}$ of the energy per nucleon in symmetric nuclear matter (SNM) exhibits the strongest correlation with the speed of sound, constrained to $Q_{\mathrm{sat}} = -69.50_{-31.93}^{+16.52} \, \mathrm{MeV}$, whose uncertainties are much smaller than those of the experiments of heavy-ion collisions. The symmetry energy parameters are determined as follows: slope $L_{\mathrm{sym}} = 34.32_{-11.85}^{+13.66} \, \mathrm{MeV}$, curvature $K_{\mathrm{sym}} = -58.45_{-89.46}^{+88.47} \, \mathrm{MeV}$, and skewness $Q_{\mathrm{sym}} = 302.28_{-231.89}^{+251.62} \, \mathrm{MeV}$. Additionally, the radii of canonical ($1.4 \, M_{\odot}$) and massive ($2.0 \, M_{\odot}$) neutron stars are predicted as $R_{1.4} = 11.85_{-0.15}^{+0.06} \, \text{km}$ and $R_{2.0} = 11.42_{-0.35}^{+0.23} \, \text{km}$, respectively, with a maximum mass of $M_{\mathrm{max}} = 2.12_{-0.05}^{+0.11} \, M_{\odot}$.  The tidal deformability is $\Lambda_{1.4} = 303.57_{-45.22}^{+47.95}$ at $1.4 \, M_\odot$, which is consistent with the analysis of the GW170817 event. The trace anomaly $\Delta$ in the high density region increases when the causality of the speed of sound is taken.
	\end{abstract}
	
    
	\maketitle
	
	\section{Introduction}\label{sec1}
     
     
     Over the past decade, as observational techniques have advanced, there has been remarkable progress in studying the properties of neutron stars. For example, massive neutron stars with masses approaching $2M_\odot$ have been discovered using the Shapiro delay effect, which has enabled relatively precise mass measurements. Notable examples include PSR J0740+6620 \cite{Cromartie2019}, PSR J1614-2230 \cite{Demorest2010}, PSR J0348+0432 \cite{Antoniadis2013}, and PSR J0952-0607 \cite{romani2022psr}. In 2017, the LIGO-Virgo Collaboration (LVC) made the first detection of gravitational waves emitted during the merger of two neutron stars (GW170817), providing important constraints on the tidal deformability of neutron stars \cite{Abbott_2017,PhysRevLett.119.161101}. Since 2019, the NICER collaboration has further advanced the field by simultaneously measuring the masses and radii of several neutron stars through detailed X‑ray observations of their surface hotspots. In particular, NICER’s pulse profile modeling has yielded precise measurements for PSR J0030+0451 with a mass around $1.4 \, M_\odot$ and a radius of approximately $11.7 \, \text{km}$, and for PSR J0740+6620, which has a mass of about $2.07 \, M_\odot$ and a radius estimated to be in the range of $11.6$-$13.8 \, \text{km}$ \cite{Riley_2019,Miller_2019,Riley_2021,Miller_2021}.
     
     Recent research has shown that the discovery of massive neutron stars with masses exceeding $2M_\odot$ significantly narrows the range of viable equations of state (EOSs) for dense matter, as only a relatively stiff EOS can support such high masses \cite{huang2020possibility}. Additionally, gravitational wave detections such as GW170817 have not only confirmed the binary neutron star merger scenario, but also provided constraints on how much neutron stars deform under tidal forces, a key parameter that directly relates to the pressure and stiffness of the stellar interior \cite{De2018,Annala2018,Most2018}. NICER’s measurements, by providing accurate mass and radius estimates through advanced modeling that incorporates relativistic light bending and Doppler effects, further constrain the EOS of ultradense nuclear matter \cite{Raaijmakers2019, Pang2021, Jiang2020}. Together, these observational breakthroughs from radio timing, gravitational-wave astronomy, and X‑ray pulse profile modeling have created a synergistic framework that offers robust and complementary constraints on the behavior of matter at supranuclear densities. This convergence of multi-messenger observations is proving instrumental in bridging the gap between nuclear physics theory and astrophysical phenomena, ultimately deepening our understanding of the internal composition of neutron stars and the fundamental properties of dense matter.
    
    Neutron star matter is primarily composed of neutrons, protons, and leptons such as electrons and muons in beta equilibrium. Traditionally, neutron star study has been approached from the perspective of nuclear many-body theory. By imposing conditions such as charge neutrality and beta equilibrium in neutron star matter, researchers derive the relationship between pressure and energy density, known as EOS. There are two main categories of methods used in this field. On the one hand, there are ab initio methods based on realistic nuclear forces. These first-principles calculations include approaches such as the Brueckner–Hartree–Fock (BHF) method \cite{PhysRevC.40.354, Hu2017}, the relativistic Dirac-Brueckner–Hartree–Fock (DBHF) method \cite{wang2020properties} and variational methods \cite{PhysRevC.58.1804}. These methods aim to directly solve the many-body problem starting from the fundamental nucleon-nucleon interactions.
    
    On the other hand, there are approaches based on effective interactions within the framework of density functional theory. Examples include the Skyrme–Hartree–Fock model \cite{stone2003nuclear,wang2024extended,duan2024new}, the Gogny–Hartree–Fock model \cite{RevModPhys.75.121,gonzalez2018new,vinas2021unified}, the relativistic mean-field (RMF) model \cite{zhang2018massive,hu2020effects,huang2022investigation,huang2024hadronic}, and the relativistic Hartree–Fock (RHF) model \cite{sun2008neutron,zhu2016delta,liu2018nuclear}. These models incorporate effective nucleon-nucleon interactions that are adjusted to reproduce known nuclear properties and are particularly useful for dealing with the complex many-body aspects of nuclear systems. Both types of methods have been widely applied not only to the study of nuclear structure but also to the investigation of neutron star properties. The insights gained from these approaches are crucial for understanding matter at supranuclear densities and for predicting the maximum mass of neutron stars.
    
    Furthermore, there exist parametrized EOS for neutron star matter that are expressed directly in exponential forms of density, i.e. in a piecewise polytropic form \cite{lindblom2010spectral,boyle2020parametrized}, or expanded as a polynomial in density via a Taylor series{, what is commonly referred to as meta-modeling} \cite{PhysRevC.97.025805, PhysRevC.97.025806, zhang2018combined,cai2025novel}. In such Taylor-expanded EOS, the energy density is typically divided into a component corresponding to symmetric nuclear matter and a density-dependent symmetry energy term. These parametrized EOS are widely used in the study of neutron star properties. They also serve as valuable tools for constraining the saturation properties of symmetric nuclear matter and the density dependence of the symmetry energy, based on observational data from neutron stars or experimental results from heavy-ion collision experiments \cite{xie2021bayesian}.

    {Recent advances in nuclear physics and astrophysics have significantly improved our understanding of neutron-rich matter, though uncertainties persist in its EOS due to the poorly constrained nuclear symmetry energy at supra-saturation densities \cite{Li2014Topical}. A comprehensive survey of 53 independent studies using both terrestrial experiments and astrophysical observations established updated constraints on the symmetry energy and its density dependence, determining \( E_{\text{sym}} = 31.7 \pm 3.2 \) MeV and $L_{\text{sym}} = 58.7 \pm 28.1$ MeV \cite{LI2013276, RevModPhys.89.015007}. These findings reflect substantial progress in characterizing the behavior of dense neutron-rich matter. The symmetry energy $E_{\mathrm{sym}}$ and its slope $L_{\mathrm{sym}}$ are pivotal for neutron-rich matter: they shape the neutron star crust \cite{PhysRevC.75.015801, bao2014influence} and intermediate mass radii\cite{Lattimer_2001} while remaining amenable to terrestrial scrutiny.  Parity-violating electron scattering has delivered seemingly incompatible constraints—PREX-II finds $E_{\mathrm{sym}} = 38.1 \pm 4.7$ MeV and $L_{\mathrm{sym}} = 106 \pm 37$ MeV from the $^{208}$Pb neutron skin \cite{PhysRevLett.108.112502, PhysRevC.85.032501, PhysRevLett.126.172502, PhysRevLett.126.172503, PhysRevLett.106.252501, PhysRevC.104.024329}, whereas CREX and subsequent Bayesian analyses of the $^{48}$Ca skin prefer markedly lower values, $E_{\mathrm{sym}} = 30.2^{+4.1}_{-3.0}$ MeV and $L_{\mathrm{sym}} = 15.3^{+46.8}_{-41.5}$ MeV \cite{PhysRevLett.129.042501, PhysRevLett.129.232501, PhysRevC.108.024317, particles6010003}. Lattimer’s analysis indicates that nuclear interactions reconciling both measurements favor $L_{\mathrm{sym}} = 53 \pm 13$ MeV\cite{particles6010003}, yet the sizeable gap between the two experimental determinations remains a formidable challenge for nuclear many-body theory.}
    
    {This study introduces joint Bayesian constraints on the crust–core transition density \(n_t\), transition pressure \(P_t\), and sub-saturation symmetry energy \(e_{\mathrm{sym}}\)(0.11 fm\(^{-3}\)). Guided by nuclear-structure analyses, we impose the ranges $0.05~\mathrm{fm}^{-3} < n_t < 0.11~\mathrm{fm}^{-3}$ \cite{bao2015impact} and $26~\mathrm{MeV} < e_{\mathrm{sym}}(0.11~\mathrm{fm}^{-3}) < 30~\mathrm{MeV}$ \cite{bao2014effects}; these cuts isolate the pivotal regime where non-uniform pasta phases dissolve into uniform matter—precisely the window that lies between the $\chi$EFT pure-neutron band $(0.04$--$0.16~\mathrm{fm}^{-3})$ and the bulk saturation region\cite{cartaxo2025completesurveytextttcompactobjectperspective, passarella2025relativisticmeanfieldpredictionsdense}.}
   
   With the growing volume of neutron star observational data, researchers have increasingly turned to data-driven approaches to study the EOS of neutron star matter. By directly using observable properties such as masses, radii, and tidal deformabilities, one can infer the underlying EOS that governs the behavior of ultra-dense matter. Recent advances in this field have been largely based on modern techniques, including Bayesian inference \cite{10.1093/mnras/stae844, 10.1093/mnras/stae2792,xie2020bayesian,malik2022bayesian,carvalho2023decoding,PhysRevC.107.055803,Li2024}, support vector machines \cite{ferreira2021unveiling}, and deep neural networks \cite{fujimoto2018methodology,zhou2023nonparametric,han2023nonparametric,zhou2024first,guo2024insights}.
   
   Bayesian inference, for example, utilizes Bayes' theorem to combine prior knowledge with observational data \cite{clark2007evidence,ozel2010astrophysical}. In this framework, the measured neutron star properties are incorporated into the likelihood function, while the parameter set governing the EOS is treated as the prior distribution. This approach allows researchers to obtain the posterior probability distribution for the EOS parameters, thereby quantifying uncertainties and correlations among them. Additionally, machine learning methods such as support vector machines have been employed for regression and classification tasks, effectively mapping complex relationships between observed neutron star features and the corresponding EOS parameters. Deep neural networks further enhance this effort by capturing non-linear dependencies and intricate patterns in high-dimensional datasets, offering a complementary route to traditional theoretical and computational methods.

   {Bayesian inference has emerged as a central framework for extracting quantitative constraints
   	on the neutron star EOS from multi-messenger observations. Analyses employing piecewise polytropic and extended Skyrme parameterizations, combined with MCMC techniques, have demonstrated that the stiffness of the EOS at intermediate densities governs both tidal deformability and mass–radius correlations  \cite{chimanski2022bayesianinferencephenomenologicaleos, Beznogov2024apj, Beznogov_2024}.  Within covariant density-functional (CDF) frameworks, the incorporation of constraints from GW170817, NICER, and PREX-II indicates that nucleonic and leptonic matter alone can sustain $M_{\rm max} \approx 2.5, M_{\odot}$, while predicting $R_{1.4} \approx 12$–$13$ km, provided that the symmetry energy remains moderately soft and causality with respect to the Fermi velocity is maintained \cite{Beznogov_2024, Malik_2022, Provid_ncia_2024, LI2025139501}.  Comprehensive Bayesian syntheses that include perturbative QCD (pQCD) constraints, heavy-ion collision data, and direct Urca cooling thresholds further tighten the high-density symmetry energy slope to $L_{\rm sym} \approx 48$–$54$ MeV and constrain the onset of the direct Urca process to $1.6$–$1.8, M_{\odot}$  \cite{PhysRevD.109.043052, Tsang_2024, Malik2022Inferring}. Meanwhile, meta-modeling approaches with density-dependent couplings reveal that extreme variations in isovector incompressibility $K_{\rm sym}$ and saturation incompressibility $K_{\rm sat}$ can induce measurable shifts in neutron star radii and proton fractions. However, these effects may remain hidden in global observables unless specific likelihood functions (e.g., Gaussian versus uniform) are employed \cite{char2023generaliseddescriptionneutronstar, li2025112}.  Lastly, observations of extreme-mass pulsars such as PSR J0952–0607 and ultra-compact objects like HESS J1731–347 challenge the limits of nucleonic CDF models, underscoring the crucial importance of precise symmetry energy parameterization and the ongoing need for refinement via Bayesian method \cite{PhysRevC.111.055804}.
 
   The analysis in this work employs the \texttt{CompactObject} package developed by Huang et al. \cite{huang2024compactobjectopensourcepythonpackage, 10.1093/mnras/stae844, 10.1093/mnras/stae2792}.  Within this framework, the neutron star EOS was constructed from the RMF FSU2R nucleonic model, and Bayesian inference was performed with a likelihood that combines NICER mass–radius posteriors for PSR J0030+0451\cite{ Riley_2019} and PSR J0740+6620\cite{ Riley_2021}. Going beyond this baseline, we present the first meta-modeling study that simultaneously incorporates updated mass–radius data for the canonical mass pulsar PSR J0030+0451 \cite{Vinciguerra_2024} and the massive pulsar PSR J0740+6620 \cite{Salmi_2024}, while employing uniform (flat) priors for every EOS parameter to ensure maximal model-agnostic flexibility. We further augment the likelihood with novel empirical bounds on the speed of sound and the subsaturation symmetry energy slope and curvature, tightening the posterior without sacrificing prior neutrality.}
   
   In this work, we will employ Bayesian inference within the framework of parametrized EOS to infer the properties of neutron star matter. { The parametric equation of state employed in this work follows the meta-modeling approach, which differs fundamentally from RMF models. The meta-modeling EOS directly parameterizes nuclear matter properties (including saturation density, symmetry energy, symmetry energy slope, and incompressibility) through empirical or semi-empirical parameters to construct an analytical form of the EOS. Typically requiring fewer parameters (5-10), the meta-modeling approach is particularly suitable for large-scale parameter scans and Bayesian analyses \cite{PhysRevC.97.025805, PhysRevC.97.025806}. In contrast, RMF models are grounded in quantum field theory, describing nucleon-nucleon interactions through meson ($\sigma$, $\omega$, $\rho$) exchange within a self-consistent mean-field approximation \cite{huang2020possibility, Huang_2022, huang2024hadronic}. While RMF models may incorporate nonlinear meson self-coupling terms (e.g., $\sigma^3$, $\sigma^4$) with more parameters, they maintain clearer physical interpretations. The meta-modeling approach offers superior flexibility, allowing parameter adjustments to accommodate diverse theoretical and experimental constraints, making it particularly effective for statistical analyses of multi-messenger astronomical data (e.g., neutron star mass-radius relations, tidal deformability) \cite{zhang2018combined, Zhang_2019, Zhang_20200410, Zhang_20201010, Zhang_2021}. However, its high-density extrapolations may lack microscopic physical foundations and require additional constraints (e.g., causality, stability). Computationally, the meta-modeling EOS demonstrates remarkable efficiency - requiring only algebraic operations rather than self-consistent field equation solutions as in RMF models (which demand substantially more computational time and resources\cite{huang2020possibility, Huang_2022, huang2024hadronic}) making it particularly suitable for rapid EOS generation and subsequent stellar structure calculations.}
   
   Unlike previous studies that primarily considered observational constraints such as neutron star mass–radius measurements, our approach also incorporates additional constraints that critically affect nuclear properties. Specifically, we account for the symmetry energy at subsaturation densities, which are strongly correlated to the properties of finite nuclei \cite{bao2014effects}, the density of the crust-core phase transition in neutron stars \cite{bao2014influence,bao2015impact}, and the causal limit at high densities, that is, the speed of sound in neutron star matter, $c^2_s/c^2<1$. These constraints differentiate our work from previous studies on the neutron star EOS,  which are the key innovations in our present work. The structure of this paper is organized as follows. In Section II, we introduce the main theoretical framework, detailing the polynomial form of the EOS, the static equations governing neutron stars, and the fundamentals of Bayesian inference methods. In Section III, we present the neutron star EOSs obtained by combining various observational and theoretical constraints, and discuss the high-density properties of nuclear matter and the resulting neutron star characteristics. In Section IV, we conclude with a summary of our findings and a discussion of future research directions.
   
	\section{Theoretical framework}\label{sec2}
    
    \subsection{The parametrized equation of state}
    In this work, a neutron star is assumed to be composed of neutrons, protons, electrons, and muons, i.e. model $npe\mu$, where the energy density is represented as the sum of the baryon energy density $\epsilon_b$ and the lepton energy density $\epsilon_l$, $\epsilon = \epsilon_b + \epsilon_l$. The neutron star matter can be considered as the infinite uniform nuclear matter. The number densities of neutrons and protons are denoted by $ n_n $ and $ n_p $, respectively. The baryon number density can be expressed as $ n_b = n_n + n_p $. The neutron plays a dominant role in a neutron star, where the isospin asymmetry is very important and is defined as $ \delta = (n_n - n_p) / n_b $.
    
    The EOS of nuclear matter must be a function of both baryon density $n_b$ and isospin asymmetry $\delta$. The expression can be expanded as a Taylor series around symmetric nuclear matter (SNM, $\delta = 0$) and nuclear saturation density ($n_0 = 0.155 \pm 0.005\ \mathrm{fm^{-3}}$). The ratio of energy density $\varepsilon$ to number density $n_b$ is equivalent to the energy per nucleon,  $E/A$, where $A$ represents the number of nucleons, that is, the sum of neutrons and protons. It can be represented as \cite{zhang2018combined}
    \begin{equation} \label{E/A}
    	\frac{E}{A}\left(n_b,\delta\right)=e_\mathrm{sat}\left(n_b,0\right)+e_\mathrm{sym}\left(n_b,0\right)\delta^2+\mathcal{O}(\delta^4),
    \end{equation}
    where the higher order terms related to $\delta$ are ignored, and
    \begin{equation}
    	e_\mathrm{sat}\left(n_b,0\right)=\frac{E}{A}\left(n_b,0\right),
    	\quad
    	e_\mathrm{sym}\left(n_b,0\right)=\frac{1}{2}\frac{\partial^2 E\big/A\left(n_b,\delta\right)}{\partial \delta^2}\Bigg|_{\delta = 0}.
    \end{equation}
    The $e_\mathrm{sat}\left(n_b,0\right)$ is the energy per nucleon of the SNM, and $e_\mathrm{sym}\left(n_b,0\right)$ is the symmetry energy coefficient. Conventionally, a variable $x$ is defined as $x = (n_b - n_0) / 3n_0$. Expanding $e_\mathrm{sat}$ and $e_\mathrm{sym}$ at $n = n_0$ (i.e., at $x = 0$) as polynomial functions of $x$ and retaining $x$ terms up to third order, we obtain
    \begin{equation} \label{ee}
    	\begin{split}
            e_\mathrm{sat}(x)
            &=E_\mathrm{sat}+\frac{1}{2}K_\mathrm{sat}x^2+\frac{1}{6}Q_\mathrm{sat}x^3,\\
            e_\mathrm{sym}(x)
            &=E_\mathrm{sym}+L_\mathrm{sym}x+\frac{1}{2}K_\mathrm{sym}x^2+\frac{1}{6}Q_\mathrm{sym}x^3,
    	\end{split}
    \end{equation}
    where $E_\mathrm{sat}$, $K_\mathrm{sat}$, and $Q_\mathrm{sat}$ represent the saturation energy, incompressibility, and the skewness of the energy per nucleon of the SNM, respectively. $E_\mathrm{sym}$ represents the symmetry energy at the nuclear saturation density. $L_\mathrm{sym}$, $K_\mathrm{sym}$, and $Q_\mathrm{sym}$ denote the slope, curvature, and skewness of the symmetry energy coefficient, respectively.
    
    Ultimately, the energy density of the core of a neutron star can be expressed as:
    \begin{equation} \label{eq:e_b}
    	\epsilon_b\left(x,\delta\right)=n_0\left(3x+1\right)\left( m_N c^2+\frac{E}{A} \left(x,\delta\right) \right),
    \end{equation}
    where $m_N = 939$ MeV/$c^2$ is the nucleon mass.

    The leptons in the neutron star core are modeled as a non-interacting Fermi gas at zero temperature, with all particles occupying states below the Fermi surface. For free electrons (including spin degeneracy), the Fermi momentum $k_F$ relates to the lepton density $n_l$ by $k_F=\left( 3 \pi^2 n_l \right)^{1/3}$. The lepton energy density is derived from the relativistic dispersion relation,
    \begin{equation} \label{eq:e_l}
    	\begin{split}
    		\epsilon_l
    		&=\int_{0}^{k_F}\left( \hbar^2 k^2 c^2 + m_l^2 c^4 \right)^{1/2} \, \mathrm{d} n_l\\
    		&=\frac{m_l^4 c^5}{8 \pi^2 \hbar^3}\left[ t \sqrt{1+t^2} \left( 1+2t^2 \right) - \ln\left( t+\sqrt{1+t^2} \right)\right],
    	\end{split}
    \end{equation}
    where $t=\hbar k_F\big/m_l c$, and $m_l$ denotes the lepton mass ($m_e$ or $m_{\mu}$).
    Eq. \eqref{eq:e_l} defines the lepton energy density $\epsilon_l$ as a function of $n_l$.
    To relate \(n_b\), \(\delta\), and \(n_l\) in the neutron star core EOS, we impose $\beta$-equilibrium,
    \begin{equation} \label{mu}
    	\mu_n - \mu_p = \mu_e = \mu_\mu,
    \end{equation}
    where $\mu_i$ ($i=n,p,e,\mu$) denotes the chemical potential of each particle species.
    In particular, muons emerge only when the electron chemical potential exceeds the muon mass: $\mu_e > m_\mu = 106.55\ \text{MeV}$.
    Charge neutrality also requires,
    \begin{equation} \label{charge neutrality}
    	n_p = n_e + n_\mu,
    \end{equation}
    where $n_p$, $n_e$, and $n_\mu$ denote proton, electron, and muon number densities, respectively.
    Through these constraints, the energy density $\epsilon$ can be reduced from four independent variables ($n_b$, $\delta$, $n_e$, $n_\mu$) to a single-variable function $\epsilon(n_b)$.

    The difference between neutron and proton chemical potentials is
    \begin{equation} \label{eq:mu_n-mu_p}
    	\begin{split}
    		\mu_n - \mu_p
            &=\frac{\partial \epsilon_b}{\partial n_n} - \frac{\partial \epsilon_b}{\partial n_p} \\
    		&=\left(\frac{\partial \epsilon_b}{\partial n_b}\frac{\partial n_b}{\partial n_n}+\frac{\partial \epsilon_b}{\partial \delta}\frac{\partial \delta}{\partial n_n}\right) - \left(\frac{\partial \epsilon_b}{\partial n_b}\frac{\partial n_b}{\partial n_p}+\frac{\partial \epsilon_b}{\partial \delta}\frac{\partial \delta}{\partial n_p}\right)\\
    		&=4\delta e_{\rm sym}(n_b,0).
    	\end{split}
    \end{equation}
    The lepton chemical potential $\mu_l$ ($l=e, \mu$) is
    \begin{equation} \label{mu_l}
    \mu_l = \sqrt{\hbar^2 k_F^2 c^2 + m_l^2 c^4},
    \end{equation}
    by combining Eqs. \eqref{mu}, \eqref{eq:mu_n-mu_p}, and \eqref{mu_l}, we derive the lepton number density:
    \begin{equation} \label{eq:n_l}
    	n_l = \frac{\left( 16 \delta^2 e_{\rm sym}^2(n_b,0) - m_l^2 c^4 \right)^{3/2}}{3 \pi^2 \hbar^3 c^3},
    \end{equation}
    the proton density follows,
    \begin{equation} \label{n_p}
    	n_p = \frac{1 - \delta}{2} n_b.
    \end{equation}
    The isospin asymmetry $\delta(n_b)$ is determined by solving the condition of charge neutrality with Eqs. \eqref{eq:n_l} and \eqref{n_p}.
    Muons appear only when $\mu_n-\mu_p \ge m_\mu$. Below this threshold, $n_\mu=0$; above it, both electrons and muons contribute to charge neutrality.

    With $\delta(n_b)$ determined, the baryon energy density $\epsilon_b(x)$ follows from Eqs. (\ref{E/A}), (\ref{ee}), and (\ref{eq:e_b}), while the leptonic contribution $\epsilon_l(x)$ derives from Eqs. (\ref{eq:e_l}) and (\ref{eq:n_l}).
    The corresponding pressures are obtained via the thermodynamic identity
    \begin{equation} \label{eq:P_b}
    	\begin{split}
    		P_b(x)
    		&=n_{b}^{2}\left(\frac{\partial(\epsilon_{b}/n_{b})}{\partial n_{b}}\right)\\
    		&=\frac{n_{0}}{3}(3x+1)^{2}\left[(K_{\mathrm{sat}}x+\frac{1}{2}Q_{\mathrm{sat}}x^{2})+\delta^{2}(L_{\mathrm{sym}}+K_{\mathrm{sym}}x+\frac{1}{2}Q_{\mathrm{sym}}x^{2})+2e_{\mathrm{sym}}\delta\delta^{\prime}\right],
    	\end{split}
    \end{equation}
    where $x = \left( n_b/n_0 - 1 \right)/3$, and $\delta' = {\rm d} \delta / {\rm d}x$.  The leptonic pressure is
    \begin{equation}
    	P_l(x) 
        =n_{l}^{2}\left(\frac{\partial(\epsilon_{l}/n_{l})}{\partial n_{l}}\right) 
        = n_l\mu_l - \epsilon_l.
    \end{equation}
    The total energy density $\epsilon$ and pressure $P$ of the neutron star core are given by $\epsilon = \epsilon_b + \epsilon_l, P = P_b + P_l$.

    
    The transition from the homogeneous outer core to the inner crust marks the onset of nuclear inhomogeneity, necessitating a nonuniform matter EOS. The phase transition density $n_t$ is determined by the thermodynamic instability condition \cite{kubis2007nuclear,lattimer2007neutron}
    \begin{equation} \label{n_t}
    	K_\mu = \ n^{2} \frac{\mathrm{d}^{2} e_\mathrm{sat}}{\mathrm{d} n^{2}}+2 n \frac{\mathrm{d} e_\mathrm{sat}}{\mathrm{d} n}
    		+\delta^{2} \left[n^{2} \frac{\mathrm{d}^{2} e_\mathrm{sym}}{\mathrm{d} n^{2}}+2 n \frac{\mathrm{d} e_\mathrm{sym}}{\mathrm{d} n}-2 e_\mathrm{sym}^{-1}\left(n \frac{\mathrm{d} e_\mathrm{sym}}{\mathrm{d} n}\right)^{2}\right]=0.
    \end{equation}
    The EOS for the crust of the neutron star was selected in a parametrized form with the SLy4 interaction \cite{douchin2001unified}. By combining the EOSs for the core and the crust of the neutron star at the crust-core phase transition density, we can obtain a complete EOS at the density region of the neutron star.

    \subsection{The theoretical framework of neutron stars}
    
   The mass-radius relation of a neutron star is governed by the Tolman-Oppenheimer-Volkoff (TOV) equation \cite{tolman1939static,oppenheimer1939massive}, which describes a spherically symmetric star in hydrostatic equilibrium within the framework of general relativity
    \begin{equation} \label{TOV}
    \begin{split}
        \frac{\mathrm{d} P(r)}{\mathrm{d} r}
        &=- \frac{GM(r)\epsilon(r)}{c^2r^2}
        \left[1+\frac{P(r)}{\epsilon(r)}\right]
        \left[1+\frac{4{\pi}r^3P(r)}{c^2M(r)}\right]
        \left[1-\frac{2GM(r)}{c^2r}\right]^{-1}, \\
        \frac{\mathrm{d} M(r)}{\mathrm{d} r}
        &=\frac{4 \pi  r^2 \epsilon(r)}{c^2},
    \end{split}
    \end{equation}
    where $P(r)$ and $M(r)$ are the pressure and mass of the neutron star at $r$, respectively. Additionally, another physical quantity that characterizes the properties of neutron stars is tidal deformability \cite{hinderer2010,read2013}, which describes the deformation of a compact object in the external field produced by another star. It can be calculated in a dimensionless form,
    \begin{equation}
    	\Lambda = \frac{2}{3} k_2 C^{-5},
    \end{equation}
    where $C = GM/Rc^2$ is the compactness parameter. The Love number $k_2$ is  given by
    \begin{equation}
    	\begin{aligned}
    		k_{2}= & \frac{8 C^{5}}{5}(1-2 C)^{2}\left[2-y_{R}+2 C\left(y_{R}-1\right)\right] \\
    		& \times \left\{2 C\left[6-3 y_{R}+3 C\left(5 y_{R}-8\right)\right]\right. \\
    		& +4 C^{3}\left[13-11 y_{R}+C\left(3 y_{R}-2\right)+2 C^{2}\left(1+y_{R}\right)\right] \\
    		& \left. +3(1-2 C)^{2}\left[2-y_{R}+2 C\left(y_{R}-1\right) \ln (1-2 C)\right]\right\}^{-1},
    	\end{aligned}
    \end{equation}
    and $y_R = y(R)$. $y(r)$ is governed by the following differential equation
    \begin{equation}
    	r \frac{\mathrm{d} y(r)}{\mathrm{d} r}+y^{2}(r)+y(r) F(r)+r^{2} Q(r)=0,
    \end{equation}
    with the initial condition $y(0) = 2$. Here $F(r)$ and $Q(r)$ are functions of $P(r)$, $\epsilon(r)$, and $M(r)$, and are defined as follows
    \begin{equation}
    	F(r)=\left[1-\frac{4 \pi r^{2} G}{c^{4}}\bigg(\epsilon(r)-P(r)\bigg)\right]\left(1-\frac{2 M(r) G}{r c^{2}}\right)^{-1},
    \end{equation}
    \begin{equation}
    	\begin{aligned}
    		r^{2} Q(r)= & \frac{4 \pi r^{2} G}{c^{4}}\left[5 \epsilon(r)+9 P(r)+\frac{\epsilon(r)+P(r)}{\partial P(r) \big/ \partial \epsilon(r)}\right] \left(1-\frac{2 M(r) G}{r c^{2}}\right)^{-1} \\
    		& -6\left(1-\frac{2 M(r) G}{r c^{2}}\right)^{-1} \\
    		& -\frac{4 M^{2}(r) G^{2}}{r^{2} c^{4}}\left(1+\frac{4 \pi r^{3} P(r)}{M(r) c^{2}}\right)^{2}\left(1-\frac{2 M(r) G}{r c^{2}}\right)^{-2}.
    	\end{aligned}
    \end{equation}
    which can be solved with the TOV equation, simultaneously. 

    \subsection{Bayesian inference method}
    There are several parameters in the above parametrized EOS, which are related to the isoscalar and isovector saturation properties of symmetric nuclear matter and high-density behavior $E_\mathrm{sat}, ~K_\mathrm{sat}, ~Q_\mathrm{sat}, ~E_\mathrm{sym}, ~L_\mathrm{sym}, ~K_\mathrm{sym}, ~Q_\mathrm{sym}$. The binding energy, incompressibility and symmetry energy, $E_\mathrm{sat}, ~K_\mathrm{sat}$, and $E_\mathrm{sym}$ at nuclear saturation density are almost well constrained by existing experimental data from finite nuclei. However, the remaining parameters, particularly $Q_\mathrm{sat},~L_\mathrm{sym}, ~K_\mathrm{sym}, ~Q_\mathrm{sym}$, remain poorly constrained. To better constrain these parameters, we must incorporate observational data from neutron stars along with other empirical constraints. The Bayesian inference method is an effective tool for addressing this class of problems.
    
    We begin by assuming that the neutron star EOS and its resulting macroscopic properties can be fully characterized by a parameter vector \(\bm{\vartheta}\). 
    The likelihood of observing specific astrophysical data, conditioned on a given EOS, is expressed as \(p(\mathcal{D} \mid \bm{\vartheta})\). 
    Through Bayes' theorem, the posterior probability distribution for the parameters \(\bm{\vartheta}\) conditioned on the observed data $\mathcal{D}$ is given by
    \begin{equation}
    p(\bm{\vartheta} \mid \mathcal{D}) = \frac{p(\mathcal{D} \mid \bm{\vartheta}) \, p(\bm{\vartheta})}{p(\mathcal{D})},
    \end{equation}
    where \(p(\bm{\vartheta})\) denotes the prior distribution encoding our initial knowledge of the parameters, and $p(\mathcal{D})=\int p(\mathcal{D}\mid\boldsymbol{\vartheta})p(\boldsymbol{\vartheta})\text{d}\boldsymbol{\vartheta}$ is the evidence serving as the normalization constant.
    
    In this work, the parameter space is characterized by four key parameters: \( Q_{\text{sat}} \), \( L_{\text{sym}} \), \( K_{\text{sym}} \), and \( Q_{\text{sym}} \). Specifically, \( Q_{\text{sat}} \) is defined as a uniform prior over the range \(-200\) to \(200 \, \text{MeV}\). The parameter \( L_{\text{sym}} \) is uniformly sampled in the interval from \(0\) to \(80 \, \text{MeV}\). Similarly, \( K_{\text{sym}} \) is uniformly sampled in the range of \(-200\) to \(200 \, \text{MeV}\). Lastly, \( Q_{\text{sym}} \) is uniformly sampled within the broader range of \(-200\) to \(800 \, \text{MeV}\). This uniform sampling approach ensures a thorough exploration of the parameter space, thereby facilitating a detailed analysis of its impact on the system's properties.
    
    To constrain the parameters in our parametrized EOS, we incorporate multiple physical conditions: (1) astronomical observables from neutron stars, (2) the crust-core phase transition density, (3) symmetry energy constraints at subsaturation density, and (4) the causality condition on the speed of sound in neutron star matter.
    
    \begin{itemize}
    
    \item Astronomical Observables. The masses and radii of two neutron stars from NICER are taken, for PSR J0030+0451, $M = 1.4_{-0.12}^{+0.13} \, M_{\odot}$, $R = 11.71_{-0.83}^{+0.88} \, \mathrm{km}$ \cite{Vinciguerra_2024}, and for PSR J0740+6620, $M = 2.073_{-0.069}^{+0.069} \, M_{\odot}$, $R = 12.49_{-0.88}^{+1.28} \, \mathrm{km}$ \cite{Salmi_2024}.
    	
    \item Crust-Core Phase Transition Density. We have constrained the crust-core phase transition density in neutron stars to the range \( 0.05 \, \text{fm}^{-3} < n_t < 0.11 \, \text{fm}^{-3} \) \cite{bao2015impact}. Additionally, the phase transition pressure is constrained to \( P_t > 0 \,\text{MeV} \), although in some regions of the parameter space, the calculated phase transition pressure may be negative{, since this requirement was not imposed in \cite{char2023generaliseddescriptionneutronstar}}.
    
    \item Symmetry Energy at Subsaturation Density. Density functional theory reveals that the symmetry energy at subsaturation density \( n_b \sim 0.11 \, \text{fm}^{-3} \) is closely related to the ground state properties of finite nuclei \cite{bao2014effects}. Consequently, the symmetry energy at the subsaturation density \( n_b = 0.11 \, \text{fm}^{-3} \) is constrained to the range \( 26 \, \text{MeV} < e_{\text{sub}} < 30 \, \text{MeV} \). {Our symmetry energy constraints at \( 0.11 \, \text{fm}^{-3} \) maintain consistency with \(\chi\)EFT predictions when properly extrapolated to lower densities. For example, the symmetry energy range in our study (\(26\)--\(30 \, \text{MeV}\)) aligns well with \(\chi\)EFT-derived values shown in Ref. \cite{li2025112}, where \(e_{\text{sym}}(0.1 \, \text{fm}^{-3}) \approx 24\)--\(28 \, \text{MeV}\). While \(\chi\)EFT provides a more fundamental theoretical foundation, our current methodology offers a practical approach to constrain the crucial crust--core transition region.}

    \item Causal Limit. In the mass-radius curve of neutron stars, before reaching the maximum mass, the squared sound speed \( c_s^2 = \partial P/\partial \epsilon \) must remain subluminal, i.e., \( c_s^2/c^2 \leq 1 \). Additionally, the isospin asymmetry \( \delta \) must not reach unity. {Through an extensive literature review, we observe that although numerous meta-modeling studies \cite{zhang2018combined, Zhang_2019, Zhang_20200410, Zhang_20201010, Zhang_2021, xie2019bayesian, xie2020bayesian} constrain the EOS with empirical and observational data, they seldom explicitly examine the variation of the speed of sound within neutron-star matter. Works such as Refs. \cite{zhang2018combined} and \cite{xie2020bayesian} focus primarily on observational constraints (e.g., mass--radius relations, tidal deformability) without a detailed assessment of the speed of sound or its causal limit. Therefore, in the present study, we explicitly enforce the requirement $c_s^{2}/c^{2}\le 1$ as an innovative aspect of our meta-modeling parameterization of the EOS.}
    
    \end{itemize}

    This study employs the nested sampling method for Bayesian inference, implemented via the slice sampler within the \texttt{UltraNest} package \cite{2021JOSS....6.3001B}. Each inference run utilizes 100,000 live points. The sampling scheme, executed through the \texttt{CompactObject} package developed by Huang et al.~\cite{huang2024compactobjectopensourcepythonpackage}, establishes a statistically rigorous foundation for Bayesian constraints on the equation of state.Research based on this package includes works by \cite{10.1093/mnras/stae844, 10.1093/mnras/stae2792}.
    
    In subsequent discussions, the constraints from the crust-core phase transition and the symmetry energy at subsaturation density are collectively referred to as the \textbf{low-density limit}, representing a novel approach in our work compared to previous studies. Meanwhile, the causal limit is termed the \textbf{high-density limit}.

	\section{Results and discussions}\label{sec3}
\begin{figure}[htbp]
    \centering
    \includegraphics[width=0.6\textwidth]{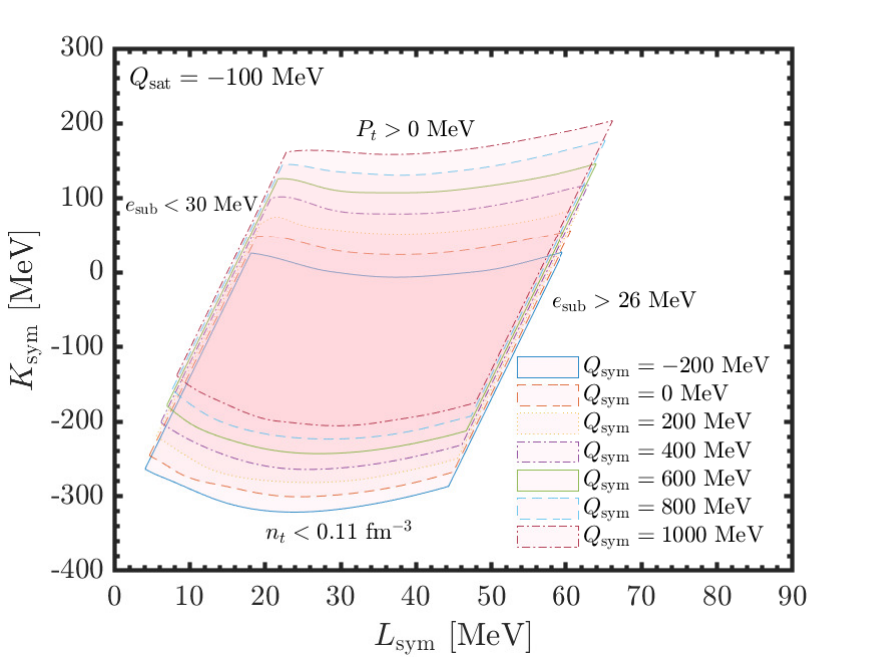} 
    \caption{The \( K_{\text{sym}} \)-\( L_{\text{sym}} \) parameter spaces constrained by the symmetry energy at nuclear subsaturation density and the properties of the crust-core phase transition for different values of \( Q_{\text{sym}} \).}
    \label{fig1}
    \end{figure}

First, we investigate the constraining effects of the parameter space on the crust-core phase transition and the symmetry energy at nuclear subsaturation density. The binding energy, incompressibility and symmetry energy at nuclear saturation density are fixed at $E_{\text{sat}} = -15.9 \, \text{MeV}$, $K_{\text{sat}} = 230 \, \text{MeV}$, and $E_{\text{sym}} = 31.6 \, \text{MeV}$, respectively, based on empirical data extracted from finite nuclei systems \cite{Li2024}. In Fig.~\ref{fig1}, we plot the possible regions between \(K_{\text{sym}}\) and \(L_{\text{sym}}\) constrained by the crust-core phase density (\(n_t<0.11\,\text{fm}^{-3}\)) and the positive phase transition pressure (\(P_t>0\,\text{MeV}\)), as well as the symmetry energy at the nuclear subsaturation density (\(26\,\text{MeV}<e_{\text{sub}}<30\,\text{MeV}\)), with \(Q_{\text{sat}}=-100 \, \text{MeV}\) and \(Q_{\text{sym}}\) ranging from \(-200\) to \(1000\,\text{MeV}\). A strong linear correlation between \(K_{\text{sym}}\) and \(L_{\text{sym}}\) is observed due to the constraints on the symmetry energy. From Eq.~(\ref{ee}), it is evident that the higher-order terms related to \(Q_{\text{sym}}\) can be neglected in the low-density region, and the symmetry energy at nuclear saturation density, \(E_{\text{sym}}\) is fixed. Therefore, \(K_{\text{sym}}\) and \(L_{\text{sym}}\) exhibit a linear dependence at a given density.

Meanwhile, the properties of the crust-core phase transition, \( n_t \) and \( P_t \), will constrain the lower and upper limits of \( K_{\text{sym}} \) through the thermodynamical approach in Eq.~(\ref{n_t}). Their magnitudes are strongly dependent on \( Q_{\text{sym}} \). When \( Q_{\text{sym}} = -200 \, \text{MeV}\), \( K_{\text{sym}} \) ranges between \(-300 \, \text{MeV}\) and \(0 \, \text{MeV}\). This range increases with \( Q_{\text{sym}} \) and is approximately \(-200\)--\(200 \, \text{MeV}\) for \( Q_{\text{sym}} = 1000 \, \text{MeV}\). Additionally, an increase in \( Q_{\text{sat}} \) can also slightly increase the magnitude of \( K_{\text{sym}} \). Correspondingly, the slope of the symmetry energy \( L_{\text{sym}} \) is about \(0\)--\(50 \, \text{MeV}\) at \( K_{\text{sym}} = -300 \, \text{MeV}\) and is \(20\)--\(70 \, \text{MeV}\) at \( K_{\text{sym}} = 200 \, \text{MeV}\). These novel physical constraints further constrain the parameter space of the EOS.

The higher-order expansion coefficients, $Q_\text{sat}$ and $Q_\text{sym}$ are not sensitive to the constraints at low density, which should be inferred from the properties of neutron star. Therefore, we apply Bayesian inference to determine their values. The masses and radii of two neutron stars, PSR J0030+0451 and PSR J0740+6620, from NICER are chosen as the likelihood function \cite{Vinciguerra_2024, Salmi_2024}. The prior distributions for \( K_{\text{sym}}, L_{\text{sym}}, Q_{\text{sat}}, Q_{\text{sym}} \) are assumed to be uniform in the ranges \([-200, 200]\), \([0, 80]\), \([-200, 200]\), and \([-200, 800] \, \text{MeV}\), respectively, based on available investigations. The posterior distributions are derived through the Bayes' theorem using a standard nested sampling algorithm.

\begin{figure}[htbp]
        \centering
        \includegraphics[width=0.8\textwidth]{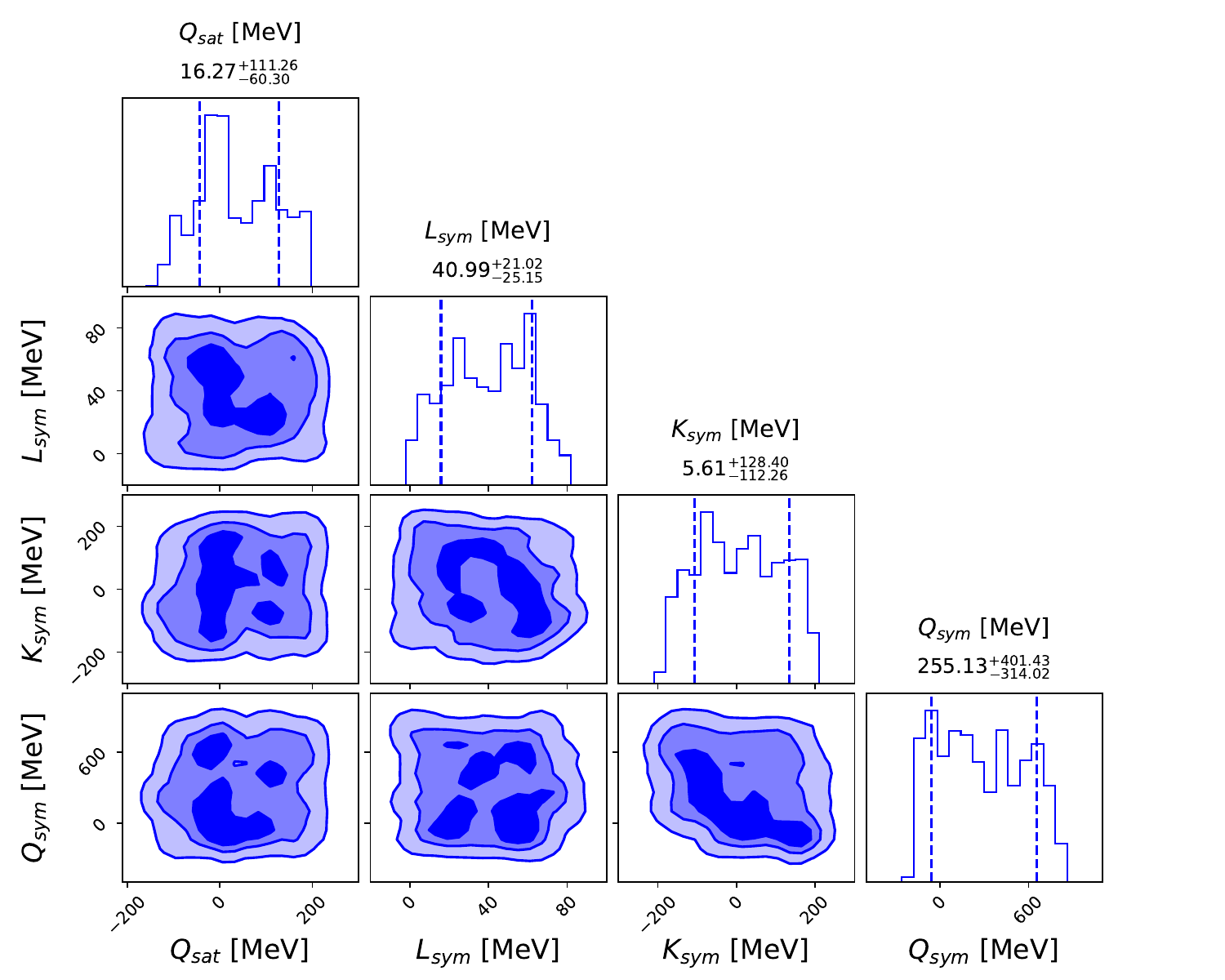} 
        \caption{The posterior distribution of parameters, constrained by observations of the neutron stars PSR J0030+0451 and PSR J0740+6620. The regions from dark to light represent the \( 1\sigma, \, 2\sigma \), and \( 3\sigma \) credible levels, respectively.}
        \label{fig2}
    \end{figure}

The posterior distributions of \( Q_{\text{sat}},\, L_{\text{sym}},\, K_{\text{sym}},\, Q_{\text{sym}} \) are shown in Fig.~\ref{fig2}. The three regions, from dark to light, represent the credible levels of \( 1\sigma, \, 2\sigma \) and \( 3\sigma \), respectively. The skewness of the energy per nucleon of the SNM is inferred to be \( 16.27^{+111.26}_{-60.30} \, \text{MeV}\) at the \( 1\sigma \) level. This result is slightly higher than the -180 $\pm$ 110 MeV obtained by Xie et al. \cite{xie2021bayesian} using a Gaussian prior probability distribution function of -200 $\pm$ 200 MeV, and is somewhat higher than those obtained from astronomical observations based on the RMF models and the SHF models \cite{cai2017constraints,chen2011higher}. Additionally, the slope, curvature, and skewness of the symmetry energy are predicted to be \( 40.99^{+21.02}_{-25.15} \, \text{MeV}\), \( 5.61^{+128.40}_{-112.26} \, \text{MeV}\), and \( 255.13^{+401.43}_{-314.02} \, \text{MeV}\), respectively. These values fall within the reasonable range as reported in previous studies \cite{xie2019bayesian,li2019towards,Li2024}.

\begin{figure}[htbp]
        \centering
        \includegraphics[width=0.9\textwidth]{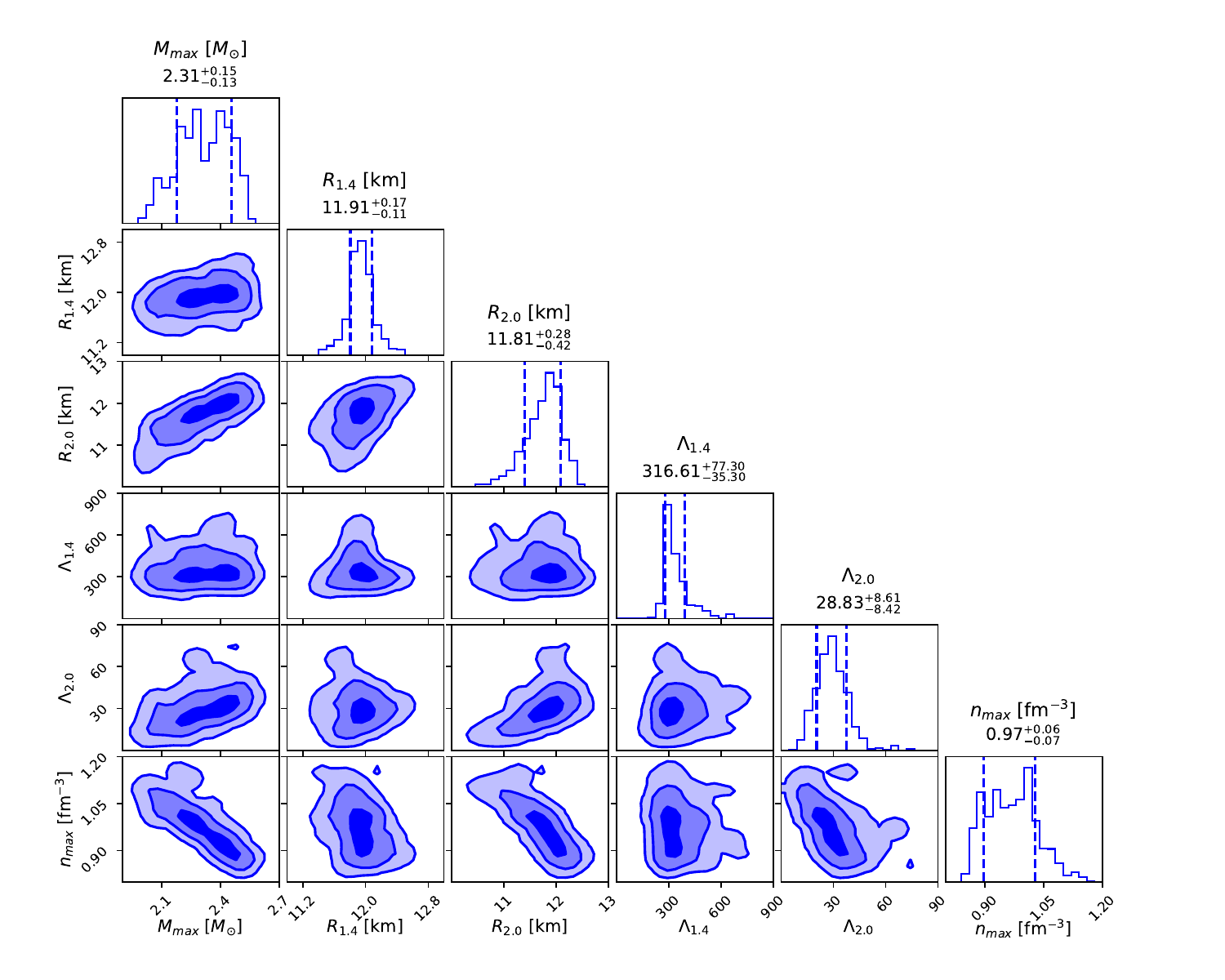} 
        \caption{The posterior distributions of neutron star properties constrained by observations of the neutron stars PSR J0030+0451 and PSR J0740+6620. The regions from dark to light also represent the \( 1\sigma, \, 2\sigma \), and \( 3\sigma \) credible levels, respectively.}
        \label{fig3}
    \end{figure}

Once the parameter spaces are fixed, the EOS of neutron star matter can be determined, and the properties of neutron stars can be solved using the TOV equation. The posterior distributions of the maximum mass \( M_{\text{max}} \), the radii of neutron stars at \( 1.4 \, M_{\odot} \) and \( 2.0 \, M_{\odot} \) (\( R_{1.4} \) and \( R_{2.0} \)), the dimensionless tidal deformabilities at \( 1.4 \, M_{\odot} \) and \( 2.0 \, M_{\odot} \) (\( \Lambda_{1.4} \) and \( \Lambda_{2.0} \)), and the central density of neutron stars \( n_{\text{max}} \) are plotted in Fig.~\ref{fig3}. The maximum mass of neutron stars is found to be around \( M_{\text{max}} = 2.31^{+0.15}_{-0.13} \, M_{\odot} \), which supports the current heaviest known neutron star measured, PSR J0952-0607 \cite{romani2022psr}. Notably, the data from this neutron star were not included in the prior distribution.

The dimensionless tidal deformability at \( 1.4 \, M_{\odot} \) is inferred to be \( \Lambda_{1.4} = 316.61^{+77.30}_{-35.30} \), which is in high agreement with the value obtained from the GW170817 event, \( \Lambda_{1.4} = 190^{+390}_{-120} \). The tidal deformability decreases with increasing mass and is \( 28.83^{+8.61}_{-8.42} \) at \( 2.0 \, M_{\odot} \), which is smaller than the results from the RMF model \cite{huang2020possibility} with a harder EOS. The central densities for the maximum masses of neutron stars are around \( 1.0 \, \text{fm}^{-3}\), which is approximately \( 6n_0 \), where the skewness coefficients \( Q_{\text{sat}} \) and \( Q_{\text{sym}} \) play a dominant role. Additionally, the maximum mass has strong correlations with the radius at \( 2.0 \, M_{\odot} \) and the central density from the posterior distributions.

\begin{figure}[htbp]
        \includegraphics[width=0.8\textwidth]{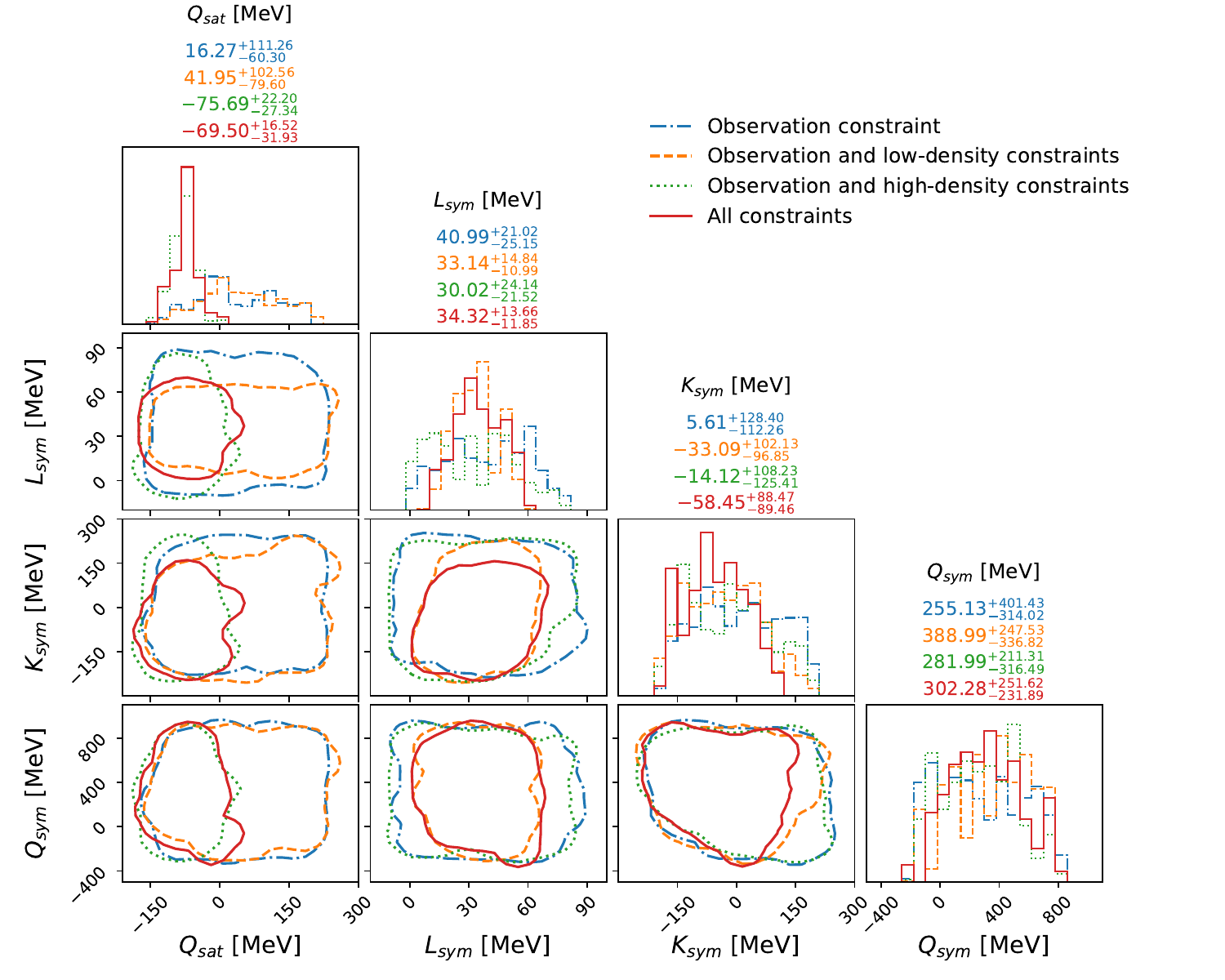} 
        \caption{The posterior distributions of parameters under various constraints. The contours delineate the \(3\sigma\) credible regions, while the histograms data represent the \(1\sigma\) marginal distributions.}
        \label{fig4}
    \end{figure}

Based on constraints from NICER observations, the crust-core phase transition density and the speed of sound are incorporated into the prior distributions to infer parameters in the polynomial EOSs. The posterior distributions of the parameters (\( Q_{\text{sat}},\, L_{\text{sym}},\, K_{\text{sym}},\, Q_{\text{sym}} \)) are presented in Fig.~\ref{fig4}, with distinct line styles differentiating each parameter. It is evident that \( L_{\text{sym}} \) and \( K_{\text{sym}} \) are significantly influenced by the crust-core phase transition limit. Specifically, \( L_{\text{sym}} \) shifts from \( 40.99^{+21.02}_{-25.15} \, \text{MeV}\) to \( 33.14^{+14.84}_{-10.99} \, \text{MeV}\), and \( K_{\text{sym}} \) shifts from \( 5.61^{+128.40}_{-112.26} \, \text{MeV}\) to \( -33.09^{+102.13}_{-96.85} \, \text{MeV}\). Meanwhile, \( Q_{\text{sat}} \) and \( Q_{\text{sym}} \) are mainly affected by the causal limit. The value of \( Q_{\text{sat}} \) is significantly reduced when the constraint \( c_s^2/c^2 < 1 \) is incorporated, shifting from \( 16.27^{+111.26}_{-60.30} \, \text{MeV}\) to \( -75.69^{+22.20}_{-27.34} \, \text{MeV}\). The narrow range of $Q_{\text{sat}}$ sufficiently demonstrated that the Bayesian inference is strongly dependent on the EOS model. We should consider all possible constraints as we have done. In fact, in many previous investigations, the region of $Q_{\text{sat}}$ were quit narrow. Such as, the study using the SHF model in Ref. \cite{PhysRevC.107.055803} shows that the posterior distribution of $ Q_{\text{sat}}$ (denoted as $Q_0$ in the paper) is concentrated in the range of -400 to -200 MeV (with a narrower 1$\sigma$ confidence interval). Similarly, the posterior distributions of $Q_{\text{sat}}$ in Refs. \cite{Beznogov2024apj,Beznogov_2024} have their 68\% confidence intervals confined to different ranges within 50-150 MeV. Similarly, \( Q_{\text{sym}} \) shifts from \([-58.89,\,656.56]\,\text{MeV}\) to \([-34.50,\,493.30]\,\text{MeV}\). Finally, all constraints---including those from neutron star observations, crust-core phase transition, and the causal limit---are considered together. The prior distributions and the posterior values at the \( 1\sigma \) credible level for different constraints on the parameters \( L_{\text{sym}},\, K_{\text{sym}},\, Q_{\text{sat}},\, Q_{\text{sym}} \) are listed in Table \ref{tab1}. All these values are consistent with previous investigations from heavy-ion collisions~\cite{xie2021bayesian} ($Q_{\text{sat}} = -180^{+100}_{-110}$ MeV), gravitational wave events~\cite{universe7060182} ($L_{\text{sym}} = 57.7 \pm 19$ MeV and $K_{\text{sym}} = -107 \pm 88$ MeV), and neutron star observables~\cite{xie2020bayesian} (for the Case 3 scenario with $200\,\mathrm{MeV} \leq Q_{\text{sym}} \leq 800\,\mathrm{MeV}$ providing $Q_{\text{sat}} = -100^{+20}_{-70}$ MeV and $K_{\text{sym}} = -30^{+80}_{-70}$ MeV) and neutron star observables~\cite{Zhang_2019} ($Q_{\text{sat}}$ to a lower limit of $-150$ MeV).

    \begingroup
    \setlength{\tabcolsep}{6pt}         
    \renewcommand{\arraystretch}{1.5}   
    \begin{table}[htbp]
    	\centering
    	\caption{The inferred parameters from various constraint at $1\sigma$ credible region.}
    	\begin{tabular}{cccccccc}
    		\hline \hline
    		-- & $Q_{\text{sat}}$ [MeV] & $L_{\text{sym}}$ [MeV] & $K_{\text{sym}}$ [MeV] & $Q_{\text{sym}}$ [MeV] \\
    		\hline 
    		Prior distribution & -200 to 200 & 0 to 80 & -200 to 200 & -200 to 800 \\
    		Observation limit  & $16.27_{-60.30}^{+111.26}$ & $40.99_{-25.15}^{+21.02}$ & $5.61_{-112.26}^{+128.40}$   & $255.13_{-314.02}^{+401.43}$ \\
    		Low-density limit  & $41.95_{-79.60}^{+102.56}$ & $33.14_{-10.99}^{+14.84}$ & $-33.09_{-96.85}^{+102.13}$  & $388.99_{-336.82}^{+247.53}$ \\
    		High-density limit & $-75.69_{-27.34}^{+22.20}$ & $30.02_{-21.52}^{+24.14}$ & $-14.12_{-125.41}^{+108.23}$ & $281.99_{-316.49}^{+211.31}$ \\
    		All limitations    & $-69.50_{-31.93}^{+16.52}$ & $34.32_{-11.85}^{+13.66}$ & $-58.45_{-89.46}^{+88.47}$   & $302.28_{-231.89}^{+251.62}$ \\
    		\hline\hline 
    	\end{tabular}
    	\label{tab1}
    \end{table}
    \endgroup  

\begin{figure}[htbp]
    	\centering
    	\includegraphics[width=0.6\textwidth]{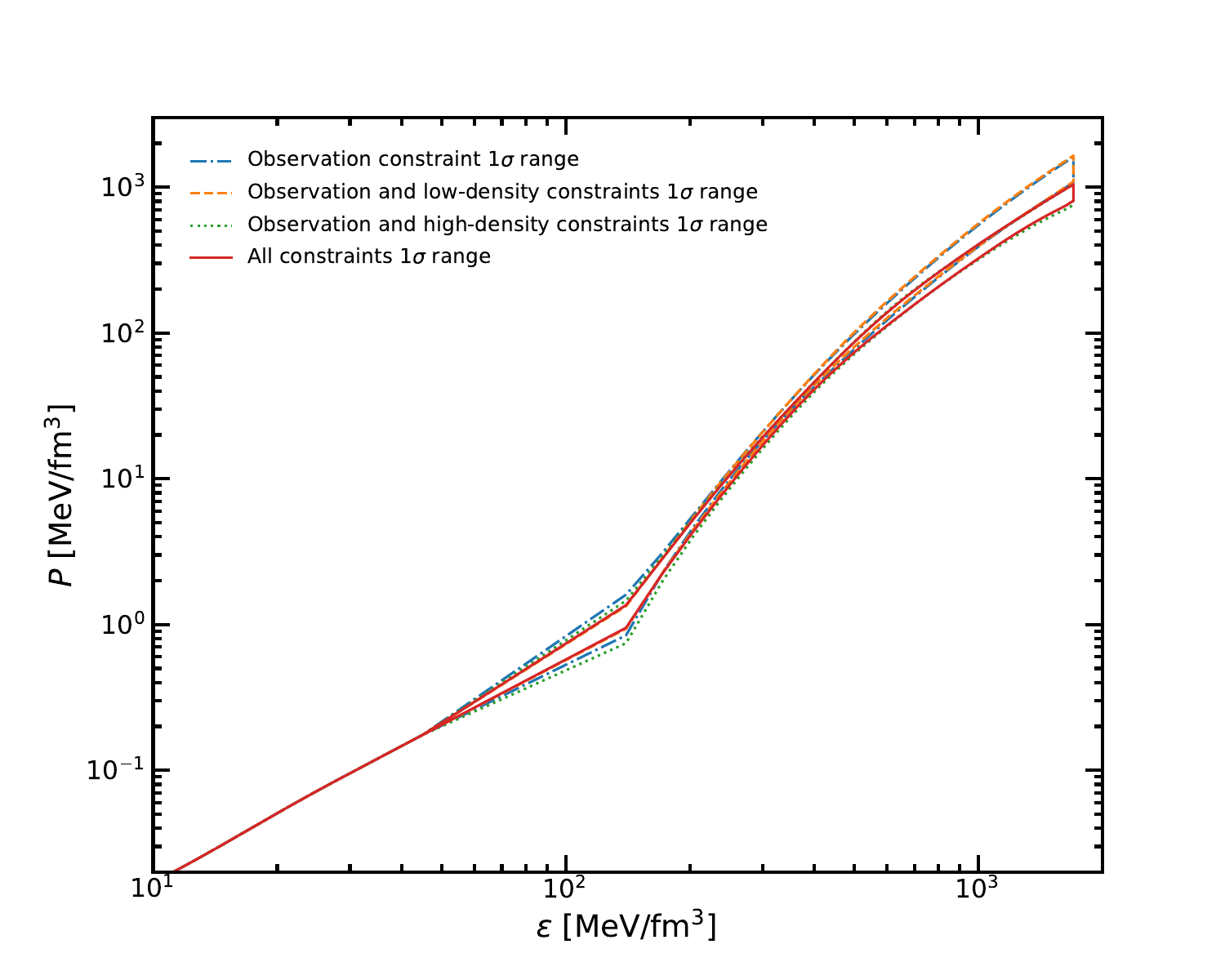} 
    	\caption{The EOSs of neutron star matter, \( P - \epsilon \), from various constraints.}
    	\label{fig5}
    \end{figure}

With the inferred saturation properties of nuclear matter, the parametrized EOSs, with pressure as a function of energy density (\(P-\epsilon\)), are plotted in Fig.~\ref{fig5} under various constraints. The dashed-dotted line represents the EOS derived from the mass-radius measurements of neutron stars, which exhibits the largest pressure at a fixed energy density in the low-density region. This EOS is particularly stiff at high densities, facilitating the formation of massive neutron stars. Meanwhile, incorporating constraints from the crust-core phase transition and symmetry energy at subsaturation density reduces the pressure. The causal limit further softens the EOS at high densities.

\begin{figure}[htbp]
    	\centering
    	\includegraphics[width=0.6\textwidth]{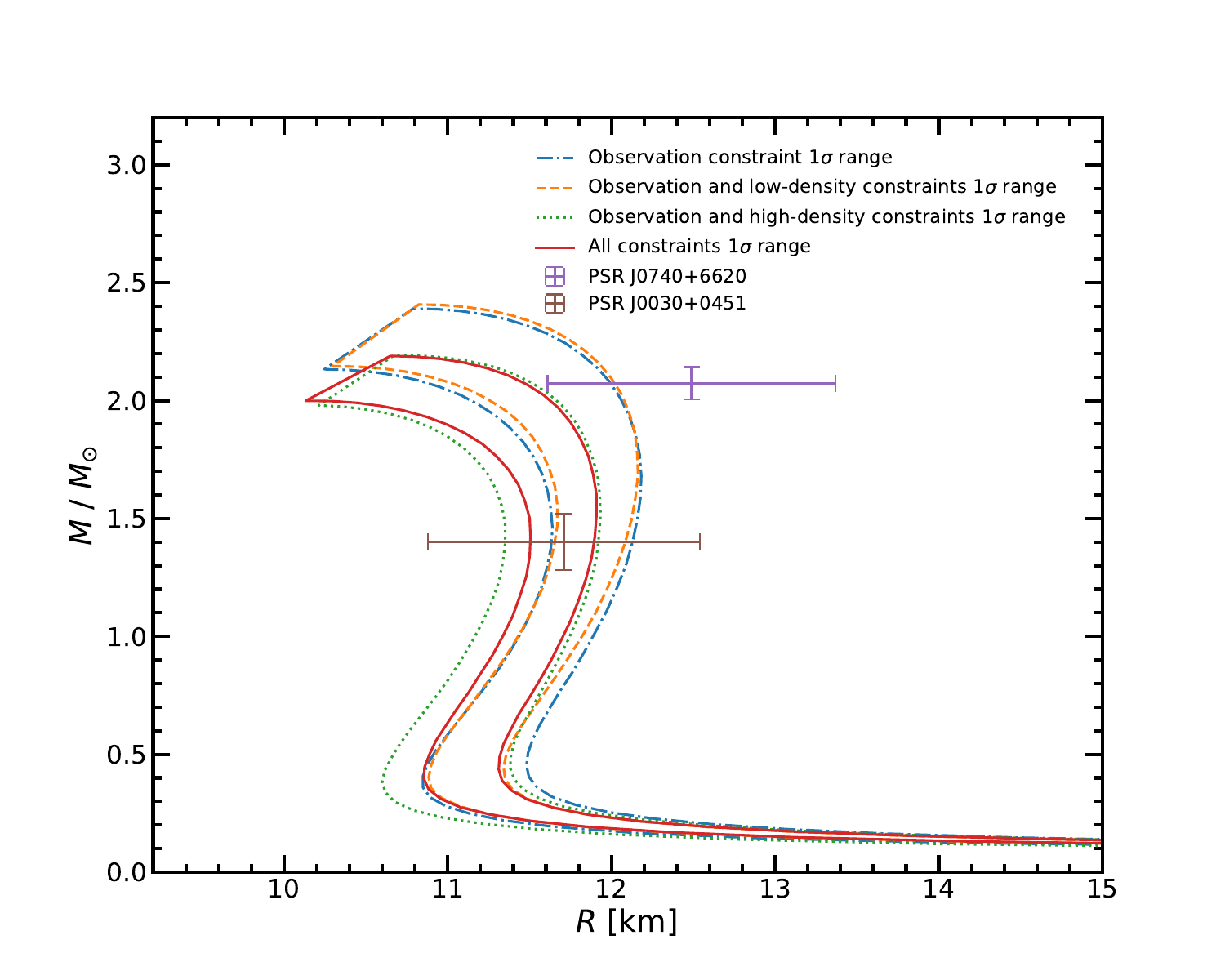} 
    	\caption{The mass-radius relations of neutron stars from various constraints. The purple and brown error bars represent the mass-radius distributions of PSR J0740+6620 and PSR J0030+0451, respectively.}
    	\label{fig6}
    \end{figure}

The mass-radius posterior distributions of the neutron star derived from the above EOSs is calculated by solving the TOV equation and is shown in Fig.~\ref{fig6}. Constraints at low density result in a smaller radius for neutron stars in the intermediate-mass region, while the constraints on the speed of sound reduce the maximum mass and radius of the neutron star. The masses and radii of PSR J0740+6620 and PSR J0030+0451, as observed by NICER, are also compared.

\begin{figure}[htbp]
	\centering
	\includegraphics[width=0.6\textwidth]{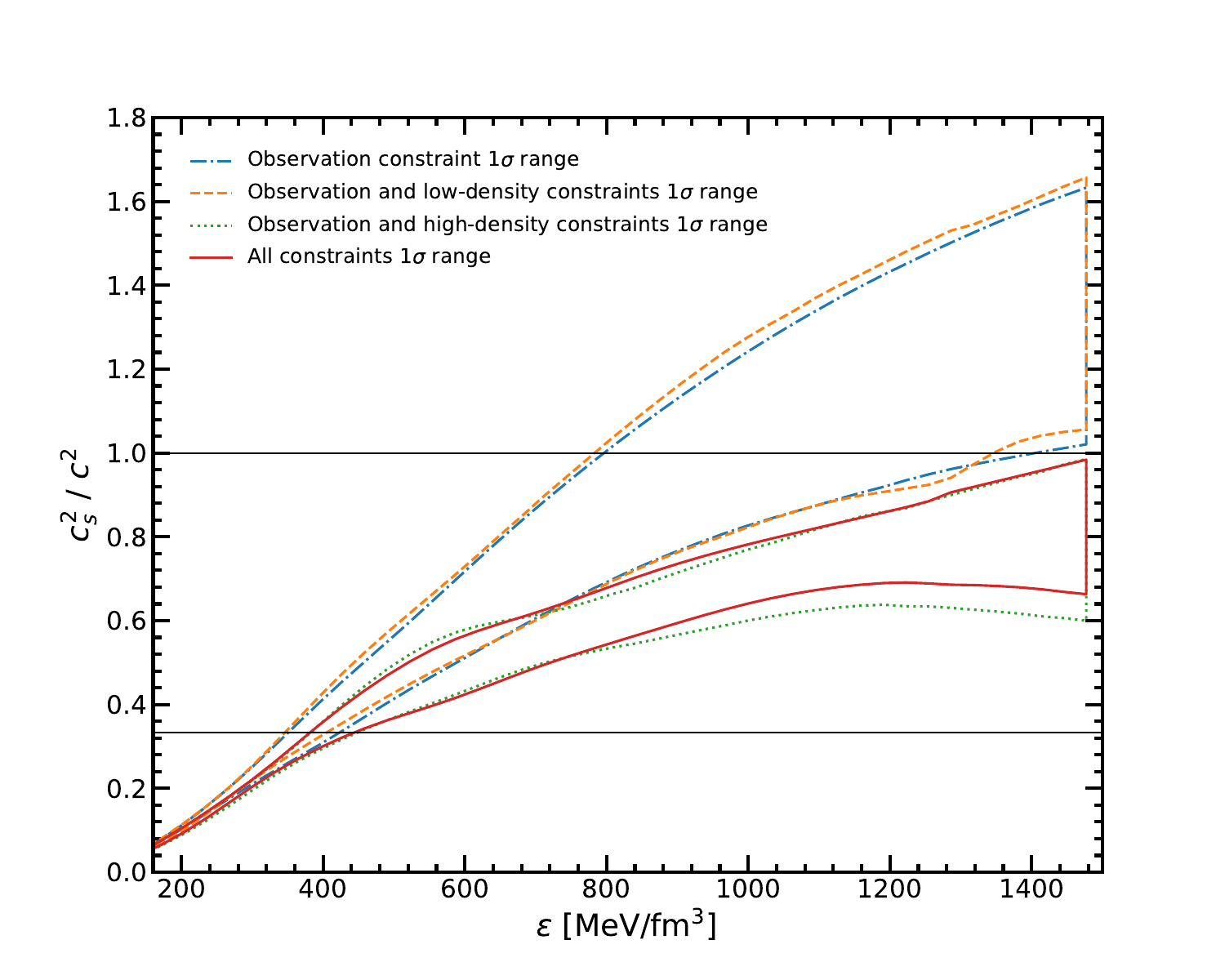} 
	\caption{The speeds of sound in neutron star matter under various constraints.}
	\label{fig7}
\end{figure}

The speeds of sound of neutron star matter under various constraints are plotted in Fig.~\ref{fig7}. If we only consider the mass-radius and low-density constraints in the parametrized EOSs, the speed of sound for \(\epsilon > 800 \, \text{MeV}/\text{fm}^{3}\) can easily exceed the speed of light, violating causality. However, when the causal constraint is included, the speed of sound remains less than \(1\). The effect of the low-density limit on the speed of sound can be completely ignored. In fact, the roles of these four constraints are very similar for the dimensionless tidal deformability, \(\Lambda\).

\begin{figure}[htbp]
	\centering
	\includegraphics[width=0.6\textwidth]{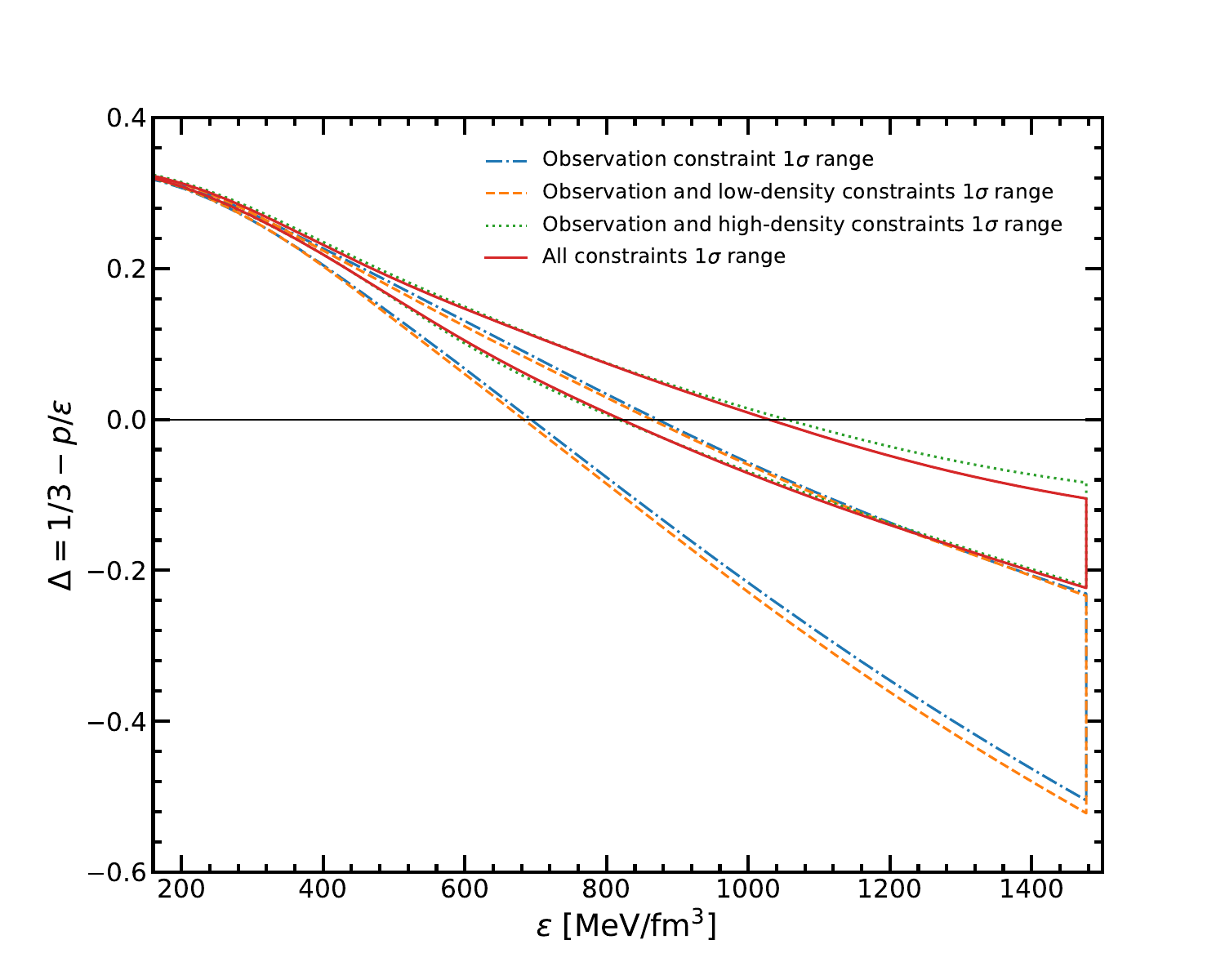} 
	\caption{The trace anomaly of neutron star matter under various constraints.}
	\label{fig8}
\end{figure}

The trace anomaly of neutron star matter, defined as \(\Delta = 1/3 - P/\epsilon\), indicates the restoration of conformability, approaching \(0\) at the conformal limit. When the constraint on the speed of sound is considered, the trace anomaly in the high-density region is closer to \(0\) compared to the results obtained with constraints from neutron star mass-radius measurements and crust-core phase transitions in Fig.~\ref{fig8}. This behavior is consistent with other calculations from perturbative QCD and neutron star mass-radius observables \cite{fujimoto2022trace}.

\begin{figure}[htbp]
	\centering
	\includegraphics[width=0.9\textwidth]{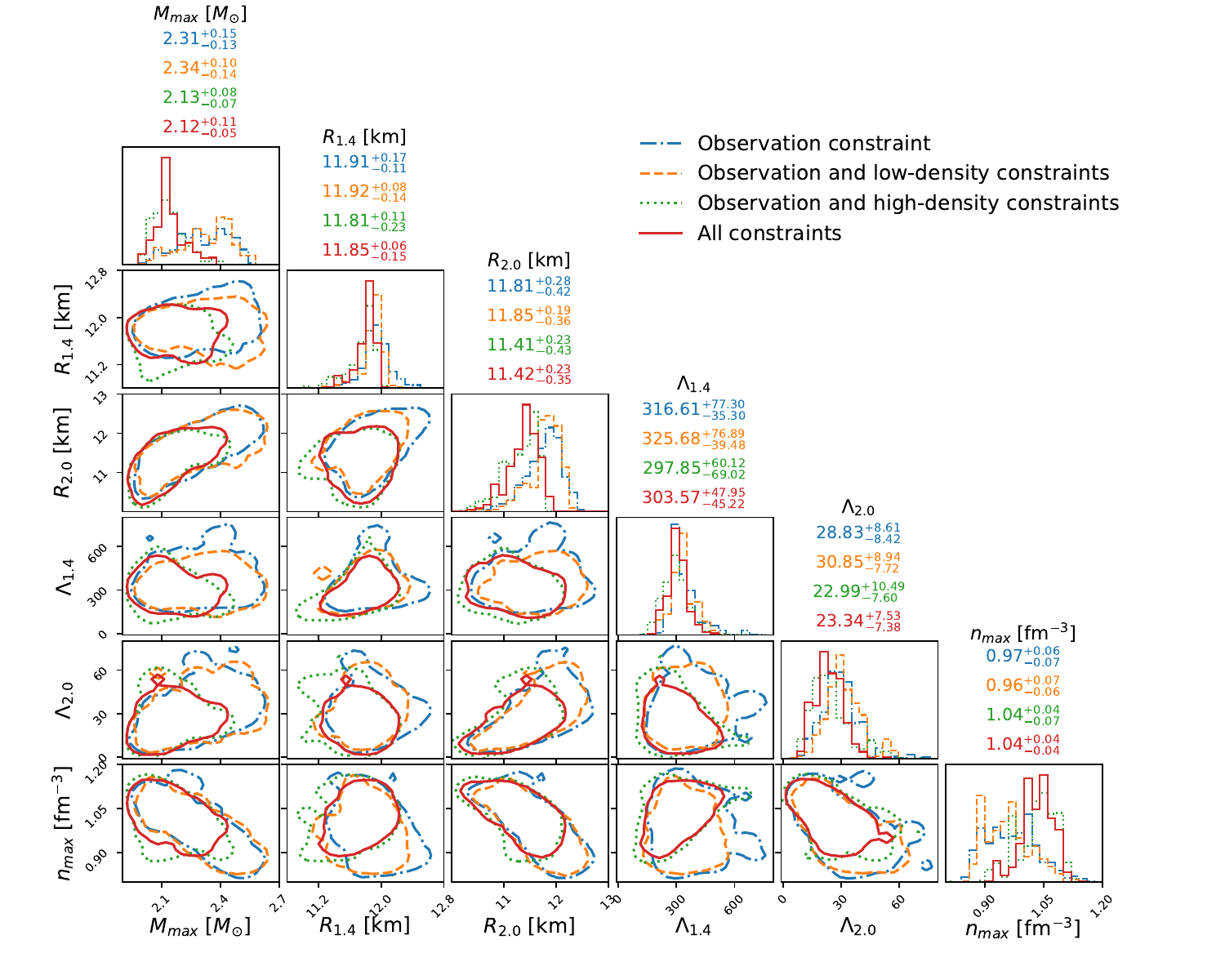} 
	\caption{Posterior distributions of neutron star properties under four kind constraint scenarios.}
	\label{fig9}
\end{figure}

The posterior distributions of neutron star properties derived from EOSs constrained by various constraints are compared in Fig.~\ref{fig9}, highlighting the significant impact of the speed of sound in neutron star matter. The maximum mass of neutron stars, \( M_{\text{max}} \), decreases by approximately \( 10\% \) after incorporating the speed of sound limit, resulting in \( M_{\text{max}} = 2.13^{+0.08}_{-0.07} \, M_{\odot} \), which shows excellent agreement with the result of \( M_{\text{max}} = 2.144^{+0.133}_{-0.092} M_{\odot} \) reported in Ref. \cite{Malik_2022}, while being notably lower than the value of \( M_{\text{max}} = 2.25^{+0.08}_{-0.07} M_{\odot} \) from Ref. \cite{PhysRevD.109.043052}. The crust-core phase transition and causal limit have minimal influence on the radius of neutron stars at \( 1.4 \, M_{\odot} \) (\( R_{1.4} \)), which is primarily determined by the EOSs at densities around \( 2n_0 \sim 3n_0 \). After including low-density constraints, we obtain a remarkably stable result for the canonical 1.4 \( M_{\odot} \) neutron star radius (\( R_{1.4} = 11.92^{+0.08}_{-0.14} \) km), that is consistent with Refs. \cite{chimanski2022bayesianinferencephenomenologicaleos, Traversi_2020} (\( R_{1.4} \approx 12 \) km). However, the radius of heavier neutron stars at \( 2.0 \, M_{\odot} \) (\( R_{2.0} \)) is significantly affected by high-density behaviors. For the 2.0 \( M_{\odot} \) configuration, we generate its radius  \( R_{2.0} = 11.85^{+0.19}_{-0.36} \) km, which is lower than the values reported in Ref. \cite{Altiparmak_2022} (\( R_{2.0} = 12.12^{+1.11}_{-1.23} \) km), Ref. \cite{char2023generaliseddescriptionneutronstar} (\( R_{2.0} = 12.58^{+0.68}_{-0.67} \) km), and Scenario B (\( R_{2.0} = 12.04^{+0.91}_{-1.03} \) km) and Scenario F (\( R_{2.0} = 12.60^{+0.88}_{-1.14} \) km) in Refs. \cite{LI2025139501, PhysRevC.111.055804, li2025112}. These systematic offsets likely originate from the stricter constraints on the symmetry energy at subsaturation points that we have imposed in this work. Correspondingly, the dimensionless tidal deformabilities at \( 1.4 \, M_{\odot} \) and \( 2.0 \, M_{\odot} \) (\( \Lambda_{1.4} \) and \( \Lambda_{2.0} \)) decrease. Specifically, \( \Lambda_{1.4} \) is \( 303.57_{-45.22}^{+47.95} \), which is consistent with data extracted from the GW170817 event \cite{PhysRevLett.119.161101, Annala2018}. The central density of the maximum mass neutron star increases from \( 0.97^{+0.06}_{-0.07} \, \text{fm}^{-3} \) to \( 1.04^{+0.04}_{-0.04} \, \text{fm}^{-3} \). All these values are summarized in Table \ref{tab2}.

    \begingroup
    \setlength{\tabcolsep}{3pt}
    \renewcommand{\arraystretch}{1.5}
    \begin{table}
    	\centering
        \caption{Properties of neutron star derived from EOSs inferred by various constraints.}
    	\begin{tabular}{ccccccc}
    		\hline 
    		-- & $M_{\rm max}$ [$M_{\odot}$] & $R_{1.4}$ [km] & $R_{2.0}$ [km] & $\Lambda_{1.4}$ & $\Lambda_{2.0}$ & $n_{\rm max}$ [${\rm fm}^{-3}$] \\
    		\hline 
    		Observation limit  & $2.31_{-0.13}^{+0.15}$ & $11.91_{-0.11}^{+0.17}$ & $11.81_{-0.42}^{+0.28}$ & $316.61_{-35.30}^{+77.30}$ & $28.83_{-8.42}^{+8.61}$  & $0.97_{-0.07}^{+0.06}$ \\
    		Low-density limit  & $2.34_{-0.14}^{+0.10}$ & $11.92_{-0.14}^{+0.08}$ & $11.85_{-0.36}^{+0.19}$ & $325.68_{-39.48}^{+76.89}$ & $30.85_{-7.72}^{+8.94}$  & $0.96_{-0.06}^{+0.07}$ \\
    		High-density limit & $2.13_{-0.07}^{+0.08}$ & $11.81_{-0.23}^{+0.11}$ & $11.41_{-0.43}^{+0.23}$ & $297.85_{-69.02}^{+60.12}$ & $22.99_{-7.60}^{+10.49}$ & $1.04_{-0.07}^{+0.04}$ \\
    		All limitations    & $2.12_{-0.05}^{+0.11}$ & $11.85_{-0.15}^{+0.06}$ & $11.42_{-0.35}^{+0.23}$ & $303.57_{-45.22}^{+47.95}$ & $23.34_{-7.38}^{+7.53}$  & $1.04_{-0.04}^{+0.04}$ \\
    		\hline 
    	\end{tabular}
    	\label{tab2}
    \end{table}
    \endgroup

	\section{Summary}\label{sec4}
    
    The parametrized equations of state (EOSs) of neutron star matter, which are related to the saturation properties of symmetric nuclear matter and symmetry energy, were investigated using Bayesian inference. Constraints were applied from the neutron star mass-radius observations, the crust-core phase transition density, the symmetry energy at subsaturation density, and the causal limit. These constraints distinguish our work from previous parameter-space constraints on the neutron star EOS. The energy per nucleon (\(E/A\)), symmetry energy (\(E_{\text{sym}}\)), and incompressibility (\(K_{\text{sat}}\)) at nuclear saturation density in the parametrized EOS were fixed based on empirical data extracted from the properties of finite nuclei.

    The skewness of symmetric nuclear matter, \( Q_{\text{sat}} \), and the slope, curvature, and skewness of symmetry energy, \( L_{\text{sym}} \), \( K_{\text{sym}} \), and \( Q_{\text{sym}} \), were inferred from the above constraints using Bayes' theorem. \( L_{\text{sym}} \) and \( K_{\text{sym}} \) exhibit a strong linear correlation due to the consideration of the symmetry energy at the nuclear subsaturation density, which is well-examined by the finite nuclei system. This constraint was often neglected in previous investigations. Their values are also influenced by the density region of the crust-core phase transition. The slope and curvature of the symmetry energy were inferred as \( L_{\text{sym}} = 34.32_{-11.85}^{+13.66} \, \text{MeV} \) and \( K_{\text{sym}} = -58.45_{-89.46}^{+88.47} \, \text{MeV} \), respectively. Meanwhile, \( Q_{\text{sat}} \) and \( Q_{\text{sym}} \) are restricted by the properties of massive neutron stars and the causal limit, and were estimated as \( -69.50_{-31.93}^{+16.52} \, \text{MeV} \) and \( 302.28_{-231.89}^{+251.62} \, \text{MeV} \), respectively. These inferred results are entirely consistent with other studies from neutron star observables and heavy-ion collision experiments.
    
    With the inferred EOSs, the properties of neutron star can be calculated using the Tolman-Oppenheimer-Volkoff (TOV) equation. The maximum neutron star mass decreases by about \(10\%\) after incorporating the constraint on the speed of sound. However, the radius of the neutron star at \(1.4 \, M_{\odot}\) remains almost unchanged at \(11.85^{+0.06}_{-0.15} \, \text{km}\). The dimensionless tidal deformability at \(1.4 \, M_{\odot}\) is \(303.57^{+47.95}_{-45.22}\), consistent with the analysis of GW170817 event. Furthermore, due to the speed of sound constraint in neutron stars, the trace anomaly, which indicates the deviation from the conformal limit, approaches \(0\) in the central region of the neutron star. In this work, the parametrized EOSs of neutron stars were estimated using properties derived from both neutron stars and finite nuclei. The experimental data of heavy ion collision and hadron-quark phase transition will be considered in future work. Future work will consider experimental data from heavy-ion collisions and hadron-quark phase transitions.

	\section{Acknowledgments}
    Xieyuan Dong special thanks to Chun Huang (Washington University in St. Louis) for his guidance on the \texttt{CompactObject} package implementation.This work was supported in part by the National Natural Science Foundation of China under Grants Nos. 12175109 and 12475149, and by the Natural Science Foundation of Guangdong Province (Grant No. 2024A1515010911).

	\bibliographystyle{apsrev4-1}
	\bibliography{references}

\begin{thebibliography}{112}%
\makeatletter
\providecommand \@ifxundefined [1]{%
 \@ifx{#1\undefined}
}%
\providecommand \@ifnum [1]{%
 \ifnum #1\expandafter \@firstoftwo
 \else \expandafter \@secondoftwo
 \fi
}%
\providecommand \@ifx [1]{%
 \ifx #1\expandafter \@firstoftwo
 \else \expandafter \@secondoftwo
 \fi
}%
\providecommand \natexlab [1]{#1}%
\providecommand \enquote  [1]{``#1''}%
\providecommand \bibnamefont  [1]{#1}%
\providecommand \bibfnamefont [1]{#1}%
\providecommand \citenamefont [1]{#1}%
\providecommand \href@noop [0]{\@secondoftwo}%
\providecommand \href [0]{\begingroup \@sanitize@url \@href}%
\providecommand \@href[1]{\@@startlink{#1}\@@href}%
\providecommand \@@href[1]{\endgroup#1\@@endlink}%
\providecommand \@sanitize@url [0]{\catcode `\\12\catcode `\$12\catcode
  `\&12\catcode `\#12\catcode `\^12\catcode `\_12\catcode `\%12\relax}%
\providecommand \@@startlink[1]{}%
\providecommand \@@endlink[0]{}%
\providecommand \url  [0]{\begingroup\@sanitize@url \@url }%
\providecommand \@url [1]{\endgroup\@href {#1}{\urlprefix }}%
\providecommand \urlprefix  [0]{URL }%
\providecommand \Eprint [0]{\href }%
\providecommand \doibase [0]{http://dx.doi.org/}%
\providecommand \selectlanguage [0]{\@gobble}%
\providecommand \bibinfo  [0]{\@secondoftwo}%
\providecommand \bibfield  [0]{\@secondoftwo}%
\providecommand \translation [1]{[#1]}%
\providecommand \BibitemOpen [0]{}%
\providecommand \bibitemStop [0]{}%
\providecommand \bibitemNoStop [0]{.\EOS\space}%
\providecommand \EOS [0]{\spacefactor3000\relax}%
\providecommand \BibitemShut  [1]{\csname bibitem#1\endcsname}%
\let\auto@bib@innerbib\@empty
\bibitem [{\citenamefont {Cromartie}\ \emph {et~al.}(2020)\citenamefont
  {Cromartie}, \citenamefont {Fonseca}, \citenamefont {Ransom} \emph
  {et~al.}}]{Cromartie2019}%
  \BibitemOpen
  \bibfield  {author} {\bibinfo {author} {\bibfnamefont {H.~T.}\ \bibnamefont
  {Cromartie}}, \bibinfo {author} {\bibfnamefont {E.}~\bibnamefont {Fonseca}},
  \bibinfo {author} {\bibfnamefont {S.~M.}\ \bibnamefont {Ransom}},  \emph
  {et~al.},\ }\href {\doibase 10.1038/s41550-019-0880-2} {\bibfield  {journal}
  {\bibinfo  {journal} {Nature Astronomy}\ }\textbf {\bibinfo {volume} {4}},\
  \bibinfo {pages} {72} (\bibinfo {year} {2020})}\BibitemShut {NoStop}%
\bibitem [{\citenamefont {Demorest}\ \emph {et~al.}(2010)\citenamefont
  {Demorest}, \citenamefont {Pennucci}, \citenamefont {Ransom}, \citenamefont
  {Roberts},\ and\ \citenamefont {Hessels}}]{Demorest2010}%
  \BibitemOpen
  \bibfield  {author} {\bibinfo {author} {\bibfnamefont {P.~B.}\ \bibnamefont
  {Demorest}}, \bibinfo {author} {\bibfnamefont {T.}~\bibnamefont {Pennucci}},
  \bibinfo {author} {\bibfnamefont {S.~M.}\ \bibnamefont {Ransom}}, \bibinfo
  {author} {\bibfnamefont {M.~S.~E.}\ \bibnamefont {Roberts}}, \ and\ \bibinfo
  {author} {\bibfnamefont {J.~W.~T.}\ \bibnamefont {Hessels}},\ }\href
  {\doibase 10.1038/nature09466} {\bibfield  {journal} {\bibinfo  {journal}
  {Nature}\ }\textbf {\bibinfo {volume} {467}},\ \bibinfo {pages} {1081}
  (\bibinfo {year} {2010})}\BibitemShut {NoStop}%
\bibitem [{\citenamefont {Antoniadis}\ \emph {et~al.}(2013)\citenamefont
  {Antoniadis}, \citenamefont {Freire}, \citenamefont {Wex}, \citenamefont
  {Tauris}, \citenamefont {Lynch} \emph {et~al.}}]{Antoniadis2013}%
  \BibitemOpen
  \bibfield  {author} {\bibinfo {author} {\bibfnamefont {J.}~\bibnamefont
  {Antoniadis}}, \bibinfo {author} {\bibfnamefont {P.~C.~C.}\ \bibnamefont
  {Freire}}, \bibinfo {author} {\bibfnamefont {N.}~\bibnamefont {Wex}},
  \bibinfo {author} {\bibfnamefont {T.~M.}\ \bibnamefont {Tauris}}, \bibinfo
  {author} {\bibfnamefont {R.~S.}\ \bibnamefont {Lynch}},  \emph {et~al.},\
  }\href {\doibase 10.1126/science.1233232} {\bibfield  {journal} {\bibinfo
  {journal} {Science}\ }\textbf {\bibinfo {volume} {340}},\ \bibinfo {pages}
  {1233232} (\bibinfo {year} {2013})}\BibitemShut {NoStop}%
\bibitem [{\citenamefont {Romani}\ \emph {et~al.}(2022)\citenamefont {Romani},
  \citenamefont {Kandel}, \citenamefont {Filippenko}, \citenamefont {Brink},\
  and\ \citenamefont {Zheng}}]{romani2022psr}%
  \BibitemOpen
  \bibfield  {author} {\bibinfo {author} {\bibfnamefont {R.~W.}\ \bibnamefont
  {Romani}}, \bibinfo {author} {\bibfnamefont {D.}~\bibnamefont {Kandel}},
  \bibinfo {author} {\bibfnamefont {A.~V.}\ \bibnamefont {Filippenko}},
  \bibinfo {author} {\bibfnamefont {T.~G.}\ \bibnamefont {Brink}}, \ and\
  \bibinfo {author} {\bibfnamefont {W.}~\bibnamefont {Zheng}},\ }\href@noop {}
  {\bibfield  {journal} {\bibinfo  {journal} {The Astrophysical Journal
  Letters}\ }\textbf {\bibinfo {volume} {934}},\ \bibinfo {pages} {L17}
  (\bibinfo {year} {2022})}\BibitemShut {NoStop}%
\bibitem [{\citenamefont {Abbott}\ \emph
  {et~al.}(2017{\natexlab{a}})\citenamefont {Abbott}, \citenamefont {Abbott},
  \citenamefont {Abbott}, \citenamefont {Acernese}, \citenamefont {Ackley}
  \emph {et~al.}}]{Abbott_2017}%
  \BibitemOpen
  \bibfield  {author} {\bibinfo {author} {\bibfnamefont {B.~P.}\ \bibnamefont
  {Abbott}}, \bibinfo {author} {\bibfnamefont {R.}~\bibnamefont {Abbott}},
  \bibinfo {author} {\bibfnamefont {T.~D.}\ \bibnamefont {Abbott}}, \bibinfo
  {author} {\bibfnamefont {F.}~\bibnamefont {Acernese}}, \bibinfo {author}
  {\bibfnamefont {K.}~\bibnamefont {Ackley}},  \emph {et~al.},\ }\href
  {\doibase 10.3847/2041-8213/aa91c9} {\bibfield  {journal} {\bibinfo
  {journal} {The Astrophysical Journal Letters}\ }\textbf {\bibinfo {volume}
  {848}},\ \bibinfo {pages} {L12} (\bibinfo {year}
  {2017}{\natexlab{a}})}\BibitemShut {NoStop}%
\bibitem [{\citenamefont {Abbott}\ \emph
  {et~al.}(2017{\natexlab{b}})\citenamefont {Abbott}, \citenamefont {Abbott},
  \citenamefont {Abbott}, \citenamefont {Acernese}, \citenamefont {Ackley}
  \emph {et~al.}}]{PhysRevLett.119.161101}%
  \BibitemOpen
  \bibfield  {author} {\bibinfo {author} {\bibfnamefont {B.~P.}\ \bibnamefont
  {Abbott}}, \bibinfo {author} {\bibfnamefont {R.}~\bibnamefont {Abbott}},
  \bibinfo {author} {\bibfnamefont {T.~D.}\ \bibnamefont {Abbott}}, \bibinfo
  {author} {\bibfnamefont {F.}~\bibnamefont {Acernese}}, \bibinfo {author}
  {\bibfnamefont {K.}~\bibnamefont {Ackley}},  \emph {et~al.} (\bibinfo
  {collaboration} {LIGO Scientific Collaboration and Virgo Collaboration}),\
  }\href {\doibase 10.1103/PhysRevLett.119.161101} {\bibfield  {journal}
  {\bibinfo  {journal} {Phys. Rev. Lett.}\ }\textbf {\bibinfo {volume} {119}},\
  \bibinfo {pages} {161101} (\bibinfo {year} {2017}{\natexlab{b}})}\BibitemShut
  {NoStop}%
\bibitem [{\citenamefont {Riley}\ \emph {et~al.}(2019)\citenamefont {Riley},
  \citenamefont {Watts}, \citenamefont {Bogdanov}, \citenamefont {Ray},
  \citenamefont {Ludlam} \emph {et~al.}}]{Riley_2019}%
  \BibitemOpen
  \bibfield  {author} {\bibinfo {author} {\bibfnamefont {T.~E.}\ \bibnamefont
  {Riley}}, \bibinfo {author} {\bibfnamefont {A.~L.}\ \bibnamefont {Watts}},
  \bibinfo {author} {\bibfnamefont {S.}~\bibnamefont {Bogdanov}}, \bibinfo
  {author} {\bibfnamefont {P.~S.}\ \bibnamefont {Ray}}, \bibinfo {author}
  {\bibfnamefont {R.~M.}\ \bibnamefont {Ludlam}},  \emph {et~al.},\ }\href
  {\doibase 10.3847/2041-8213/ab481c} {\bibfield  {journal} {\bibinfo
  {journal} {The Astrophysical Journal Letters}\ }\textbf {\bibinfo {volume}
  {887}},\ \bibinfo {pages} {L21} (\bibinfo {year} {2019})}\BibitemShut
  {NoStop}%
\bibitem [{\citenamefont {Miller}\ \emph {et~al.}(2019)\citenamefont {Miller},
  \citenamefont {Lamb}, \citenamefont {Dittmann}, \citenamefont {Bogdanov},
  \citenamefont {Arzoumanian} \emph {et~al.}}]{Miller_2019}%
  \BibitemOpen
  \bibfield  {author} {\bibinfo {author} {\bibfnamefont {M.~C.}\ \bibnamefont
  {Miller}}, \bibinfo {author} {\bibfnamefont {F.~K.}\ \bibnamefont {Lamb}},
  \bibinfo {author} {\bibfnamefont {A.~J.}\ \bibnamefont {Dittmann}}, \bibinfo
  {author} {\bibfnamefont {S.}~\bibnamefont {Bogdanov}}, \bibinfo {author}
  {\bibfnamefont {Z.}~\bibnamefont {Arzoumanian}},  \emph {et~al.},\ }\href
  {\doibase 10.3847/2041-8213/ab50c5} {\bibfield  {journal} {\bibinfo
  {journal} {The Astrophysical Journal Letters}\ }\textbf {\bibinfo {volume}
  {887}},\ \bibinfo {pages} {L24} (\bibinfo {year} {2019})}\BibitemShut
  {NoStop}%
\bibitem [{\citenamefont {Riley}\ \emph {et~al.}(2021)\citenamefont {Riley},
  \citenamefont {Watts}, \citenamefont {Ray}, \citenamefont {Bogdanov},
  \citenamefont {Guillot} \emph {et~al.}}]{Riley_2021}%
  \BibitemOpen
  \bibfield  {author} {\bibinfo {author} {\bibfnamefont {T.~E.}\ \bibnamefont
  {Riley}}, \bibinfo {author} {\bibfnamefont {A.~L.}\ \bibnamefont {Watts}},
  \bibinfo {author} {\bibfnamefont {P.~S.}\ \bibnamefont {Ray}}, \bibinfo
  {author} {\bibfnamefont {S.}~\bibnamefont {Bogdanov}}, \bibinfo {author}
  {\bibfnamefont {S.}~\bibnamefont {Guillot}},  \emph {et~al.},\ }\href
  {\doibase 10.3847/2041-8213/ac0a81} {\bibfield  {journal} {\bibinfo
  {journal} {The Astrophysical Journal Letters}\ }\textbf {\bibinfo {volume}
  {918}},\ \bibinfo {pages} {L27} (\bibinfo {year} {2021})}\BibitemShut
  {NoStop}%
\bibitem [{\citenamefont {Miller}\ \emph {et~al.}(2021)\citenamefont {Miller},
  \citenamefont {Lamb}, \citenamefont {Dittmann}, \citenamefont {Bogdanov},
  \citenamefont {Arzoumanian} \emph {et~al.}}]{Miller_2021}%
  \BibitemOpen
  \bibfield  {author} {\bibinfo {author} {\bibfnamefont {M.~C.}\ \bibnamefont
  {Miller}}, \bibinfo {author} {\bibfnamefont {F.~K.}\ \bibnamefont {Lamb}},
  \bibinfo {author} {\bibfnamefont {A.~J.}\ \bibnamefont {Dittmann}}, \bibinfo
  {author} {\bibfnamefont {S.}~\bibnamefont {Bogdanov}}, \bibinfo {author}
  {\bibfnamefont {Z.}~\bibnamefont {Arzoumanian}},  \emph {et~al.},\ }\href
  {\doibase 10.3847/2041-8213/ac089b} {\bibfield  {journal} {\bibinfo
  {journal} {The Astrophysical Journal Letters}\ }\textbf {\bibinfo {volume}
  {918}},\ \bibinfo {pages} {L28} (\bibinfo {year} {2021})}\BibitemShut
  {NoStop}%
\bibitem [{\citenamefont {Huang}\ \emph {et~al.}(2020)\citenamefont {Huang},
  \citenamefont {Hu}, \citenamefont {Zhang},\ and\ \citenamefont
  {Shen}}]{huang2020possibility}%
  \BibitemOpen
  \bibfield  {author} {\bibinfo {author} {\bibfnamefont {K.}~\bibnamefont
  {Huang}}, \bibinfo {author} {\bibfnamefont {J.}~\bibnamefont {Hu}}, \bibinfo
  {author} {\bibfnamefont {Y.}~\bibnamefont {Zhang}}, \ and\ \bibinfo {author}
  {\bibfnamefont {H.}~\bibnamefont {Shen}},\ }\href@noop {} {\bibfield
  {journal} {\bibinfo  {journal} {The Astrophysical Journal}\ }\textbf
  {\bibinfo {volume} {904}},\ \bibinfo {pages} {39} (\bibinfo {year}
  {2020})}\BibitemShut {NoStop}%
\bibitem [{\citenamefont {De}\ \emph {et~al.}(2018)\citenamefont {De},
  \citenamefont {Finstad}, \citenamefont {Lattimer}, \citenamefont {Brown},
  \citenamefont {Berger},\ and\ \citenamefont {Biwer}}]{De2018}%
  \BibitemOpen
  \bibfield  {author} {\bibinfo {author} {\bibfnamefont {S.}~\bibnamefont
  {De}}, \bibinfo {author} {\bibfnamefont {D.}~\bibnamefont {Finstad}},
  \bibinfo {author} {\bibfnamefont {J.~M.}\ \bibnamefont {Lattimer}}, \bibinfo
  {author} {\bibfnamefont {D.~A.}\ \bibnamefont {Brown}}, \bibinfo {author}
  {\bibfnamefont {E.}~\bibnamefont {Berger}}, \ and\ \bibinfo {author}
  {\bibfnamefont {C.~M.}\ \bibnamefont {Biwer}},\ }\href {\doibase
  10.1103/PhysRevLett.121.091102} {\bibfield  {journal} {\bibinfo  {journal}
  {Phys. Rev. Lett.}\ }\textbf {\bibinfo {volume} {121}},\ \bibinfo {pages}
  {091102} (\bibinfo {year} {2018})}\BibitemShut {NoStop}%
\bibitem [{\citenamefont {Annala}\ \emph {et~al.}(2018)\citenamefont {Annala},
  \citenamefont {Gorda}, \citenamefont {Kurkela},\ and\ \citenamefont
  {Vuorinen}}]{Annala2018}%
  \BibitemOpen
  \bibfield  {author} {\bibinfo {author} {\bibfnamefont {E.}~\bibnamefont
  {Annala}}, \bibinfo {author} {\bibfnamefont {T.}~\bibnamefont {Gorda}},
  \bibinfo {author} {\bibfnamefont {A.}~\bibnamefont {Kurkela}}, \ and\
  \bibinfo {author} {\bibfnamefont {A.}~\bibnamefont {Vuorinen}},\ }\href
  {\doibase 10.1103/PhysRevLett.120.172703} {\bibfield  {journal} {\bibinfo
  {journal} {Phys. Rev. Lett.}\ }\textbf {\bibinfo {volume} {120}},\ \bibinfo
  {pages} {172703} (\bibinfo {year} {2018})}\BibitemShut {NoStop}%
\bibitem [{\citenamefont {Most}\ \emph {et~al.}(2018)\citenamefont {Most},
  \citenamefont {Weih}, \citenamefont {Rezzolla},\ and\ \citenamefont
  {Schaffner-Bielich}}]{Most2018}%
  \BibitemOpen
  \bibfield  {author} {\bibinfo {author} {\bibfnamefont {E.~R.}\ \bibnamefont
  {Most}}, \bibinfo {author} {\bibfnamefont {L.~R.}\ \bibnamefont {Weih}},
  \bibinfo {author} {\bibfnamefont {L.}~\bibnamefont {Rezzolla}}, \ and\
  \bibinfo {author} {\bibfnamefont {J.}~\bibnamefont {Schaffner-Bielich}},\
  }\href {\doibase 10.1103/PhysRevLett.120.261103} {\bibfield  {journal}
  {\bibinfo  {journal} {Phys. Rev. Lett.}\ }\textbf {\bibinfo {volume} {120}},\
  \bibinfo {pages} {261103} (\bibinfo {year} {2018})}\BibitemShut {NoStop}%
\bibitem [{\citenamefont {Raaijmakers}\ \emph {et~al.}(2019)\citenamefont
  {Raaijmakers}, \citenamefont {Riley}, \citenamefont {Watts}, \citenamefont
  {Greif}, \citenamefont {Morsink} \emph {et~al.}}]{Raaijmakers2019}%
  \BibitemOpen
  \bibfield  {author} {\bibinfo {author} {\bibfnamefont {G.}~\bibnamefont
  {Raaijmakers}}, \bibinfo {author} {\bibfnamefont {T.~E.}\ \bibnamefont
  {Riley}}, \bibinfo {author} {\bibfnamefont {A.~L.}\ \bibnamefont {Watts}},
  \bibinfo {author} {\bibfnamefont {S.~K.}\ \bibnamefont {Greif}}, \bibinfo
  {author} {\bibfnamefont {S.~M.}\ \bibnamefont {Morsink}},  \emph {et~al.},\
  }\href {\doibase 10.3847/2041-8213/ab4519} {\bibfield  {journal} {\bibinfo
  {journal} {The Astrophysical Journal Letters}\ }\textbf {\bibinfo {volume}
  {887}},\ \bibinfo {pages} {L22} (\bibinfo {year} {2019})}\BibitemShut
  {NoStop}%
\bibitem [{\citenamefont {Pang}\ \emph {et~al.}(2021)\citenamefont {Pang},
  \citenamefont {Tews}, \citenamefont {Coughlin}, \citenamefont {Bulla},
  \citenamefont {Van Den~Broeck},\ and\ \citenamefont {Dietrich}}]{Pang2021}%
  \BibitemOpen
  \bibfield  {author} {\bibinfo {author} {\bibfnamefont {P.~T.~H.}\
  \bibnamefont {Pang}}, \bibinfo {author} {\bibfnamefont {I.}~\bibnamefont
  {Tews}}, \bibinfo {author} {\bibfnamefont {M.~W.}\ \bibnamefont {Coughlin}},
  \bibinfo {author} {\bibfnamefont {M.}~\bibnamefont {Bulla}}, \bibinfo
  {author} {\bibfnamefont {C.}~\bibnamefont {Van Den~Broeck}}, \ and\ \bibinfo
  {author} {\bibfnamefont {T.}~\bibnamefont {Dietrich}},\ }\href {\doibase
  10.3847/2041-8213/ac2fad} {\bibfield  {journal} {\bibinfo  {journal} {The
  Astrophysical Journal Letters}\ }\textbf {\bibinfo {volume} {922}},\ \bibinfo
  {pages} {L14} (\bibinfo {year} {2021})}\BibitemShut {NoStop}%
\bibitem [{\citenamefont {Jiang}\ \emph {et~al.}(2020)\citenamefont {Jiang},
  \citenamefont {Tang}, \citenamefont {Wang}, \citenamefont {Fan},\ and\
  \citenamefont {Wei}}]{Jiang2020}%
  \BibitemOpen
  \bibfield  {author} {\bibinfo {author} {\bibfnamefont {J.-L.}\ \bibnamefont
  {Jiang}}, \bibinfo {author} {\bibfnamefont {S.-P.}\ \bibnamefont {Tang}},
  \bibinfo {author} {\bibfnamefont {Y.-Z.}\ \bibnamefont {Wang}}, \bibinfo
  {author} {\bibfnamefont {Y.-Z.}\ \bibnamefont {Fan}}, \ and\ \bibinfo
  {author} {\bibfnamefont {D.-M.}\ \bibnamefont {Wei}},\ }\href {\doibase
  10.3847/1538-4357/ab77cf} {\bibfield  {journal} {\bibinfo  {journal} {The
  Astrophysical Journal}\ }\textbf {\bibinfo {volume} {892}},\ \bibinfo {pages}
  {1} (\bibinfo {year} {2020})}\BibitemShut {NoStop}%
\bibitem [{\citenamefont {Jaminon}\ and\ \citenamefont
  {Mahaux}(1989)}]{PhysRevC.40.354}%
  \BibitemOpen
  \bibfield  {author} {\bibinfo {author} {\bibfnamefont {M.}~\bibnamefont
  {Jaminon}}\ and\ \bibinfo {author} {\bibfnamefont {C.}~\bibnamefont
  {Mahaux}},\ }\href {\doibase 10.1103/PhysRevC.40.354} {\bibfield  {journal}
  {\bibinfo  {journal} {Phys. Rev. C}\ }\textbf {\bibinfo {volume} {40}},\
  \bibinfo {pages} {354} (\bibinfo {year} {1989})}\BibitemShut {NoStop}%
\bibitem [{\citenamefont {Hu}\ \emph {et~al.}(2017)\citenamefont {Hu},
  \citenamefont {Zhang}, \citenamefont {Epelbaum}, \citenamefont {Mei\ss~ner},\
  and\ \citenamefont {Meng}}]{Hu2017}%
  \BibitemOpen
  \bibfield  {author} {\bibinfo {author} {\bibfnamefont {J.}~\bibnamefont
  {Hu}}, \bibinfo {author} {\bibfnamefont {Y.}~\bibnamefont {Zhang}}, \bibinfo
  {author} {\bibfnamefont {E.}~\bibnamefont {Epelbaum}}, \bibinfo {author}
  {\bibfnamefont {U.-G.}\ \bibnamefont {Mei\ss~ner}}, \ and\ \bibinfo {author}
  {\bibfnamefont {J.}~\bibnamefont {Meng}},\ }\href {\doibase
  10.1103/PhysRevC.96.034307} {\bibfield  {journal} {\bibinfo  {journal} {Phys.
  Rev. C}\ }\textbf {\bibinfo {volume} {96}},\ \bibinfo {pages} {034307}
  (\bibinfo {year} {2017})}\BibitemShut {NoStop}%
\bibitem [{\citenamefont {Wang}\ \emph {et~al.}(2020)\citenamefont {Wang},
  \citenamefont {Hu}, \citenamefont {Zhang},\ and\ \citenamefont
  {Shen}}]{wang2020properties}%
  \BibitemOpen
  \bibfield  {author} {\bibinfo {author} {\bibfnamefont {C.}~\bibnamefont
  {Wang}}, \bibinfo {author} {\bibfnamefont {J.}~\bibnamefont {Hu}}, \bibinfo
  {author} {\bibfnamefont {Y.}~\bibnamefont {Zhang}}, \ and\ \bibinfo {author}
  {\bibfnamefont {H.}~\bibnamefont {Shen}},\ }\href@noop {} {\bibfield
  {journal} {\bibinfo  {journal} {The Astrophysical Journal}\ }\textbf
  {\bibinfo {volume} {897}},\ \bibinfo {pages} {96} (\bibinfo {year}
  {2020})}\BibitemShut {NoStop}%
\bibitem [{\citenamefont {Akmal}\ \emph {et~al.}(1998)\citenamefont {Akmal},
  \citenamefont {Pandharipande},\ and\ \citenamefont
  {Ravenhall}}]{PhysRevC.58.1804}%
  \BibitemOpen
  \bibfield  {author} {\bibinfo {author} {\bibfnamefont {A.}~\bibnamefont
  {Akmal}}, \bibinfo {author} {\bibfnamefont {V.~R.}\ \bibnamefont
  {Pandharipande}}, \ and\ \bibinfo {author} {\bibfnamefont {D.~G.}\
  \bibnamefont {Ravenhall}},\ }\href {\doibase 10.1103/PhysRevC.58.1804}
  {\bibfield  {journal} {\bibinfo  {journal} {Phys. Rev. C}\ }\textbf {\bibinfo
  {volume} {58}},\ \bibinfo {pages} {1804} (\bibinfo {year}
  {1998})}\BibitemShut {NoStop}%
\bibitem [{\citenamefont {Stone}\ \emph {et~al.}(2003)\citenamefont {Stone},
  \citenamefont {Miller}, \citenamefont {Koncewicz}, \citenamefont
  {Stevenson},\ and\ \citenamefont {Strayer}}]{stone2003nuclear}%
  \BibitemOpen
  \bibfield  {author} {\bibinfo {author} {\bibfnamefont {J.~R.}\ \bibnamefont
  {Stone}}, \bibinfo {author} {\bibfnamefont {J.}~\bibnamefont {Miller}},
  \bibinfo {author} {\bibfnamefont {R.}~\bibnamefont {Koncewicz}}, \bibinfo
  {author} {\bibfnamefont {P.}~\bibnamefont {Stevenson}}, \ and\ \bibinfo
  {author} {\bibfnamefont {M.}~\bibnamefont {Strayer}},\ }\href@noop {}
  {\bibfield  {journal} {\bibinfo  {journal} {Phys. Rev. C}\ }\textbf {\bibinfo
  {volume} {68}},\ \bibinfo {pages} {034324} (\bibinfo {year}
  {2003})}\BibitemShut {NoStop}%
\bibitem [{\citenamefont {Wang}\ \emph {et~al.}(2024)\citenamefont {Wang},
  \citenamefont {Wang}, \citenamefont {Ye},\ and\ \citenamefont
  {Chen}}]{wang2024extended}%
  \BibitemOpen
  \bibfield  {author} {\bibinfo {author} {\bibfnamefont {S.-P.}\ \bibnamefont
  {Wang}}, \bibinfo {author} {\bibfnamefont {R.}~\bibnamefont {Wang}}, \bibinfo
  {author} {\bibfnamefont {J.-T.}\ \bibnamefont {Ye}}, \ and\ \bibinfo {author}
  {\bibfnamefont {L.-W.}\ \bibnamefont {Chen}},\ }\href@noop {} {\bibfield
  {journal} {\bibinfo  {journal} {Ph C}\ }\textbf {\bibinfo {volume} {109}},\
  \bibinfo {pages} {054623} (\bibinfo {year} {2024})}\BibitemShut {NoStop}%
\bibitem [{\citenamefont {Duan}\ and\ \citenamefont
  {Urban}(2024)}]{duan2024new}%
  \BibitemOpen
  \bibfield  {author} {\bibinfo {author} {\bibfnamefont {M.}~\bibnamefont
  {Duan}}\ and\ \bibinfo {author} {\bibfnamefont {M.}~\bibnamefont {Urban}},\
  }\href@noop {} {\bibfield  {journal} {\bibinfo  {journal} {Phys. Rev. C}\
  }\textbf {\bibinfo {volume} {110}},\ \bibinfo {pages} {065806} (\bibinfo
  {year} {2024})}\BibitemShut {NoStop}%
\bibitem [{\citenamefont {Bender}\ \emph {et~al.}(2003)\citenamefont {Bender},
  \citenamefont {Heenen},\ and\ \citenamefont {Reinhard}}]{RevModPhys.75.121}%
  \BibitemOpen
  \bibfield  {author} {\bibinfo {author} {\bibfnamefont {M.}~\bibnamefont
  {Bender}}, \bibinfo {author} {\bibfnamefont {P.-H.}\ \bibnamefont {Heenen}},
  \ and\ \bibinfo {author} {\bibfnamefont {P.-G.}\ \bibnamefont {Reinhard}},\
  }\href {\doibase 10.1103/RevModPhys.75.121} {\bibfield  {journal} {\bibinfo
  {journal} {Rev. Mod. Phys.}\ }\textbf {\bibinfo {volume} {75}},\ \bibinfo
  {pages} {121} (\bibinfo {year} {2003})}\BibitemShut {NoStop}%
\bibitem [{\citenamefont {Gonzalez-Boquera}\ \emph {et~al.}(2018)\citenamefont
  {Gonzalez-Boquera}, \citenamefont {Centelles}, \citenamefont {Vi{\~n}as},\
  and\ \citenamefont {Robledo}}]{gonzalez2018new}%
  \BibitemOpen
  \bibfield  {author} {\bibinfo {author} {\bibfnamefont {C.}~\bibnamefont
  {Gonzalez-Boquera}}, \bibinfo {author} {\bibfnamefont {M.}~\bibnamefont
  {Centelles}}, \bibinfo {author} {\bibfnamefont {X.}~\bibnamefont
  {Vi{\~n}as}}, \ and\ \bibinfo {author} {\bibfnamefont {L.}~\bibnamefont
  {Robledo}},\ }\href@noop {} {\bibfield  {journal} {\bibinfo  {journal}
  {Physics Letters B}\ }\textbf {\bibinfo {volume} {779}},\ \bibinfo {pages}
  {195} (\bibinfo {year} {2018})}\BibitemShut {NoStop}%
\bibitem [{\citenamefont {Vi{\~n}as}\ \emph {et~al.}(2021)\citenamefont
  {Vi{\~n}as}, \citenamefont {Gonzalez-Boquera}, \citenamefont {Centelles},
  \citenamefont {Mondal},\ and\ \citenamefont {Robledo}}]{vinas2021unified}%
  \BibitemOpen
  \bibfield  {author} {\bibinfo {author} {\bibfnamefont {X.}~\bibnamefont
  {Vi{\~n}as}}, \bibinfo {author} {\bibfnamefont {C.}~\bibnamefont
  {Gonzalez-Boquera}}, \bibinfo {author} {\bibfnamefont {M.}~\bibnamefont
  {Centelles}}, \bibinfo {author} {\bibfnamefont {C.}~\bibnamefont {Mondal}}, \
  and\ \bibinfo {author} {\bibfnamefont {L.~M.}\ \bibnamefont {Robledo}},\
  }\href@noop {} {\bibfield  {journal} {\bibinfo  {journal} {Symmetry}\
  }\textbf {\bibinfo {volume} {13}},\ \bibinfo {pages} {1613} (\bibinfo {year}
  {2021})}\BibitemShut {NoStop}%
\bibitem [{\citenamefont {Zhang}\ \emph
  {et~al.}(2018{\natexlab{a}})\citenamefont {Zhang}, \citenamefont {Hu},\ and\
  \citenamefont {Liu}}]{zhang2018massive}%
  \BibitemOpen
  \bibfield  {author} {\bibinfo {author} {\bibfnamefont {Y.}~\bibnamefont
  {Zhang}}, \bibinfo {author} {\bibfnamefont {J.}~\bibnamefont {Hu}}, \ and\
  \bibinfo {author} {\bibfnamefont {P.}~\bibnamefont {Liu}},\ }\href@noop {}
  {\bibfield  {journal} {\bibinfo  {journal} {Phys. Rev. C}\ }\textbf {\bibinfo
  {volume} {97}},\ \bibinfo {pages} {015805} (\bibinfo {year}
  {2018}{\natexlab{a}})}\BibitemShut {NoStop}%
\bibitem [{\citenamefont {Hu}\ \emph {et~al.}(2020)\citenamefont {Hu},
  \citenamefont {Bao}, \citenamefont {Zhang}, \citenamefont {Nakazato},
  \citenamefont {Sumiyoshi},\ and\ \citenamefont {Shen}}]{hu2020effects}%
  \BibitemOpen
  \bibfield  {author} {\bibinfo {author} {\bibfnamefont {J.}~\bibnamefont
  {Hu}}, \bibinfo {author} {\bibfnamefont {S.}~\bibnamefont {Bao}}, \bibinfo
  {author} {\bibfnamefont {Y.}~\bibnamefont {Zhang}}, \bibinfo {author}
  {\bibfnamefont {K.}~\bibnamefont {Nakazato}}, \bibinfo {author}
  {\bibfnamefont {K.}~\bibnamefont {Sumiyoshi}}, \ and\ \bibinfo {author}
  {\bibfnamefont {H.}~\bibnamefont {Shen}},\ }\href@noop {} {\bibfield
  {journal} {\bibinfo  {journal} {Progress of Theoretical and Experimental
  Physics}\ }\textbf {\bibinfo {volume} {2020}},\ \bibinfo {pages} {043D01}
  (\bibinfo {year} {2020})}\BibitemShut {NoStop}%
\bibitem [{\citenamefont {Huang}\ \emph
  {et~al.}(2022{\natexlab{a}})\citenamefont {Huang}, \citenamefont {Hu},
  \citenamefont {Zhang},\ and\ \citenamefont {Shen}}]{huang2022investigation}%
  \BibitemOpen
  \bibfield  {author} {\bibinfo {author} {\bibfnamefont {K.}~\bibnamefont
  {Huang}}, \bibinfo {author} {\bibfnamefont {J.}~\bibnamefont {Hu}}, \bibinfo
  {author} {\bibfnamefont {Y.}~\bibnamefont {Zhang}}, \ and\ \bibinfo {author}
  {\bibfnamefont {H.}~\bibnamefont {Shen}},\ }\href@noop {} {\bibfield
  {journal} {\bibinfo  {journal} {Nucl. Phys. Rev.}\ }\textbf {\bibinfo
  {volume} {39}},\ \bibinfo {pages} {135} (\bibinfo {year}
  {2022}{\natexlab{a}})}\BibitemShut {NoStop}%
\bibitem [{\citenamefont {Huang}\ \emph
  {et~al.}(2024{\natexlab{a}})\citenamefont {Huang}, \citenamefont {Shen},
  \citenamefont {Hu},\ and\ \citenamefont {Zhang}}]{huang2024hadronic}%
  \BibitemOpen
  \bibfield  {author} {\bibinfo {author} {\bibfnamefont {K.}~\bibnamefont
  {Huang}}, \bibinfo {author} {\bibfnamefont {H.}~\bibnamefont {Shen}},
  \bibinfo {author} {\bibfnamefont {J.}~\bibnamefont {Hu}}, \ and\ \bibinfo
  {author} {\bibfnamefont {Y.}~\bibnamefont {Zhang}},\ }\href@noop {}
  {\bibfield  {journal} {\bibinfo  {journal} {Phys. Rev. D}\ }\textbf {\bibinfo
  {volume} {109}},\ \bibinfo {pages} {043036} (\bibinfo {year}
  {2024}{\natexlab{a}})}\BibitemShut {NoStop}%
\bibitem [{\citenamefont {Sun}\ \emph {et~al.}(2008)\citenamefont {Sun},
  \citenamefont {Long}, \citenamefont {Meng},\ and\ \citenamefont
  {Lombardo}}]{sun2008neutron}%
  \BibitemOpen
  \bibfield  {author} {\bibinfo {author} {\bibfnamefont {B.~Y.}\ \bibnamefont
  {Sun}}, \bibinfo {author} {\bibfnamefont {W.~H.}\ \bibnamefont {Long}},
  \bibinfo {author} {\bibfnamefont {J.}~\bibnamefont {Meng}}, \ and\ \bibinfo
  {author} {\bibfnamefont {U.}~\bibnamefont {Lombardo}},\ }\href@noop {}
  {\bibfield  {journal} {\bibinfo  {journal} {Phys. Rev. C}\ }\textbf {\bibinfo
  {volume} {78}},\ \bibinfo {pages} {065805} (\bibinfo {year}
  {2008})}\BibitemShut {NoStop}%
\bibitem [{\citenamefont {Zhu}\ \emph {et~al.}(2016)\citenamefont {Zhu},
  \citenamefont {Li}, \citenamefont {Hu},\ and\ \citenamefont
  {Sagawa}}]{zhu2016delta}%
  \BibitemOpen
  \bibfield  {author} {\bibinfo {author} {\bibfnamefont {Z.}~\bibnamefont
  {Zhu}}, \bibinfo {author} {\bibfnamefont {A.}~\bibnamefont {Li}}, \bibinfo
  {author} {\bibfnamefont {J.}~\bibnamefont {Hu}}, \ and\ \bibinfo {author}
  {\bibfnamefont {H.}~\bibnamefont {Sagawa}},\ }\href@noop {} {\bibfield
  {journal} {\bibinfo  {journal} {Phys. Rev. C}\ }\textbf {\bibinfo {volume}
  {94}},\ \bibinfo {pages} {045803} (\bibinfo {year} {2016})}\BibitemShut
  {NoStop}%
\bibitem [{\citenamefont {Liu}\ \emph {et~al.}(2018)\citenamefont {Liu},
  \citenamefont {Qian}, \citenamefont {Xing}, \citenamefont {Niu},\ and\
  \citenamefont {Sun}}]{liu2018nuclear}%
  \BibitemOpen
  \bibfield  {author} {\bibinfo {author} {\bibfnamefont {Z.~W.}\ \bibnamefont
  {Liu}}, \bibinfo {author} {\bibfnamefont {Z.}~\bibnamefont {Qian}}, \bibinfo
  {author} {\bibfnamefont {R.~Y.}\ \bibnamefont {Xing}}, \bibinfo {author}
  {\bibfnamefont {J.~R.}\ \bibnamefont {Niu}}, \ and\ \bibinfo {author}
  {\bibfnamefont {B.~Y.}\ \bibnamefont {Sun}},\ }\href@noop {} {\bibfield
  {journal} {\bibinfo  {journal} {Phys. Rev. C}\ }\textbf {\bibinfo {volume}
  {97}},\ \bibinfo {pages} {025801} (\bibinfo {year} {2018})}\BibitemShut
  {NoStop}%
\bibitem [{\citenamefont {Lindblom}(2010)}]{lindblom2010spectral}%
  \BibitemOpen
  \bibfield  {author} {\bibinfo {author} {\bibfnamefont {L.}~\bibnamefont
  {Lindblom}},\ }\href@noop {} {\bibfield  {journal} {\bibinfo  {journal}
  {Phys. Rev. D}\ }\textbf {\bibinfo {volume} {82}},\ \bibinfo {pages} {103011}
  (\bibinfo {year} {2010})}\BibitemShut {NoStop}%
\bibitem [{\citenamefont {O’Boyle}\ \emph {et~al.}(2020)\citenamefont
  {O’Boyle}, \citenamefont {Markakis}, \citenamefont {Stergioulas},\ and\
  \citenamefont {Read}}]{boyle2020parametrized}%
  \BibitemOpen
  \bibfield  {author} {\bibinfo {author} {\bibfnamefont {M.~F.}\ \bibnamefont
  {O’Boyle}}, \bibinfo {author} {\bibfnamefont {C.}~\bibnamefont {Markakis}},
  \bibinfo {author} {\bibfnamefont {N.}~\bibnamefont {Stergioulas}}, \ and\
  \bibinfo {author} {\bibfnamefont {J.~S.}\ \bibnamefont {Read}},\ }\href@noop
  {} {\bibfield  {journal} {\bibinfo  {journal} {Phys. Rev. D}\ }\textbf
  {\bibinfo {volume} {102}},\ \bibinfo {pages} {083027} (\bibinfo {year}
  {2020})}\BibitemShut {NoStop}%
\bibitem [{\citenamefont {Margueron}\ \emph
  {et~al.}(2018{\natexlab{a}})\citenamefont {Margueron}, \citenamefont
  {Hoffmann~Casali},\ and\ \citenamefont {Gulminelli}}]{PhysRevC.97.025805}%
  \BibitemOpen
  \bibfield  {author} {\bibinfo {author} {\bibfnamefont {J.}~\bibnamefont
  {Margueron}}, \bibinfo {author} {\bibfnamefont {R.}~\bibnamefont
  {Hoffmann~Casali}}, \ and\ \bibinfo {author} {\bibfnamefont {F.}~\bibnamefont
  {Gulminelli}},\ }\href {\doibase 10.1103/PhysRevC.97.025805} {\bibfield
  {journal} {\bibinfo  {journal} {Phys. Rev. C}\ }\textbf {\bibinfo {volume}
  {97}},\ \bibinfo {pages} {025805} (\bibinfo {year}
  {2018}{\natexlab{a}})}\BibitemShut {NoStop}%
\bibitem [{\citenamefont {Margueron}\ \emph
  {et~al.}(2018{\natexlab{b}})\citenamefont {Margueron}, \citenamefont
  {Hoffmann~Casali},\ and\ \citenamefont {Gulminelli}}]{PhysRevC.97.025806}%
  \BibitemOpen
  \bibfield  {author} {\bibinfo {author} {\bibfnamefont {J.}~\bibnamefont
  {Margueron}}, \bibinfo {author} {\bibfnamefont {R.}~\bibnamefont
  {Hoffmann~Casali}}, \ and\ \bibinfo {author} {\bibfnamefont {F.}~\bibnamefont
  {Gulminelli}},\ }\href {\doibase 10.1103/PhysRevC.97.025806} {\bibfield
  {journal} {\bibinfo  {journal} {Phys. Rev. C}\ }\textbf {\bibinfo {volume}
  {97}},\ \bibinfo {pages} {025806} (\bibinfo {year}
  {2018}{\natexlab{b}})}\BibitemShut {NoStop}%
\bibitem [{\citenamefont {Zhang}\ \emph
  {et~al.}(2018{\natexlab{b}})\citenamefont {Zhang}, \citenamefont {Li},\ and\
  \citenamefont {Xu}}]{zhang2018combined}%
  \BibitemOpen
  \bibfield  {author} {\bibinfo {author} {\bibfnamefont {N.-B.}\ \bibnamefont
  {Zhang}}, \bibinfo {author} {\bibfnamefont {B.-A.}\ \bibnamefont {Li}}, \
  and\ \bibinfo {author} {\bibfnamefont {J.}~\bibnamefont {Xu}},\ }\href@noop
  {} {\bibfield  {journal} {\bibinfo  {journal} {The Astrophysical Journal}\
  }\textbf {\bibinfo {volume} {859}},\ \bibinfo {pages} {90} (\bibinfo {year}
  {2018}{\natexlab{b}})}\BibitemShut {NoStop}%
\bibitem [{\citenamefont {Cai}\ and\ \citenamefont {Li}(2025)}]{cai2025novel}%
  \BibitemOpen
  \bibfield  {author} {\bibinfo {author} {\bibfnamefont {B.-J.}\ \bibnamefont
  {Cai}}\ and\ \bibinfo {author} {\bibfnamefont {B.-A.}\ \bibnamefont {Li}},\
  }\href@noop {} {\bibfield  {journal} {\bibinfo  {journal} {The European
  Physical Journal A}\ }\textbf {\bibinfo {volume} {61}},\ \bibinfo {pages}
  {55} (\bibinfo {year} {2025})}\BibitemShut {NoStop}%
\bibitem [{\citenamefont {Xie}\ and\ \citenamefont
  {Li}(2021)}]{xie2021bayesian}%
  \BibitemOpen
  \bibfield  {author} {\bibinfo {author} {\bibfnamefont {W.-J.}\ \bibnamefont
  {Xie}}\ and\ \bibinfo {author} {\bibfnamefont {B.-A.}\ \bibnamefont {Li}},\
  }\href@noop {} {\bibfield  {journal} {\bibinfo  {journal} {Journal of Physics
  G: Nuclear and Particle Physics}\ }\textbf {\bibinfo {volume} {48}},\
  \bibinfo {pages} {025110} (\bibinfo {year} {2021})}\BibitemShut {NoStop}%
\bibitem [{\citenamefont {Li}\ \emph {et~al.}(2014)\citenamefont {Li},
  \citenamefont {Ramos}, \citenamefont {Verde},\ and\ \citenamefont
  {Vidaña}}]{Li2014Topical}%
  \BibitemOpen
  \bibfield  {author} {\bibinfo {author} {\bibfnamefont {B.-A.}\ \bibnamefont
  {Li}}, \bibinfo {author} {\bibfnamefont {A.}~\bibnamefont {Ramos}}, \bibinfo
  {author} {\bibfnamefont {G.}~\bibnamefont {Verde}}, \ and\ \bibinfo {author}
  {\bibfnamefont {I.}~\bibnamefont {Vidaña}},\ }\href {\doibase
  10.1140/epja/i2014-14009-x} {\bibfield  {journal} {\bibinfo  {journal} {The
  European Physical Journal A}\ }\textbf {\bibinfo {volume} {50}},\ \bibinfo
  {pages} {9} (\bibinfo {year} {2014})}\BibitemShut {NoStop}%
\bibitem [{\citenamefont {Li}\ and\ \citenamefont {Han}(2013)}]{LI2013276}%
  \BibitemOpen
  \bibfield  {author} {\bibinfo {author} {\bibfnamefont {B.-A.}\ \bibnamefont
  {Li}}\ and\ \bibinfo {author} {\bibfnamefont {X.}~\bibnamefont {Han}},\
  }\href {\doibase https://doi.org/10.1016/j.physletb.2013.10.006} {\bibfield
  {journal} {\bibinfo  {journal} {Physics Letters B}\ }\textbf {\bibinfo
  {volume} {727}},\ \bibinfo {pages} {276} (\bibinfo {year}
  {2013})}\BibitemShut {NoStop}%
\bibitem [{\citenamefont {Oertel}\ \emph {et~al.}(2017)\citenamefont {Oertel},
  \citenamefont {Hempel}, \citenamefont {Kl\"ahn},\ and\ \citenamefont
  {Typel}}]{RevModPhys.89.015007}%
  \BibitemOpen
  \bibfield  {author} {\bibinfo {author} {\bibfnamefont {M.}~\bibnamefont
  {Oertel}}, \bibinfo {author} {\bibfnamefont {M.}~\bibnamefont {Hempel}},
  \bibinfo {author} {\bibfnamefont {T.}~\bibnamefont {Kl\"ahn}}, \ and\
  \bibinfo {author} {\bibfnamefont {S.}~\bibnamefont {Typel}},\ }\href
  {\doibase 10.1103/RevModPhys.89.015007} {\bibfield  {journal} {\bibinfo
  {journal} {Rev. Mod. Phys.}\ }\textbf {\bibinfo {volume} {89}},\ \bibinfo
  {pages} {015007} (\bibinfo {year} {2017})}\BibitemShut {NoStop}%
\bibitem [{\citenamefont {Oyamatsu}\ and\ \citenamefont
  {Iida}(2007)}]{PhysRevC.75.015801}%
  \BibitemOpen
  \bibfield  {author} {\bibinfo {author} {\bibfnamefont {K.}~\bibnamefont
  {Oyamatsu}}\ and\ \bibinfo {author} {\bibfnamefont {K.}~\bibnamefont
  {Iida}},\ }\href {\doibase 10.1103/PhysRevC.75.015801} {\bibfield  {journal}
  {\bibinfo  {journal} {Phys. Rev. C}\ }\textbf {\bibinfo {volume} {75}},\
  \bibinfo {pages} {015801} (\bibinfo {year} {2007})}\BibitemShut {NoStop}%
\bibitem [{\citenamefont {Bao}\ and\ \citenamefont
  {Shen}(2014)}]{bao2014influence}%
  \BibitemOpen
  \bibfield  {author} {\bibinfo {author} {\bibfnamefont {S.}~\bibnamefont
  {Bao}}\ and\ \bibinfo {author} {\bibfnamefont {H.}~\bibnamefont {Shen}},\
  }\href@noop {} {\bibfield  {journal} {\bibinfo  {journal} {Phys. Rev. C}\
  }\textbf {\bibinfo {volume} {89}},\ \bibinfo {pages} {045807} (\bibinfo
  {year} {2014})}\BibitemShut {NoStop}%
\bibitem [{\citenamefont {Lattimer}\ and\ \citenamefont
  {Prakash}(2001)}]{Lattimer_2001}%
  \BibitemOpen
  \bibfield  {author} {\bibinfo {author} {\bibfnamefont {J.~M.}\ \bibnamefont
  {Lattimer}}\ and\ \bibinfo {author} {\bibfnamefont {M.}~\bibnamefont
  {Prakash}},\ }\href {\doibase 10.1086/319702} {\bibfield  {journal} {\bibinfo
   {journal} {The Astrophysical Journal}\ }\textbf {\bibinfo {volume} {550}},\
  \bibinfo {pages} {426} (\bibinfo {year} {2001})}\BibitemShut {NoStop}%
\bibitem [{\citenamefont {Abrahamyan}\ \emph {et~al.}(2012)\citenamefont
  {Abrahamyan}, \citenamefont {Ahmed}, \citenamefont {Albataineh},
  \citenamefont {Aniol}, \citenamefont {Armstrong} \emph
  {et~al.}}]{PhysRevLett.108.112502}%
  \BibitemOpen
  \bibfield  {author} {\bibinfo {author} {\bibfnamefont {S.}~\bibnamefont
  {Abrahamyan}}, \bibinfo {author} {\bibfnamefont {Z.}~\bibnamefont {Ahmed}},
  \bibinfo {author} {\bibfnamefont {H.}~\bibnamefont {Albataineh}}, \bibinfo
  {author} {\bibfnamefont {K.}~\bibnamefont {Aniol}}, \bibinfo {author}
  {\bibfnamefont {D.~S.}\ \bibnamefont {Armstrong}},  \emph {et~al.} (\bibinfo
  {collaboration} {PREX Collaboration}),\ }\href {\doibase
  10.1103/PhysRevLett.108.112502} {\bibfield  {journal} {\bibinfo  {journal}
  {Phys. Rev. Lett.}\ }\textbf {\bibinfo {volume} {108}},\ \bibinfo {pages}
  {112502} (\bibinfo {year} {2012})}\BibitemShut {NoStop}%
\bibitem [{\citenamefont {Horowitz}\ \emph {et~al.}(2012)\citenamefont
  {Horowitz}, \citenamefont {Ahmed}, \citenamefont {Jen}, \citenamefont
  {Rakhman}, \citenamefont {Souder} \emph {et~al.}}]{PhysRevC.85.032501}%
  \BibitemOpen
  \bibfield  {author} {\bibinfo {author} {\bibfnamefont {C.~J.}\ \bibnamefont
  {Horowitz}}, \bibinfo {author} {\bibfnamefont {Z.}~\bibnamefont {Ahmed}},
  \bibinfo {author} {\bibfnamefont {C.-M.}\ \bibnamefont {Jen}}, \bibinfo
  {author} {\bibfnamefont {A.}~\bibnamefont {Rakhman}}, \bibinfo {author}
  {\bibfnamefont {P.~A.}\ \bibnamefont {Souder}},  \emph {et~al.},\ }\href
  {\doibase 10.1103/PhysRevC.85.032501} {\bibfield  {journal} {\bibinfo
  {journal} {Phys. Rev. C}\ }\textbf {\bibinfo {volume} {85}},\ \bibinfo
  {pages} {032501} (\bibinfo {year} {2012})}\BibitemShut {NoStop}%
\bibitem [{\citenamefont {Adhikari}\ \emph {et~al.}(2021)\citenamefont
  {Adhikari}, \citenamefont {Albataineh}, \citenamefont {Androic},
  \citenamefont {Aniol}, \citenamefont {Armstrong} \emph
  {et~al.}}]{PhysRevLett.126.172502}%
  \BibitemOpen
  \bibfield  {author} {\bibinfo {author} {\bibfnamefont {D.}~\bibnamefont
  {Adhikari}}, \bibinfo {author} {\bibfnamefont {H.}~\bibnamefont
  {Albataineh}}, \bibinfo {author} {\bibfnamefont {D.}~\bibnamefont {Androic}},
  \bibinfo {author} {\bibfnamefont {K.}~\bibnamefont {Aniol}}, \bibinfo
  {author} {\bibfnamefont {D.~S.}\ \bibnamefont {Armstrong}},  \emph {et~al.}
  (\bibinfo {collaboration} {PREX Collaboration}),\ }\href {\doibase
  10.1103/PhysRevLett.126.172502} {\bibfield  {journal} {\bibinfo  {journal}
  {Phys. Rev. Lett.}\ }\textbf {\bibinfo {volume} {126}},\ \bibinfo {pages}
  {172502} (\bibinfo {year} {2021})}\BibitemShut {NoStop}%
\bibitem [{\citenamefont {Reed}\ \emph {et~al.}(2021)\citenamefont {Reed},
  \citenamefont {Fattoyev}, \citenamefont {Horowitz},\ and\ \citenamefont
  {Piekarewicz}}]{PhysRevLett.126.172503}%
  \BibitemOpen
  \bibfield  {author} {\bibinfo {author} {\bibfnamefont {B.~T.}\ \bibnamefont
  {Reed}}, \bibinfo {author} {\bibfnamefont {F.~J.}\ \bibnamefont {Fattoyev}},
  \bibinfo {author} {\bibfnamefont {C.~J.}\ \bibnamefont {Horowitz}}, \ and\
  \bibinfo {author} {\bibfnamefont {J.}~\bibnamefont {Piekarewicz}},\ }\href
  {\doibase 10.1103/PhysRevLett.126.172503} {\bibfield  {journal} {\bibinfo
  {journal} {Phys. Rev. Lett.}\ }\textbf {\bibinfo {volume} {126}},\ \bibinfo
  {pages} {172503} (\bibinfo {year} {2021})}\BibitemShut {NoStop}%
\bibitem [{\citenamefont {Roca-Maza}\ \emph {et~al.}(2011)\citenamefont
  {Roca-Maza}, \citenamefont {Centelles}, \citenamefont {Vi\~nas},\ and\
  \citenamefont {Warda}}]{PhysRevLett.106.252501}%
  \BibitemOpen
  \bibfield  {author} {\bibinfo {author} {\bibfnamefont {X.}~\bibnamefont
  {Roca-Maza}}, \bibinfo {author} {\bibfnamefont {M.}~\bibnamefont
  {Centelles}}, \bibinfo {author} {\bibfnamefont {X.}~\bibnamefont {Vi\~nas}},
  \ and\ \bibinfo {author} {\bibfnamefont {M.}~\bibnamefont {Warda}},\ }\href
  {\doibase 10.1103/PhysRevLett.106.252501} {\bibfield  {journal} {\bibinfo
  {journal} {Phys. Rev. Lett.}\ }\textbf {\bibinfo {volume} {106}},\ \bibinfo
  {pages} {252501} (\bibinfo {year} {2011})}\BibitemShut {NoStop}%
\bibitem [{\citenamefont {Piekarewicz}(2021)}]{PhysRevC.104.024329}%
  \BibitemOpen
  \bibfield  {author} {\bibinfo {author} {\bibfnamefont {J.}~\bibnamefont
  {Piekarewicz}},\ }\href {\doibase 10.1103/PhysRevC.104.024329} {\bibfield
  {journal} {\bibinfo  {journal} {Phys. Rev. C}\ }\textbf {\bibinfo {volume}
  {104}},\ \bibinfo {pages} {024329} (\bibinfo {year} {2021})}\BibitemShut
  {NoStop}%
\bibitem [{\citenamefont {Adhikari}\ \emph {et~al.}(2022)\citenamefont
  {Adhikari}, \citenamefont {Albataineh}, \citenamefont {Androic},
  \citenamefont {Aniol}, \citenamefont {Armstrong} \emph
  {et~al.}}]{PhysRevLett.129.042501}%
  \BibitemOpen
  \bibfield  {author} {\bibinfo {author} {\bibfnamefont {D.}~\bibnamefont
  {Adhikari}}, \bibinfo {author} {\bibfnamefont {H.}~\bibnamefont
  {Albataineh}}, \bibinfo {author} {\bibfnamefont {D.}~\bibnamefont {Androic}},
  \bibinfo {author} {\bibfnamefont {K.~A.}\ \bibnamefont {Aniol}}, \bibinfo
  {author} {\bibfnamefont {D.~S.}\ \bibnamefont {Armstrong}},  \emph {et~al.}
  (\bibinfo {collaboration} {CREX Collaboration}),\ }\href {\doibase
  10.1103/PhysRevLett.129.042501} {\bibfield  {journal} {\bibinfo  {journal}
  {Phys. Rev. Lett.}\ }\textbf {\bibinfo {volume} {129}},\ \bibinfo {pages}
  {042501} (\bibinfo {year} {2022})}\BibitemShut {NoStop}%
\bibitem [{\citenamefont {Reinhard}\ \emph {et~al.}(2022)\citenamefont
  {Reinhard}, \citenamefont {Roca-Maza},\ and\ \citenamefont
  {Nazarewicz}}]{PhysRevLett.129.232501}%
  \BibitemOpen
  \bibfield  {author} {\bibinfo {author} {\bibfnamefont {P.-G.}\ \bibnamefont
  {Reinhard}}, \bibinfo {author} {\bibfnamefont {X.}~\bibnamefont {Roca-Maza}},
  \ and\ \bibinfo {author} {\bibfnamefont {W.}~\bibnamefont {Nazarewicz}},\
  }\href {\doibase 10.1103/PhysRevLett.129.232501} {\bibfield  {journal}
  {\bibinfo  {journal} {Phys. Rev. Lett.}\ }\textbf {\bibinfo {volume} {129}},\
  \bibinfo {pages} {232501} (\bibinfo {year} {2022})}\BibitemShut {NoStop}%
\bibitem [{\citenamefont {Zhang}\ and\ \citenamefont
  {Chen}(2023)}]{PhysRevC.108.024317}%
  \BibitemOpen
  \bibfield  {author} {\bibinfo {author} {\bibfnamefont {Z.}~\bibnamefont
  {Zhang}}\ and\ \bibinfo {author} {\bibfnamefont {L.-W.}\ \bibnamefont
  {Chen}},\ }\href {\doibase 10.1103/PhysRevC.108.024317} {\bibfield  {journal}
  {\bibinfo  {journal} {Phys. Rev. C}\ }\textbf {\bibinfo {volume} {108}},\
  \bibinfo {pages} {024317} (\bibinfo {year} {2023})}\BibitemShut {NoStop}%
\bibitem [{\citenamefont {Lattimer}(2023)}]{particles6010003}%
  \BibitemOpen
  \bibfield  {author} {\bibinfo {author} {\bibfnamefont {J.~M.}\ \bibnamefont
  {Lattimer}},\ }\href {\doibase 10.3390/particles6010003} {\bibfield
  {journal} {\bibinfo  {journal} {Particles}\ }\textbf {\bibinfo {volume}
  {6}},\ \bibinfo {pages} {30} (\bibinfo {year} {2023})}\BibitemShut {NoStop}%
\bibitem [{\citenamefont {Bao}\ and\ \citenamefont
  {Shen}(2015)}]{bao2015impact}%
  \BibitemOpen
  \bibfield  {author} {\bibinfo {author} {\bibfnamefont {S.}~\bibnamefont
  {Bao}}\ and\ \bibinfo {author} {\bibfnamefont {H.}~\bibnamefont {Shen}},\
  }\href@noop {} {\bibfield  {journal} {\bibinfo  {journal} {Phys. Rev. C}\
  }\textbf {\bibinfo {volume} {91}},\ \bibinfo {pages} {015807} (\bibinfo
  {year} {2015})}\BibitemShut {NoStop}%
\bibitem [{\citenamefont {Bao}\ \emph {et~al.}(2014)\citenamefont {Bao},
  \citenamefont {Hu}, \citenamefont {Zhang},\ and\ \citenamefont
  {Shen}}]{bao2014effects}%
  \BibitemOpen
  \bibfield  {author} {\bibinfo {author} {\bibfnamefont {S.}~\bibnamefont
  {Bao}}, \bibinfo {author} {\bibfnamefont {J.}~\bibnamefont {Hu}}, \bibinfo
  {author} {\bibfnamefont {Z.}~\bibnamefont {Zhang}}, \ and\ \bibinfo {author}
  {\bibfnamefont {H.}~\bibnamefont {Shen}},\ }\href@noop {} {\bibfield
  {journal} {\bibinfo  {journal} {Phys. Rev. C}\ }\textbf {\bibinfo {volume}
  {90}},\ \bibinfo {pages} {045802} (\bibinfo {year} {2014})}\BibitemShut
  {NoStop}%
\bibitem [{\citenamefont {Cartaxo}\ \emph {et~al.}(2025)\citenamefont
  {Cartaxo}, \citenamefont {Huang}, \citenamefont {Malik}, \citenamefont
  {Sourav}, \citenamefont {Yuan}, \citenamefont {Zhou}, \citenamefont {Liu},\
  and\ \citenamefont
  {Providência}}]{cartaxo2025completesurveytextttcompactobjectperspective}%
  \BibitemOpen
  \bibfield  {author} {\bibinfo {author} {\bibfnamefont {J.}~\bibnamefont
  {Cartaxo}}, \bibinfo {author} {\bibfnamefont {C.}~\bibnamefont {Huang}},
  \bibinfo {author} {\bibfnamefont {T.}~\bibnamefont {Malik}}, \bibinfo
  {author} {\bibfnamefont {S.}~\bibnamefont {Sourav}}, \bibinfo {author}
  {\bibfnamefont {W.-L.}\ \bibnamefont {Yuan}}, \bibinfo {author}
  {\bibfnamefont {T.}~\bibnamefont {Zhou}}, \bibinfo {author} {\bibfnamefont
  {X.}~\bibnamefont {Liu}}, \ and\ \bibinfo {author} {\bibfnamefont
  {C.}~\bibnamefont {Providência}},\ }\href {https://arxiv.org/abs/2506.03112}
  {\enquote {\bibinfo {title} {A complete survey from the
  $\texttt{CompactObject}$ perspective on equation of state cross-comparison
  using observational and nuclear experimental constraints},}\ } (\bibinfo
  {year} {2025}),\ \Eprint {http://arxiv.org/abs/2506.03112} {arXiv:2506.03112
  [nucl-th]} \BibitemShut {NoStop}%
\bibitem [{\citenamefont {Passarella}\ \emph {et~al.}(2025)\citenamefont
  {Passarella}, \citenamefont {Margueron},\ and\ \citenamefont
  {Pagliara}}]{passarella2025relativisticmeanfieldpredictionsdense}%
  \BibitemOpen
  \bibfield  {author} {\bibinfo {author} {\bibfnamefont {L.}~\bibnamefont
  {Passarella}}, \bibinfo {author} {\bibfnamefont {J.}~\bibnamefont
  {Margueron}}, \ and\ \bibinfo {author} {\bibfnamefont {G.}~\bibnamefont
  {Pagliara}},\ }\href {https://arxiv.org/abs/2503.23028} {\enquote {\bibinfo
  {title} {Relativistic mean-field predictions for dense matter equation of
  state and application to neutron stars},}\ } (\bibinfo {year} {2025}),\
  \Eprint {http://arxiv.org/abs/2503.23028} {arXiv:2503.23028 [nucl-th]}
  \BibitemShut {NoStop}%
\bibitem [{\citenamefont {Huang}\ \emph
  {et~al.}(2024{\natexlab{b}})\citenamefont {Huang}, \citenamefont
  {Raaijmakers}, \citenamefont {Watts}, \citenamefont {Tolos},\ and\
  \citenamefont {Providência}}]{10.1093/mnras/stae844}%
  \BibitemOpen
  \bibfield  {author} {\bibinfo {author} {\bibfnamefont {C.}~\bibnamefont
  {Huang}}, \bibinfo {author} {\bibfnamefont {G.}~\bibnamefont {Raaijmakers}},
  \bibinfo {author} {\bibfnamefont {A.~L.}\ \bibnamefont {Watts}}, \bibinfo
  {author} {\bibfnamefont {L.}~\bibnamefont {Tolos}}, \ and\ \bibinfo {author}
  {\bibfnamefont {C.}~\bibnamefont {Providência}},\ }\href {\doibase
  10.1093/mnras/stae844} {\bibfield  {journal} {\bibinfo  {journal} {Monthly
  Notices of the Royal Astronomical Society}\ }\textbf {\bibinfo {volume}
  {529}},\ \bibinfo {pages} {4650} (\bibinfo {year}
  {2024}{\natexlab{b}})}\BibitemShut {NoStop}%
\bibitem [{\citenamefont {Huang}\ \emph
  {et~al.}(2024{\natexlab{c}})\citenamefont {Huang}, \citenamefont {Tolos},
  \citenamefont {Providência},\ and\ \citenamefont
  {Watts}}]{10.1093/mnras/stae2792}%
  \BibitemOpen
  \bibfield  {author} {\bibinfo {author} {\bibfnamefont {C.}~\bibnamefont
  {Huang}}, \bibinfo {author} {\bibfnamefont {L.}~\bibnamefont {Tolos}},
  \bibinfo {author} {\bibfnamefont {C.}~\bibnamefont {Providência}}, \ and\
  \bibinfo {author} {\bibfnamefont {A.}~\bibnamefont {Watts}},\ }\href
  {\doibase 10.1093/mnras/stae2792} {\bibfield  {journal} {\bibinfo  {journal}
  {Monthly Notices of the Royal Astronomical Society}\ }\textbf {\bibinfo
  {volume} {536}},\ \bibinfo {pages} {3262} (\bibinfo {year}
  {2024}{\natexlab{c}})}\BibitemShut {NoStop}%
\bibitem [{\citenamefont {Xie}\ and\ \citenamefont
  {Li}(2020)}]{xie2020bayesian}%
  \BibitemOpen
  \bibfield  {author} {\bibinfo {author} {\bibfnamefont {W.-J.}\ \bibnamefont
  {Xie}}\ and\ \bibinfo {author} {\bibfnamefont {B.-A.}\ \bibnamefont {Li}},\
  }\href@noop {} {\bibfield  {journal} {\bibinfo  {journal} {The Astrophysical
  Journal}\ }\textbf {\bibinfo {volume} {899}},\ \bibinfo {pages} {4} (\bibinfo
  {year} {2020})}\BibitemShut {NoStop}%
\bibitem [{\citenamefont {Malik}\ and\ \citenamefont
  {Provid{\^e}ncia}(2022)}]{malik2022bayesian}%
  \BibitemOpen
  \bibfield  {author} {\bibinfo {author} {\bibfnamefont {T.}~\bibnamefont
  {Malik}}\ and\ \bibinfo {author} {\bibfnamefont {C.}~\bibnamefont
  {Provid{\^e}ncia}},\ }\href@noop {} {\bibfield  {journal} {\bibinfo
  {journal} {Phys. Rev. D}\ }\textbf {\bibinfo {volume} {106}},\ \bibinfo
  {pages} {063024} (\bibinfo {year} {2022})}\BibitemShut {NoStop}%
\bibitem [{\citenamefont {Carvalho}\ \emph {et~al.}(2023)\citenamefont
  {Carvalho}, \citenamefont {Ferreira}, \citenamefont {Malik},\ and\
  \citenamefont {Provid{\^e}ncia}}]{carvalho2023decoding}%
  \BibitemOpen
  \bibfield  {author} {\bibinfo {author} {\bibfnamefont {V.}~\bibnamefont
  {Carvalho}}, \bibinfo {author} {\bibfnamefont {M.}~\bibnamefont {Ferreira}},
  \bibinfo {author} {\bibfnamefont {T.}~\bibnamefont {Malik}}, \ and\ \bibinfo
  {author} {\bibfnamefont {C.}~\bibnamefont {Provid{\^e}ncia}},\ }\href@noop {}
  {\bibfield  {journal} {\bibinfo  {journal} {Phys. Rev. D}\ }\textbf {\bibinfo
  {volume} {108}},\ \bibinfo {pages} {043031} (\bibinfo {year}
  {2023})}\BibitemShut {NoStop}%
\bibitem [{\citenamefont {Zhou}\ \emph
  {et~al.}(2023{\natexlab{a}})\citenamefont {Zhou}, \citenamefont {Xu},\ and\
  \citenamefont {Papakonstantinou}}]{PhysRevC.107.055803}%
  \BibitemOpen
  \bibfield  {author} {\bibinfo {author} {\bibfnamefont {J.}~\bibnamefont
  {Zhou}}, \bibinfo {author} {\bibfnamefont {J.}~\bibnamefont {Xu}}, \ and\
  \bibinfo {author} {\bibfnamefont {P.}~\bibnamefont {Papakonstantinou}},\
  }\href {\doibase 10.1103/PhysRevC.107.055803} {\bibfield  {journal} {\bibinfo
   {journal} {Phys. Rev. C}\ }\textbf {\bibinfo {volume} {107}},\ \bibinfo
  {pages} {055803} (\bibinfo {year} {2023}{\natexlab{a}})}\BibitemShut
  {NoStop}%
\bibitem [{\citenamefont {Li}\ \emph {et~al.}(2024)\citenamefont {Li},
  \citenamefont {Grundler}, \citenamefont {Xie},\ and\ \citenamefont
  {Zhang}}]{Li2024}%
  \BibitemOpen
  \bibfield  {author} {\bibinfo {author} {\bibfnamefont {B.-A.}\ \bibnamefont
  {Li}}, \bibinfo {author} {\bibfnamefont {X.}~\bibnamefont {Grundler}},
  \bibinfo {author} {\bibfnamefont {W.-J.}\ \bibnamefont {Xie}}, \ and\
  \bibinfo {author} {\bibfnamefont {N.-B.}\ \bibnamefont {Zhang}},\ }\href
  {\doibase 10.1103/PhysRevD.110.103040} {\bibfield  {journal} {\bibinfo
  {journal} {Phys. Rev. D}\ }\textbf {\bibinfo {volume} {110}},\ \bibinfo
  {pages} {103040} (\bibinfo {year} {2024})}\BibitemShut {NoStop}%
\bibitem [{\citenamefont {Ferreira}\ and\ \citenamefont
  {Provid{\^e}ncia}(2021)}]{ferreira2021unveiling}%
  \BibitemOpen
  \bibfield  {author} {\bibinfo {author} {\bibfnamefont {M.}~\bibnamefont
  {Ferreira}}\ and\ \bibinfo {author} {\bibfnamefont {C.}~\bibnamefont
  {Provid{\^e}ncia}},\ }\href@noop {} {\bibfield  {journal} {\bibinfo
  {journal} {Journal of Cosmology and Astroparticle Physics}\ }\textbf
  {\bibinfo {volume} {2021}},\ \bibinfo {pages} {011} (\bibinfo {year}
  {2021})}\BibitemShut {NoStop}%
\bibitem [{\citenamefont {Fujimoto}\ \emph {et~al.}(2018)\citenamefont
  {Fujimoto}, \citenamefont {Fukushima},\ and\ \citenamefont
  {Murase}}]{fujimoto2018methodology}%
  \BibitemOpen
  \bibfield  {author} {\bibinfo {author} {\bibfnamefont {Y.}~\bibnamefont
  {Fujimoto}}, \bibinfo {author} {\bibfnamefont {K.}~\bibnamefont {Fukushima}},
  \ and\ \bibinfo {author} {\bibfnamefont {K.}~\bibnamefont {Murase}},\
  }\href@noop {} {\bibfield  {journal} {\bibinfo  {journal} {Phys. Rev. D}\
  }\textbf {\bibinfo {volume} {98}},\ \bibinfo {pages} {023019} (\bibinfo
  {year} {2018})}\BibitemShut {NoStop}%
\bibitem [{\citenamefont {Zhou}\ \emph
  {et~al.}(2023{\natexlab{b}})\citenamefont {Zhou}, \citenamefont {Hu},
  \citenamefont {Zhang},\ and\ \citenamefont {Shen}}]{zhou2023nonparametric}%
  \BibitemOpen
  \bibfield  {author} {\bibinfo {author} {\bibfnamefont {W.}~\bibnamefont
  {Zhou}}, \bibinfo {author} {\bibfnamefont {J.}~\bibnamefont {Hu}}, \bibinfo
  {author} {\bibfnamefont {Y.}~\bibnamefont {Zhang}}, \ and\ \bibinfo {author}
  {\bibfnamefont {H.}~\bibnamefont {Shen}},\ }\href@noop {} {\bibfield
  {journal} {\bibinfo  {journal} {The Astrophysical Journal}\ }\textbf
  {\bibinfo {volume} {950}},\ \bibinfo {pages} {186} (\bibinfo {year}
  {2023}{\natexlab{b}})}\BibitemShut {NoStop}%
\bibitem [{\citenamefont {Han}\ \emph {et~al.}(2023)\citenamefont {Han},
  \citenamefont {Tang},\ and\ \citenamefont {Fan}}]{han2023nonparametric}%
  \BibitemOpen
  \bibfield  {author} {\bibinfo {author} {\bibfnamefont {M.-Z.}\ \bibnamefont
  {Han}}, \bibinfo {author} {\bibfnamefont {S.-P.}\ \bibnamefont {Tang}}, \
  and\ \bibinfo {author} {\bibfnamefont {Y.-Z.}\ \bibnamefont {Fan}},\
  }\href@noop {} {\bibfield  {journal} {\bibinfo  {journal} {The Astrophysical
  Journal}\ }\textbf {\bibinfo {volume} {950}},\ \bibinfo {pages} {77}
  (\bibinfo {year} {2023})}\BibitemShut {NoStop}%
\bibitem [{\citenamefont {Zhou}\ \emph {et~al.}(2024)\citenamefont {Zhou},
  \citenamefont {Shen}, \citenamefont {Hu},\ and\ \citenamefont
  {Zhang}}]{zhou2024first}%
  \BibitemOpen
  \bibfield  {author} {\bibinfo {author} {\bibfnamefont {W.}~\bibnamefont
  {Zhou}}, \bibinfo {author} {\bibfnamefont {H.}~\bibnamefont {Shen}}, \bibinfo
  {author} {\bibfnamefont {J.}~\bibnamefont {Hu}}, \ and\ \bibinfo {author}
  {\bibfnamefont {Y.}~\bibnamefont {Zhang}},\ }\href@noop {} {\bibfield
  {journal} {\bibinfo  {journal} {Phys. Rev. D}\ }\textbf {\bibinfo {volume}
  {110}},\ \bibinfo {pages} {043017} (\bibinfo {year} {2024})}\BibitemShut
  {NoStop}%
\bibitem [{\citenamefont {Guo}\ \emph {et~al.}(2024)\citenamefont {Guo},
  \citenamefont {Xiong}, \citenamefont {Ma},\ and\ \citenamefont
  {Ma}}]{guo2024insights}%
  \BibitemOpen
  \bibfield  {author} {\bibinfo {author} {\bibfnamefont {L.-J.}\ \bibnamefont
  {Guo}}, \bibinfo {author} {\bibfnamefont {J.-Y.}\ \bibnamefont {Xiong}},
  \bibinfo {author} {\bibfnamefont {Y.}~\bibnamefont {Ma}}, \ and\ \bibinfo
  {author} {\bibfnamefont {Y.-L.}\ \bibnamefont {Ma}},\ }\href@noop {}
  {\bibfield  {journal} {\bibinfo  {journal} {The Astrophysical Journal}\
  }\textbf {\bibinfo {volume} {965}},\ \bibinfo {pages} {47} (\bibinfo {year}
  {2024})}\BibitemShut {NoStop}%
\bibitem [{\citenamefont {Clark}\ \emph {et~al.}(2007)\citenamefont {Clark},
  \citenamefont {Heng}, \citenamefont {Pitkin},\ and\ \citenamefont
  {Woan}}]{clark2007evidence}%
  \BibitemOpen
  \bibfield  {author} {\bibinfo {author} {\bibfnamefont {J.}~\bibnamefont
  {Clark}}, \bibinfo {author} {\bibfnamefont {I.~S.}\ \bibnamefont {Heng}},
  \bibinfo {author} {\bibfnamefont {M.}~\bibnamefont {Pitkin}}, \ and\ \bibinfo
  {author} {\bibfnamefont {G.}~\bibnamefont {Woan}},\ }\href@noop {} {\bibfield
   {journal} {\bibinfo  {journal} {Phys. Rev. D}\ }\textbf {\bibinfo {volume}
  {76}},\ \bibinfo {pages} {043003} (\bibinfo {year} {2007})}\BibitemShut
  {NoStop}%
\bibitem [{\citenamefont {{\"O}zel}\ \emph {et~al.}(2010)\citenamefont
  {{\"O}zel}, \citenamefont {Baym},\ and\ \citenamefont
  {G{\"u}ver}}]{ozel2010astrophysical}%
  \BibitemOpen
  \bibfield  {author} {\bibinfo {author} {\bibfnamefont {F.}~\bibnamefont
  {{\"O}zel}}, \bibinfo {author} {\bibfnamefont {G.}~\bibnamefont {Baym}}, \
  and\ \bibinfo {author} {\bibfnamefont {T.}~\bibnamefont {G{\"u}ver}},\
  }\href@noop {} {\bibfield  {journal} {\bibinfo  {journal} {Phys. Rev. D}\
  }\textbf {\bibinfo {volume} {82}},\ \bibinfo {pages} {101301} (\bibinfo
  {year} {2010})}\BibitemShut {NoStop}%
\bibitem [{\citenamefont {Chimanski}\ \emph {et~al.}(2023)\citenamefont
  {Chimanski}, \citenamefont {Lobato}, \citenamefont {Goncalves},\ and\
  \citenamefont
  {Bertulani}}]{chimanski2022bayesianinferencephenomenologicaleos}%
  \BibitemOpen
  \bibfield  {author} {\bibinfo {author} {\bibfnamefont {E.~V.}\ \bibnamefont
  {Chimanski}}, \bibinfo {author} {\bibfnamefont {R.~V.}\ \bibnamefont
  {Lobato}}, \bibinfo {author} {\bibfnamefont {A.~R.}\ \bibnamefont
  {Goncalves}}, \ and\ \bibinfo {author} {\bibfnamefont {C.~A.}\ \bibnamefont
  {Bertulani}},\ }\href@noop {} {\bibfield  {journal} {\bibinfo  {journal}
  {Particles}\ }\textbf {\bibinfo {volume} {6}},\ \bibinfo {pages} {198}
  (\bibinfo {year} {2023})}\BibitemShut {NoStop}%
\bibitem [{\citenamefont {Beznogov}\ and\ \citenamefont
  {Raduta}(2024{\natexlab{a}})}]{Beznogov2024apj}%
  \BibitemOpen
  \bibfield  {author} {\bibinfo {author} {\bibfnamefont {M.~V.}\ \bibnamefont
  {Beznogov}}\ and\ \bibinfo {author} {\bibfnamefont {A.~R.}\ \bibnamefont
  {Raduta}},\ }\href {\doibase 10.3847/1538-4357/ad2f9b} {\bibfield  {journal}
  {\bibinfo  {journal} {The Astrophysical Journal}\ }\textbf {\bibinfo {volume}
  {966}},\ \bibinfo {pages} {216} (\bibinfo {year}
  {2024}{\natexlab{a}})}\BibitemShut {NoStop}%
\bibitem [{\citenamefont {Beznogov}\ and\ \citenamefont
  {Raduta}(2024{\natexlab{b}})}]{Beznogov_2024}%
  \BibitemOpen
  \bibfield  {author} {\bibinfo {author} {\bibfnamefont {M.~V.}\ \bibnamefont
  {Beznogov}}\ and\ \bibinfo {author} {\bibfnamefont {A.~R.}\ \bibnamefont
  {Raduta}},\ }\href {\doibase 10.1103/physrevc.110.035805} {\bibfield
  {journal} {\bibinfo  {journal} {Phys. Rev. C}\ }\textbf {\bibinfo {volume}
  {110}} (\bibinfo {year} {2024}{\natexlab{b}}),\
  10.1103/physrevc.110.035805}\BibitemShut {NoStop}%
\bibitem [{\citenamefont {Malik}\ \emph
  {et~al.}(2022{\natexlab{a}})\citenamefont {Malik}, \citenamefont {Ferreira},
  \citenamefont {Agrawal},\ and\ \citenamefont {Providência}}]{Malik_2022}%
  \BibitemOpen
  \bibfield  {author} {\bibinfo {author} {\bibfnamefont {T.}~\bibnamefont
  {Malik}}, \bibinfo {author} {\bibfnamefont {M.}~\bibnamefont {Ferreira}},
  \bibinfo {author} {\bibfnamefont {B.~K.}\ \bibnamefont {Agrawal}}, \ and\
  \bibinfo {author} {\bibfnamefont {C.}~\bibnamefont {Providência}},\ }\href
  {\doibase 10.3847/1538-4357/ac5d3c} {\bibfield  {journal} {\bibinfo
  {journal} {The Astrophysical Journal}\ }\textbf {\bibinfo {volume} {930}},\
  \bibinfo {pages} {17} (\bibinfo {year} {2022}{\natexlab{a}})}\BibitemShut
  {NoStop}%
\bibitem [{\citenamefont {Providência}\ \emph {et~al.}(2024)\citenamefont
  {Providência}, \citenamefont {Malik}, \citenamefont {Albino},\ and\
  \citenamefont {Ferreira}}]{Provid_ncia_2024}%
  \BibitemOpen
  \bibfield  {author} {\bibinfo {author} {\bibfnamefont {C.}~\bibnamefont
  {Providência}}, \bibinfo {author} {\bibfnamefont {T.}~\bibnamefont {Malik}},
  \bibinfo {author} {\bibfnamefont {M.~B.}\ \bibnamefont {Albino}}, \ and\
  \bibinfo {author} {\bibfnamefont {M.}~\bibnamefont {Ferreira}},\ }\enquote
  {\bibinfo {title} {Relativistic description of the neutron star equation of
  state},}\ in\ \href {\doibase 10.1201/9781003306580-5} {\emph {\bibinfo
  {booktitle} {Nuclear Theory in the Age of Multimessenger Astronomy}}}\
  (\bibinfo  {publisher} {CRC Press},\ \bibinfo {year} {2024})\ p.\ \bibinfo
  {pages} {111–143}\BibitemShut {NoStop}%
\bibitem [{\citenamefont {Li}\ \emph {et~al.}(2025{\natexlab{a}})\citenamefont
  {Li}, \citenamefont {Tian},\ and\ \citenamefont {Sedrakian}}]{LI2025139501}%
  \BibitemOpen
  \bibfield  {author} {\bibinfo {author} {\bibfnamefont {J.-J.}\ \bibnamefont
  {Li}}, \bibinfo {author} {\bibfnamefont {Y.}~\bibnamefont {Tian}}, \ and\
  \bibinfo {author} {\bibfnamefont {A.}~\bibnamefont {Sedrakian}},\ }\href
  {\doibase https://doi.org/10.1016/j.physletb.2025.139501} {\bibfield
  {journal} {\bibinfo  {journal} {Physics Letters B}\ }\textbf {\bibinfo
  {volume} {865}},\ \bibinfo {pages} {139501} (\bibinfo {year}
  {2025}{\natexlab{a}})}\BibitemShut {NoStop}%
\bibitem [{\citenamefont {Fan}\ \emph {et~al.}(2024)\citenamefont {Fan},
  \citenamefont {Han}, \citenamefont {Jiang}, \citenamefont {Shao},\ and\
  \citenamefont {Tang}}]{PhysRevD.109.043052}%
  \BibitemOpen
  \bibfield  {author} {\bibinfo {author} {\bibfnamefont {Y.-Z.}\ \bibnamefont
  {Fan}}, \bibinfo {author} {\bibfnamefont {M.-Z.}\ \bibnamefont {Han}},
  \bibinfo {author} {\bibfnamefont {J.-L.}\ \bibnamefont {Jiang}}, \bibinfo
  {author} {\bibfnamefont {D.-S.}\ \bibnamefont {Shao}}, \ and\ \bibinfo
  {author} {\bibfnamefont {S.-P.}\ \bibnamefont {Tang}},\ }\href {\doibase
  10.1103/PhysRevD.109.043052} {\bibfield  {journal} {\bibinfo  {journal}
  {Phys. Rev. D}\ }\textbf {\bibinfo {volume} {109}},\ \bibinfo {pages}
  {043052} (\bibinfo {year} {2024})}\BibitemShut {NoStop}%
\bibitem [{\citenamefont {Tsang}\ \emph {et~al.}(2024)\citenamefont {Tsang},
  \citenamefont {Tsang}, \citenamefont {Lynch}, \citenamefont {Kumar},\ and\
  \citenamefont {Horowitz}}]{Tsang_2024}%
  \BibitemOpen
  \bibfield  {author} {\bibinfo {author} {\bibfnamefont {C.~Y.}\ \bibnamefont
  {Tsang}}, \bibinfo {author} {\bibfnamefont {M.~B.}\ \bibnamefont {Tsang}},
  \bibinfo {author} {\bibfnamefont {W.~G.}\ \bibnamefont {Lynch}}, \bibinfo
  {author} {\bibfnamefont {R.}~\bibnamefont {Kumar}}, \ and\ \bibinfo {author}
  {\bibfnamefont {C.~J.}\ \bibnamefont {Horowitz}},\ }\href {\doibase
  10.1038/s41550-023-02161-z} {\bibfield  {journal} {\bibinfo  {journal}
  {Nature Astronomy}\ }\textbf {\bibinfo {volume} {8}},\ \bibinfo {pages}
  {328–336} (\bibinfo {year} {2024})}\BibitemShut {NoStop}%
\bibitem [{\citenamefont {Malik}\ \emph
  {et~al.}(2022{\natexlab{b}})\citenamefont {Malik}, \citenamefont {Agrawal},\
  and\ \citenamefont {Providência}}]{Malik2022Inferring}%
  \BibitemOpen
  \bibfield  {author} {\bibinfo {author} {\bibfnamefont {T.}~\bibnamefont
  {Malik}}, \bibinfo {author} {\bibfnamefont {B.~K.}\ \bibnamefont {Agrawal}},
  \ and\ \bibinfo {author} {\bibfnamefont {C.}~\bibnamefont {Providência}},\
  }\href {http://dx.doi.org/10.1103/PhysRevC.106.L042801} {\bibfield  {journal}
  {\bibinfo  {journal} {Phys. Rev. C}\ }\textbf {\bibinfo {volume} {106}}
  (\bibinfo {year} {2022}{\natexlab{b}})}\BibitemShut {NoStop}%
\bibitem [{\citenamefont {Char}\ \emph {et~al.}(2023)\citenamefont {Char},
  \citenamefont {Mondal}, \citenamefont {Gulminelli},\ and\ \citenamefont
  {Oertel}}]{char2023generaliseddescriptionneutronstar}%
  \BibitemOpen
  \bibfield  {author} {\bibinfo {author} {\bibfnamefont {P.}~\bibnamefont
  {Char}}, \bibinfo {author} {\bibfnamefont {C.}~\bibnamefont {Mondal}},
  \bibinfo {author} {\bibfnamefont {F.}~\bibnamefont {Gulminelli}}, \ and\
  \bibinfo {author} {\bibfnamefont {M.}~\bibnamefont {Oertel}},\ }\href
  {\doibase 10.1103/PhysRevD.108.103045} {\bibfield  {journal} {\bibinfo
  {journal} {Phys. Rev. D}\ }\textbf {\bibinfo {volume} {108}},\ \bibinfo
  {pages} {103045} (\bibinfo {year} {2023})}\BibitemShut {NoStop}%
\bibitem [{\citenamefont {Li}\ and\ \citenamefont
  {Sedrakian}(2025)}]{li2025112}%
  \BibitemOpen
  \bibfield  {author} {\bibinfo {author} {\bibfnamefont {J.-J.}\ \bibnamefont
  {Li}}\ and\ \bibinfo {author} {\bibfnamefont {A.}~\bibnamefont {Sedrakian}},\
  }\href {\doibase 10.1103/c1k3-k4l5} {\bibfield  {journal} {\bibinfo
  {journal} {Phys. Rev. C}\ }\textbf {\bibinfo {volume} {112}},\ \bibinfo
  {pages} {015802} (\bibinfo {year} {2025})}\BibitemShut {NoStop}%
\bibitem [{\citenamefont {Li}\ \emph {et~al.}(2025{\natexlab{b}})\citenamefont
  {Li}, \citenamefont {Tian},\ and\ \citenamefont
  {Sedrakian}}]{PhysRevC.111.055804}%
  \BibitemOpen
  \bibfield  {author} {\bibinfo {author} {\bibfnamefont {J.-J.}\ \bibnamefont
  {Li}}, \bibinfo {author} {\bibfnamefont {Y.}~\bibnamefont {Tian}}, \ and\
  \bibinfo {author} {\bibfnamefont {A.}~\bibnamefont {Sedrakian}},\ }\href
  {\doibase 10.1103/PhysRevC.111.055804} {\bibfield  {journal} {\bibinfo
  {journal} {Phys. Rev. C}\ }\textbf {\bibinfo {volume} {111}},\ \bibinfo
  {pages} {055804} (\bibinfo {year} {2025}{\natexlab{b}})}\BibitemShut
  {NoStop}%
\bibitem [{\citenamefont {Huang}\ \emph
  {et~al.}(2024{\natexlab{d}})\citenamefont {Huang}, \citenamefont {Malik},
  \citenamefont {Cartaxo}, \citenamefont {Sourav}, \citenamefont {Yuan} \emph
  {et~al.}}]{huang2024compactobjectopensourcepythonpackage}%
  \BibitemOpen
  \bibfield  {author} {\bibinfo {author} {\bibfnamefont {C.}~\bibnamefont
  {Huang}}, \bibinfo {author} {\bibfnamefont {T.}~\bibnamefont {Malik}},
  \bibinfo {author} {\bibfnamefont {J.}~\bibnamefont {Cartaxo}}, \bibinfo
  {author} {\bibfnamefont {S.}~\bibnamefont {Sourav}}, \bibinfo {author}
  {\bibfnamefont {W.}~\bibnamefont {Yuan}},  \emph {et~al.},\ }\href
  {https://arxiv.org/abs/2411.14615} {\enquote {\bibinfo {title}
  {Compactobject: An open-source python package for full-scope neutron star
  equation of state inference},}\ } (\bibinfo {year} {2024}{\natexlab{d}}),\
  \Eprint {http://arxiv.org/abs/2411.14615} {arXiv:2411.14615 [astro-ph.HE]}
  \BibitemShut {NoStop}%
\bibitem [{\citenamefont {Vinciguerra}\ \emph {et~al.}(2024)\citenamefont
  {Vinciguerra}, \citenamefont {Salmi}, \citenamefont {Watts}, \citenamefont
  {Choudhury}, \citenamefont {Riley} \emph {et~al.}}]{Vinciguerra_2024}%
  \BibitemOpen
  \bibfield  {author} {\bibinfo {author} {\bibfnamefont {S.}~\bibnamefont
  {Vinciguerra}}, \bibinfo {author} {\bibfnamefont {T.}~\bibnamefont {Salmi}},
  \bibinfo {author} {\bibfnamefont {A.~L.}\ \bibnamefont {Watts}}, \bibinfo
  {author} {\bibfnamefont {D.}~\bibnamefont {Choudhury}}, \bibinfo {author}
  {\bibfnamefont {T.~E.}\ \bibnamefont {Riley}},  \emph {et~al.},\ }\href
  {\doibase 10.3847/1538-4357/acfb83} {\bibfield  {journal} {\bibinfo
  {journal} {The Astrophysical Journal}\ }\textbf {\bibinfo {volume} {961}},\
  \bibinfo {pages} {62} (\bibinfo {year} {2024})}\BibitemShut {NoStop}%
\bibitem [{\citenamefont {Salmi}\ \emph {et~al.}(2024)\citenamefont {Salmi},
  \citenamefont {Choudhury}, \citenamefont {Kini}, \citenamefont {Riley},
  \citenamefont {Vinciguerra} \emph {et~al.}}]{Salmi_2024}%
  \BibitemOpen
  \bibfield  {author} {\bibinfo {author} {\bibfnamefont {T.}~\bibnamefont
  {Salmi}}, \bibinfo {author} {\bibfnamefont {D.}~\bibnamefont {Choudhury}},
  \bibinfo {author} {\bibfnamefont {Y.}~\bibnamefont {Kini}}, \bibinfo {author}
  {\bibfnamefont {T.~E.}\ \bibnamefont {Riley}}, \bibinfo {author}
  {\bibfnamefont {S.}~\bibnamefont {Vinciguerra}},  \emph {et~al.},\ }\href
  {\doibase 10.3847/1538-4357/ad5f1f} {\bibfield  {journal} {\bibinfo
  {journal} {The Astrophysical Journal}\ }\textbf {\bibinfo {volume} {974}},\
  \bibinfo {pages} {294} (\bibinfo {year} {2024})}\BibitemShut {NoStop}%
\bibitem [{\citenamefont {Huang}\ \emph
  {et~al.}(2022{\natexlab{b}})\citenamefont {Huang}, \citenamefont {Hu},
  \citenamefont {Zhang},\ and\ \citenamefont {Shen}}]{Huang_2022}%
  \BibitemOpen
  \bibfield  {author} {\bibinfo {author} {\bibfnamefont {K.}~\bibnamefont
  {Huang}}, \bibinfo {author} {\bibfnamefont {J.}~\bibnamefont {Hu}}, \bibinfo
  {author} {\bibfnamefont {Y.}~\bibnamefont {Zhang}}, \ and\ \bibinfo {author}
  {\bibfnamefont {H.}~\bibnamefont {Shen}},\ }\href {\doibase
  10.3847/1538-4357/ac7f3c} {\bibfield  {journal} {\bibinfo  {journal} {The
  Astrophysical Journal}\ }\textbf {\bibinfo {volume} {935}},\ \bibinfo {pages}
  {88} (\bibinfo {year} {2022}{\natexlab{b}})}\BibitemShut {NoStop}%
\bibitem [{\citenamefont {Zhang}\ and\ \citenamefont {Li}(2019)}]{Zhang_2019}%
  \BibitemOpen
  \bibfield  {author} {\bibinfo {author} {\bibfnamefont {N.-B.}\ \bibnamefont
  {Zhang}}\ and\ \bibinfo {author} {\bibfnamefont {B.-A.}\ \bibnamefont {Li}},\
  }\href {\doibase 10.3847/1538-4357/ab24cb} {\bibfield  {journal} {\bibinfo
  {journal} {The Astrophysical Journal}\ }\textbf {\bibinfo {volume} {879}},\
  \bibinfo {pages} {99} (\bibinfo {year} {2019})}\BibitemShut {NoStop}%
\bibitem [{\citenamefont {Zhang}\ and\ \citenamefont
  {Li}(2020{\natexlab{a}})}]{Zhang_20200410}%
  \BibitemOpen
  \bibfield  {author} {\bibinfo {author} {\bibfnamefont {N.-B.}\ \bibnamefont
  {Zhang}}\ and\ \bibinfo {author} {\bibfnamefont {B.-A.}\ \bibnamefont {Li}},\
  }\href {\doibase 10.3847/1538-4357/ab7dbc} {\bibfield  {journal} {\bibinfo
  {journal} {The Astrophysical Journal}\ }\textbf {\bibinfo {volume} {893}},\
  \bibinfo {pages} {61} (\bibinfo {year} {2020}{\natexlab{a}})}\BibitemShut
  {NoStop}%
\bibitem [{\citenamefont {Zhang}\ and\ \citenamefont
  {Li}(2020{\natexlab{b}})}]{Zhang_20201010}%
  \BibitemOpen
  \bibfield  {author} {\bibinfo {author} {\bibfnamefont {N.-B.}\ \bibnamefont
  {Zhang}}\ and\ \bibinfo {author} {\bibfnamefont {B.-A.}\ \bibnamefont {Li}},\
  }\href {\doibase 10.3847/1538-4357/abb470} {\bibfield  {journal} {\bibinfo
  {journal} {The Astrophysical Journal}\ }\textbf {\bibinfo {volume} {902}},\
  \bibinfo {pages} {38} (\bibinfo {year} {2020}{\natexlab{b}})}\BibitemShut
  {NoStop}%
\bibitem [{\citenamefont {Zhang}\ and\ \citenamefont {Li}(2021)}]{Zhang_2021}%
  \BibitemOpen
  \bibfield  {author} {\bibinfo {author} {\bibfnamefont {N.-B.}\ \bibnamefont
  {Zhang}}\ and\ \bibinfo {author} {\bibfnamefont {B.-A.}\ \bibnamefont {Li}},\
  }\href {\doibase 10.3847/1538-4357/ac1e8c} {\bibfield  {journal} {\bibinfo
  {journal} {The Astrophysical Journal}\ }\textbf {\bibinfo {volume} {921}},\
  \bibinfo {pages} {111} (\bibinfo {year} {2021})}\BibitemShut {NoStop}%
\bibitem [{\citenamefont {Kubis}(2007)}]{kubis2007nuclear}%
  \BibitemOpen
  \bibfield  {author} {\bibinfo {author} {\bibfnamefont {S.}~\bibnamefont
  {Kubis}},\ }\href@noop {} {\bibfield  {journal} {\bibinfo  {journal} {Phys.
  Rev. C}\ }\textbf {\bibinfo {volume} {76}},\ \bibinfo {pages} {025801}
  (\bibinfo {year} {2007})}\BibitemShut {NoStop}%
\bibitem [{\citenamefont {Lattimer}\ and\ \citenamefont
  {Prakash}(2007)}]{lattimer2007neutron}%
  \BibitemOpen
  \bibfield  {author} {\bibinfo {author} {\bibfnamefont {J.~M.}\ \bibnamefont
  {Lattimer}}\ and\ \bibinfo {author} {\bibfnamefont {M.}~\bibnamefont
  {Prakash}},\ }\href@noop {} {\bibfield  {journal} {\bibinfo  {journal} {Phys.
  Rep.}\ }\textbf {\bibinfo {volume} {442}},\ \bibinfo {pages} {109} (\bibinfo
  {year} {2007})}\BibitemShut {NoStop}%
\bibitem [{\citenamefont {Douchin}\ and\ \citenamefont
  {Haensel}(2001)}]{douchin2001unified}%
  \BibitemOpen
  \bibfield  {author} {\bibinfo {author} {\bibfnamefont {F.}~\bibnamefont
  {Douchin}}\ and\ \bibinfo {author} {\bibfnamefont {P.}~\bibnamefont
  {Haensel}},\ }\href@noop {} {\bibfield  {journal} {\bibinfo  {journal}
  {Astronomy \& Astrophysics}\ }\textbf {\bibinfo {volume} {380}},\ \bibinfo
  {pages} {151} (\bibinfo {year} {2001})}\BibitemShut {NoStop}%
\bibitem [{\citenamefont {Tolman}(1939)}]{tolman1939static}%
  \BibitemOpen
  \bibfield  {author} {\bibinfo {author} {\bibfnamefont {R.~C.}\ \bibnamefont
  {Tolman}},\ }\href@noop {} {\bibfield  {journal} {\bibinfo  {journal} {Phys.
  Rev.}\ }\textbf {\bibinfo {volume} {55}},\ \bibinfo {pages} {364} (\bibinfo
  {year} {1939})}\BibitemShut {NoStop}%
\bibitem [{\citenamefont {Oppenheimer}\ and\ \citenamefont
  {Volkoff}(1939)}]{oppenheimer1939massive}%
  \BibitemOpen
  \bibfield  {author} {\bibinfo {author} {\bibfnamefont {J.~R.}\ \bibnamefont
  {Oppenheimer}}\ and\ \bibinfo {author} {\bibfnamefont {G.~M.}\ \bibnamefont
  {Volkoff}},\ }\href@noop {} {\bibfield  {journal} {\bibinfo  {journal} {Phys.
  Rev.}\ }\textbf {\bibinfo {volume} {55}},\ \bibinfo {pages} {374} (\bibinfo
  {year} {1939})}\BibitemShut {NoStop}%
\bibitem [{\citenamefont {Hinderer}\ \emph {et~al.}(2010)\citenamefont
  {Hinderer}, \citenamefont {Lackey}, \citenamefont {Lang},\ and\ \citenamefont
  {Read}}]{hinderer2010}%
  \BibitemOpen
  \bibfield  {author} {\bibinfo {author} {\bibfnamefont {T.}~\bibnamefont
  {Hinderer}}, \bibinfo {author} {\bibfnamefont {B.~D.}\ \bibnamefont
  {Lackey}}, \bibinfo {author} {\bibfnamefont {R.~N.}\ \bibnamefont {Lang}}, \
  and\ \bibinfo {author} {\bibfnamefont {J.~S.}\ \bibnamefont {Read}},\ }\href
  {\doibase 10.1103/PhysRevD.81.123016} {\bibfield  {journal} {\bibinfo
  {journal} {Phys. Rev. D}\ }\textbf {\bibinfo {volume} {81}},\ \bibinfo
  {pages} {123016} (\bibinfo {year} {2010})}\BibitemShut {NoStop}%
\bibitem [{\citenamefont {Read}\ \emph {et~al.}(2013)\citenamefont {Read},
  \citenamefont {Baiotti}, \citenamefont {Creighton}, \citenamefont {Friedman},
  \citenamefont {Giacomazzo} \emph {et~al.}}]{read2013}%
  \BibitemOpen
  \bibfield  {author} {\bibinfo {author} {\bibfnamefont {J.~S.}\ \bibnamefont
  {Read}}, \bibinfo {author} {\bibfnamefont {L.}~\bibnamefont {Baiotti}},
  \bibinfo {author} {\bibfnamefont {J.~D.~E.}\ \bibnamefont {Creighton}},
  \bibinfo {author} {\bibfnamefont {J.~L.}\ \bibnamefont {Friedman}}, \bibinfo
  {author} {\bibfnamefont {B.}~\bibnamefont {Giacomazzo}},  \emph {et~al.},\
  }\href {\doibase 10.1103/PhysRevD.88.044042} {\bibfield  {journal} {\bibinfo
  {journal} {Phys. Rev. D}\ }\textbf {\bibinfo {volume} {88}},\ \bibinfo
  {pages} {044042} (\bibinfo {year} {2013})}\BibitemShut {NoStop}%
\bibitem [{\citenamefont {Xie}\ and\ \citenamefont
  {Li}(2019)}]{xie2019bayesian}%
  \BibitemOpen
  \bibfield  {author} {\bibinfo {author} {\bibfnamefont {W.-J.}\ \bibnamefont
  {Xie}}\ and\ \bibinfo {author} {\bibfnamefont {B.-A.}\ \bibnamefont {Li}},\
  }\href@noop {} {\bibfield  {journal} {\bibinfo  {journal} {The Astrophysical
  Journal}\ }\textbf {\bibinfo {volume} {883}},\ \bibinfo {pages} {174}
  (\bibinfo {year} {2019})}\BibitemShut {NoStop}%
\bibitem [{\citenamefont {{Buchner}}(2021)}]{2021JOSS....6.3001B}%
  \BibitemOpen
  \bibfield  {author} {\bibinfo {author} {\bibfnamefont {J.}~\bibnamefont
  {{Buchner}}},\ }\href {\doibase 10.21105/joss.03001} {\bibfield  {journal}
  {\bibinfo  {journal} {The Journal of Open Source Software}\ }\textbf
  {\bibinfo {volume} {6}},\ \bibinfo {eid} {3001} (\bibinfo {year}
  {2021})}\BibitemShut {NoStop}%
\bibitem [{\citenamefont {Cai}\ and\ \citenamefont
  {Chen}(2017)}]{cai2017constraints}%
  \BibitemOpen
  \bibfield  {author} {\bibinfo {author} {\bibfnamefont {B.-J.}\ \bibnamefont
  {Cai}}\ and\ \bibinfo {author} {\bibfnamefont {L.-W.}\ \bibnamefont {Chen}},\
  }\href@noop {} {\bibfield  {journal} {\bibinfo  {journal} {Nuclear Science
  and Techniques}\ }\textbf {\bibinfo {volume} {28}},\ \bibinfo {pages} {185}
  (\bibinfo {year} {2017})}\BibitemShut {NoStop}%
\bibitem [{\citenamefont {Chen}(2011)}]{chen2011higher}%
  \BibitemOpen
  \bibfield  {author} {\bibinfo {author} {\bibfnamefont {L.}~\bibnamefont
  {Chen}},\ }\href@noop {} {\bibfield  {journal} {\bibinfo  {journal} {Science
  China Physics, Mechanics and Astronomy}\ }\textbf {\bibinfo {volume} {54}},\
  \bibinfo {pages} {124} (\bibinfo {year} {2011})}\BibitemShut {NoStop}%
\bibitem [{\citenamefont {Li}\ \emph {et~al.}(2019)\citenamefont {Li},
  \citenamefont {Krastev}, \citenamefont {Wen},\ and\ \citenamefont
  {Zhang}}]{li2019towards}%
  \BibitemOpen
  \bibfield  {author} {\bibinfo {author} {\bibfnamefont {B.-A.}\ \bibnamefont
  {Li}}, \bibinfo {author} {\bibfnamefont {P.~G.}\ \bibnamefont {Krastev}},
  \bibinfo {author} {\bibfnamefont {D.-H.}\ \bibnamefont {Wen}}, \ and\
  \bibinfo {author} {\bibfnamefont {N.-B.}\ \bibnamefont {Zhang}},\ }\href@noop
  {} {\bibfield  {journal} {\bibinfo  {journal} {The European Physical Journal
  A}\ }\textbf {\bibinfo {volume} {55}},\ \bibinfo {pages} {117} (\bibinfo
  {year} {2019})}\BibitemShut {NoStop}%
\bibitem [{\citenamefont {Li}\ \emph {et~al.}(2021)\citenamefont {Li},
  \citenamefont {Cai}, \citenamefont {Xie},\ and\ \citenamefont
  {Zhang}}]{universe7060182}%
  \BibitemOpen
  \bibfield  {author} {\bibinfo {author} {\bibfnamefont {B.-A.}\ \bibnamefont
  {Li}}, \bibinfo {author} {\bibfnamefont {B.-J.}\ \bibnamefont {Cai}},
  \bibinfo {author} {\bibfnamefont {W.-J.}\ \bibnamefont {Xie}}, \ and\
  \bibinfo {author} {\bibfnamefont {N.-B.}\ \bibnamefont {Zhang}},\ }\href@noop
  {} {\bibfield  {journal} {\bibinfo  {journal} {Universe}\ }\textbf {\bibinfo
  {volume} {7}},\ \bibinfo {pages} {182} (\bibinfo {year} {2021})}\BibitemShut
  {NoStop}%
\bibitem [{\citenamefont {Fujimoto}\ \emph {et~al.}(2022)\citenamefont
  {Fujimoto}, \citenamefont {Fukushima}, \citenamefont {McLerran},\ and\
  \citenamefont {Prasza{\l}owicz}}]{fujimoto2022trace}%
  \BibitemOpen
  \bibfield  {author} {\bibinfo {author} {\bibfnamefont {Y.}~\bibnamefont
  {Fujimoto}}, \bibinfo {author} {\bibfnamefont {K.}~\bibnamefont {Fukushima}},
  \bibinfo {author} {\bibfnamefont {L.~D.}\ \bibnamefont {McLerran}}, \ and\
  \bibinfo {author} {\bibfnamefont {M.}~\bibnamefont {Prasza{\l}owicz}},\
  }\href@noop {} {\bibfield  {journal} {\bibinfo  {journal} {Phys. Rev. Lett.}\
  }\textbf {\bibinfo {volume} {129}},\ \bibinfo {pages} {252702} (\bibinfo
  {year} {2022})}\BibitemShut {NoStop}%
\bibitem [{\citenamefont {Traversi}\ \emph {et~al.}(2020)\citenamefont
  {Traversi}, \citenamefont {Char},\ and\ \citenamefont
  {Pagliara}}]{Traversi_2020}%
  \BibitemOpen
  \bibfield  {author} {\bibinfo {author} {\bibfnamefont {S.}~\bibnamefont
  {Traversi}}, \bibinfo {author} {\bibfnamefont {P.}~\bibnamefont {Char}}, \
  and\ \bibinfo {author} {\bibfnamefont {G.}~\bibnamefont {Pagliara}},\ }\href
  {\doibase 10.3847/1538-4357/ab99c1} {\bibfield  {journal} {\bibinfo
  {journal} {The Astrophysical Journal}\ }\textbf {\bibinfo {volume} {897}},\
  \bibinfo {pages} {165} (\bibinfo {year} {2020})}\BibitemShut {NoStop}%
\bibitem [{\citenamefont {Altiparmak}\ \emph {et~al.}(2022)\citenamefont
  {Altiparmak}, \citenamefont {Ecker},\ and\ \citenamefont
  {Rezzolla}}]{Altiparmak_2022}%
  \BibitemOpen
  \bibfield  {author} {\bibinfo {author} {\bibfnamefont {S.}~\bibnamefont
  {Altiparmak}}, \bibinfo {author} {\bibfnamefont {C.}~\bibnamefont {Ecker}}, \
  and\ \bibinfo {author} {\bibfnamefont {L.}~\bibnamefont {Rezzolla}},\ }\href
  {\doibase 10.3847/2041-8213/ac9b2a} {\bibfield  {journal} {\bibinfo
  {journal} {The Astrophysical Journal Letters}\ }\textbf {\bibinfo {volume}
  {939}},\ \bibinfo {pages} {L34} (\bibinfo {year} {2022})}\BibitemShut
  {NoStop}%
\end{thebibliography}%
	
\end{document}